\documentclass[twocolumn,showpacs,pra,floatfix]{revtex4}
\usepackage{bm}
\usepackage{mathtools}  
\usepackage{amsmath,amssymb,graphicx,hyperref}
\usepackage{psfrag}
\usepackage{latexsym}
\usepackage{exscale}
\usepackage{color}
\usepackage{xcolor}
\usepackage{float}
\usepackage{ulem}
\usepackage{braket}
\usepackage{pdflscape}  % For landscape orientation
\usepackage{rotating}
\usepackage{afterpage}
\usepackage[font=footnotesize]{caption} % reducing fonts and spaces of captions
\begin{document}
% ============================================
% REDUCE SPACING ABOVE AND BELOW ALL EQUATIONS
% ============================================

% Method 1: Reduce display skips globally
\setlength{\abovedisplayskip}{6pt plus 2pt minus 2pt}
\setlength{\belowdisplayskip}{6pt plus 2pt minus 2pt}
\setlength{\abovedisplayshortskip}{0pt plus 1pt}
\setlength{\belowdisplayshortskip}{3pt plus 1pt minus 1pt}
%
%\Large
%\bf
%\boldmath{}
%
\title{Unified Phase-Space Mapping of Quantum Observables in a Multi-Driven Vapor: Resonance Fluorescence as an Electrometry Probe and Correlation Witness}
\author{M. S. Ateto$^{1}$ and Emad K. Jaradat$^{2}$ and S. M. Abo-Dahab$^{1}$} 
\affiliation{$^1$Mathematics Department, 
	Faculty of Science at Qena, South Valley University, 83523 Qena, Egypt
	\\
	mohamed.ali11@sci.svu.edu.eg; mohammed.ateto@gmail.com
	\\
	elsayed.khodeari@sci.svu.edu.eg
	\\
	$^2$Department of physics, Faculty of Science, Imam Mohammad Ibn Saud Islamic University (IMSIU), Riyadh 11623
	\\
	ekyjaradat@imamu.edu.sa
}
%\date{\today}

\begin{abstract}
We present a unified, geometry-resolved framework analyzing absorption, resonance fluorescence, entanglement negativity, and phase-space quasiprobability in Doppler-broadened four-level atomic vapor. Within a density-matrix formalism incorporating thermal velocity averaging and exact dressed-state diagonalization, these diverse observables are shown to constitute complementary projections of a common coherence-driven phase-space structure governed by multi-photon interference. Central to this framework, we introduce the Stratonovich-Weyl (SW) Wigner function as a unified phase-space representation that encodes both population and coherence contributions on equal footing. A direct comparison reveals a near one-to-one correspondence between SW quasiprobability structures and entanglement landscapes, with geometry-dependent features -- such as exact hyperbolic dispersion asymptotes and split resonance ridges -- faithfully preserved across representations. Crucially, isolating the Doppler-odd component yields an eigenvalue-free proxy that captures the exact coherence geometry underlying entanglement negativity, acting as a non-invasive quantum correlation witness. Concurrently, resonance fluorescence emerges as a thermally robust, population-driven observable: unlike fragile susceptibility-based absorption, its additive pole weighting preserves sharp Autler-Townes (AT) spectral profiles under intense driving. This geometric resilience allows fluorescence to reliably track uncoupled bright dressed states and entanglement extrema under severe Doppler dephasing, establishing its dual role as a high-sensitivity electrometry probe and a non-local correlation witness. By combining a coherence-resolved description with a Doppler-sensitive phase-space representation, this framework provides an experimentally accessible, eigenvalue-free route for state characterization, enabling Doppler-resilient, fluorescence-based quantum sensing and precise field electrometry in warm atomic media.
\end{abstract}

%\begin{document}

\maketitle

\section{Introduction}
\label{sec.1}

Room-temperature vapor-cell platforms are attractive for practical quantum technologies because they avoid the complexity of laser cooling and ultra-high-vacuum systems. In recent years, thermal atomic vapors have been widely employed in atom-based sensing and electrometry, where spectroscopic interrogation of atomic states provides precise and transferable measurement standards \cite{HSGDA17,GSPH22,SSKLP12}. A major challenge in such systems is Doppler broadening arising from the thermal velocity distribution of atoms \cite{RRPBM23}. Doppler-induced frequency shifts broaden spectral lines, introduce decoherence, and weaken quantum interference effects that are essential for high-resolution spectroscopy and Rydberg-based sensing \cite{BNMGB23,KTWZZ07,VIAJ14,NAWA25,BHJHZJ17}. Consequently, realistic modeling of thermal Rydberg media must explicitly account for Doppler averaging.

Considerable efforts have been devoted to suppressing Doppler effects through laser cooling, wavelength matching, optimized propagation geometries, velocity-selective excitation, and Doppler-free spectroscopic techniques \cite{MEST99,OOKAST04,SAST76,PBRBSA23,PBRAS24,FLIM05,LKAHL11,VIAJ14,NAWA25,RRPBM23,XXMYY24,THOMSO85}. Although these approaches can significantly reduce thermal broadening, Doppler effects remain an important limitation in multi-photon Rydberg excitation schemes and practical room-temperature sensors.

Electromagnetically induced transparency (EIT) has become a cornerstone of Rydberg spectroscopy and electrometry \cite{MOJA07}. Originating from quantum interference between competing excitation pathways, EIT enables strong modifications of absorption and dispersion and has been extensively investigated in both three-level and multi-level systems \cite{OLMA08,LKAHL11,KTWZZ07,VINA15,VIAJNA16,CHLIHE18,BBBN20,HGJSA14,HSKH18}. Such systems exhibit a rich variety of coherent phenomena, including EIT, electromagnetically induced absorption (EIA), Autler--Townes (AT) splitting, and velocity-selective resonances \cite{SCZU97,FICSWA05}. These effects underpin numerous applications ranging from slow light and quantum memory to precision metrology and quantum information processing \cite{HHDB99,MASLTA17,VANIER05,KITCHING18,BMRS04,LISLTA17}. Moreover, EIT-based protocols can generate and manipulate entanglement, providing an important resource for quantum communication networks and repeater architectures \cite{DLCZ01,LAFI17,TSSRD19}.

Despite its success, EIT-based electrometry faces practical challenges, including sensitivity to polarization alignment, photon shot noise, and degradation arising from Rydberg-state decay and dephasing \cite{CLHY00,CLCY02,SSKLP12,MANG05,KFKJS17}. These limitations motivate the development of alternative observables that remain robust under thermal conditions.

Fluorescence-based detection has recently emerged as a powerful alternative to conventional transmission measurements in Rydberg electrometry \cite{GALL94,FAUBPIC19,RAGPIFCB19,ZLLXJWMW25,ATETO26}. Unlike absorption, fluorescence directly probes spontaneous emission from dressed atomic states and therefore carries direct information about the underlying coherence dynamics. This provides enhanced sensitivity to weak electromagnetic fields, broad dynamic range extending into the strong-field regime, and access to spatially resolved field distributions, as demonstrated in electromagnetic visible imaging and fluorescence thermography \cite{FAUBPIC19,RAGPIFCB19,ZLLXJWMW25}.

In Doppler-broadened media, fluorescence exhibits behavior that differs qualitatively from conventional absorption observables. While lower-level systems often display strong directional and anisotropic responses \cite{WAGA97,PPRF00}, appropriately engineered four-level configurations can preserve narrow resonances and coherent signatures even at room temperature \cite{NSKVR04}. Furthermore, Doppler averaging can reshape coherent structures such as Autler--Townes splitting and induce nontrivial spectral redistribution \cite{SOALVI22}, suggesting that thermal motion can serve not only as a source of degradation but also as a mechanism for controlling spectroscopic response.

From a fundamental perspective, quantum sensing relies on coherence and entanglement as key resources for enhanced measurement precision \cite{SSRSS24}. In driven multi-level systems, the density-matrix coherences governing absorption and fluorescence simultaneously generate quantum correlations between effective internal-state subspaces. Consequently, spectral signatures such as transparency windows, linewidth narrowing, and dressed-state splittings can be interpreted as manifestations of the same coherence-driven processes responsible for entanglement generation \cite{FICSWA05,YUEBER06,ATETO26,AGARWAL06,ZHUSCU96,FLIM05}. Fluorescence dynamics and entanglement therefore represent complementary manifestations of a common underlying quantum-interference mechanism.

A particularly useful framework for visualizing these coherence effects is provided by the Stratonovich--Weyl (SW) Wigner function \cite{STRATONOCICH57,TIEVSMN16,RUNDLE17,ATETO26}. As a phase-space representation of the density operator, the SW function treats populations and coherences on equal footing and can exhibit negative quasiprobability regions that signal nonclassical behavior \cite{WEYL27,HICOSCWI84,KENZYC04,FERRIE11}. Recent studies have further linked Wigner negativity to quantum correlations and entanglement resources \cite{ALGEPAFE18,KWTAVOJE19,XLGGTHW22,ATETJARDAH26}. Consequently, phase-space structure offers a physically transparent picture of the coherence processes that simultaneously govern fluorescence, absorption, and entanglement.

Although fluorescence spectroscopy, entanglement measures, and phase-space quasiprobability distributions have each been studied extensively, a unified description connecting these observables in Doppler-broadened Rydberg systems remains largely unexplored. In particular, the relationship between fluorescence signatures, entanglement generation, and SW phase-space structure under realistic propagation geometries and thermal averaging has not been systematically established.

Building on this motivation, the present study develops a unified framework linking absorption, fluorescence, entanglement, and phase-space quasiprobability through their common dependence on dressed-state coherences in a Doppler-broadened four-level V+$\Theta$ system. By incorporating velocity averaging, propagation geometry, wavelength mismatch, and strong-field dressing, these observables are shown to represent complementary projections of a shared coherence structure. Within this framework, fluorescence serves as a robust experimentally accessible probe, while the SW Wigner function provides a phase-space reference for interpreting and calibrating entanglement-related features.

To place fluorescence-based electrometry within a broader and experimentally relevant context, we examine absorption, fluorescence, and entanglement under several realistic control-field configurations. Particular attention is devoted to weak and strong control-field regimes, counter-propagating geometries for Doppler compensation, and matched or mismatched wavelength conditions. These experimentally accessible scenarios provide practical routes for manipulating Doppler response and optimizing coherence-based sensing in thermal Rydberg media.

The remainder of this paper is organized as follows. Section~(\ref{sec.2}) outlines the experimental configuration and excitation scheme of the four-level V+$\Theta$ system. Section~(\ref{sec.3}) develops the density-matrix formalism. Analytical derivations of the absorption and fluorescence spectra are presented in Appendices~(\ref{app:A}) and~(\ref{app:B}), respectively. Section~(\ref{sec.5}) introduces the negativity measure of entanglement under Doppler averaging, while Appendix~(\ref{app:C}) derives the corresponding Stratonovich--Weyl Wigner representation. Section~(\ref{sec.6}) presents a detailed analysis of Doppler-resolved and Doppler-averaged observables under different control parameters and propagation geometries. Section~(\ref{sec.7_2}) introduces the Doppler-odd projection of the SW function as an eigenvalue-free entanglement proxy and establishes its validity through analytical and numerical arguments. Finally, Secs.~(\ref{sec.7_3}) and~(\ref{sec.8}) summarize the principal findings and conclusions.

\section{Experimental Setup and Excitation Path}
\label{sec.2}

\subsection{Proposed experimental implementation}

\begin{figure}[H]
    \centering
    \includegraphics[width=\columnwidth]{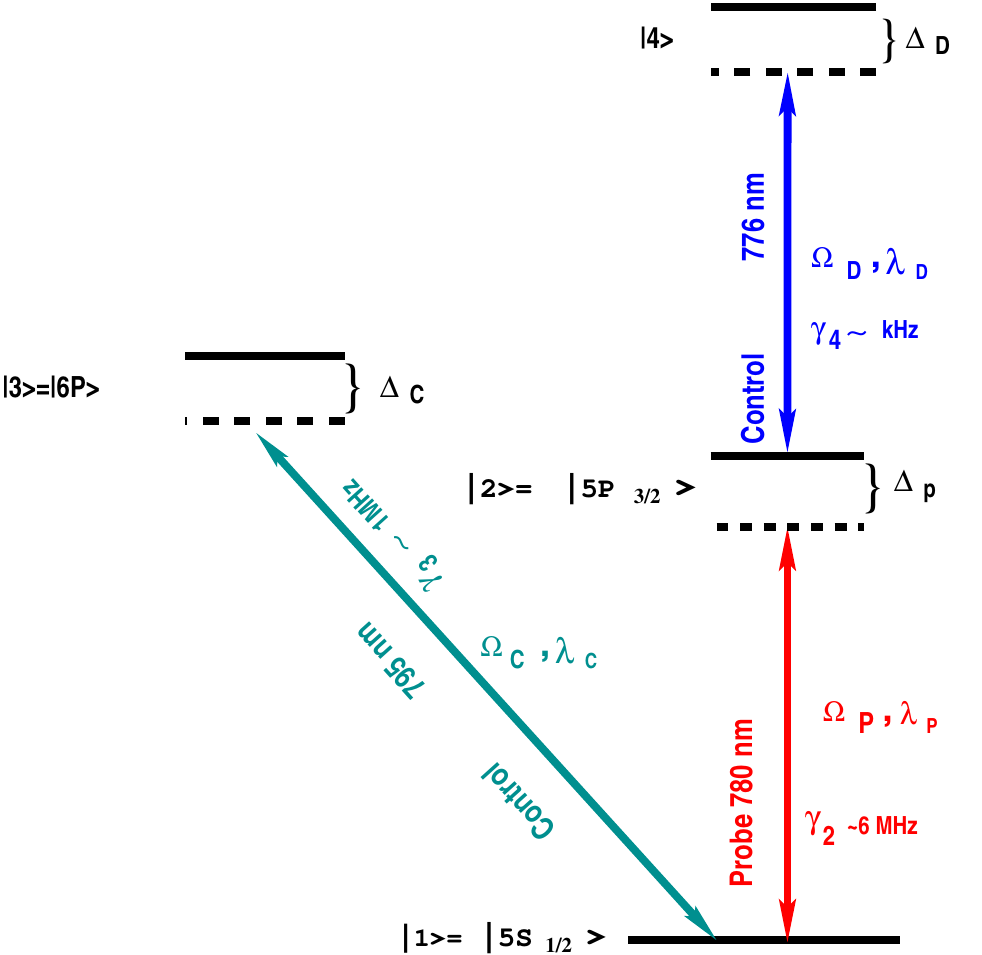}
    \caption{The four-level V+$\Theta$ system under consideration.}
    \label{scheme}
\end{figure}

The proposed scheme can be implemented in Doppler-broadened alkali vapors such as $^{85}$Rb, $^{87}$Rb, or $^{133}$Cs, using a closed four-level configuration that preserves the essential physics while incorporating experimentally established control techniques to enhance fluorescence sensitivity and coherence visibility. The level structure realizes a hybrid V+ladder ($\Theta$) configuration~\cite{VINA15,VIAJNA16,BBBN20}, where a weak probe field at $\lambda_P=780\,\mathrm{nm}$ drives $|1\rangle\equiv|5S_{1/2},F=1\rangle \leftrightarrow |2\rangle\equiv|5P_{3/2},F=2\rangle$, while two stronger coupling fields at $\lambda_C=795\,\mathrm{nm}$ and $\lambda_D=776\,\mathrm{nm}$ couple $|1\rangle\rightarrow|3\rangle\equiv|5P_{1/2},F=2\rangle$ and $|2\rangle\rightarrow|4\rangle$, respectively (Fig.~\ref{scheme}). The Rydberg state $\ket{4}$ may be chosen as $nS_{1/2}$(for $^{87}$Rb) or $nD_{3/2,5/2}$(for both isotopes), with principal quantum numbers $n\sim 26\text{--}124$ \cite{PMGWJA10, PEJOMI11, PFLHGPL12, HGSRGEW13}. On the relevant energy scales, hyperfine splittings in Rydberg levels are negligible. This establishes a dressed four-level manifold in which the simultaneous action of probe and control fields enables fluorescence-based probing of coherence and entanglement under Doppler broadening.

The vapor is operated at densities below $10^{10}\,\mathrm{cm^{-3}}$ to suppress collisional dephasing~\cite{NVRN16}, while laser frequencies are stabilized via sideband locking to a high-finesse Fabry-P\'{e}rot cavity, achieving sub-10-kHz linewidths and stable two-photon resonance conditions~\cite{DREVER83,NTHS17}. The relevant decay rates,
\footnotesize
\[
\gamma_1 = 0,\quad 
\gamma_2 = 2\pi\times6.1~\mathrm{MHz},\quad
\gamma_3 = 2\pi\times5.9~\mathrm{MHz},\quad
\gamma_4 = 2\pi\times0.68~\mathrm{MHz},
\]
\normalsize
are consistent with those used in Sec.~\ref{sec.6}.

% Fixed figure placement - uses [H] to place exactly here

Operation relies on preparing a well-defined initial population in $|1\rangle$ via polarization-selective optical pumping with weak repumping to empty other hyperfine levels. This suppresses incoherent background and enhances the contrast of interference-induced fluorescence features, which can be verified through baseline fluorescence prior to activating strong fields~\cite{HARRAI06}. The system is inherently Doppler broadened~\cite{SSKLP12,COHEN98,KLBTT18}; however, arranging the probe and at least one coupling beam in a counter-propagating geometry enables partial velocity selection and reduces the effective two-photon Doppler width. Additional optimization -- such as minimizing wavelength mismatch, using narrow-linewidth lasers, and operating in a moderate intensity regime -- preserves coherence while limiting power broadening. As in the theoretical analysis, these measures do not remove Doppler effects but isolate velocity-selected subensembles where interference remains observable.

Fluorescence detection provides a direct experimental probe of the dressed-state dynamics. In this configuration, the dominant signal originates from spontaneous decay of $|3\rangle$, whose population is indirectly controlled by both the V- and ladder excitation pathways. Although the upper state $|4\rangle$ is weakly radiative, its coupling modifies population flow and coherence, which are reflected in the steady-state occupation of $|3\rangle$. The emitted fluorescence on the $|3\rangle\rightarrow|1\rangle$ transition thus encodes the underlying dressed-state structure, with spectral features -- such as linewidth narrowing, asymmetric splitting, and transparency windows -- serving as signatures of coherence and entanglement. Efficient photon collection using high-numerical-aperture optics, combined with spectral and polarization filtering, enhances signal-to-noise ratio, while interference-based detection schemes (e.g., heterodyne or Mach-Zehnder) can further suppress technical noise~\cite{HGJSA14,Lukin2001,KFKSS16,KFKJS17,FLIM05}.

Sensitivity to off-diagonal density-matrix elements can be further enhanced through controlled phase modulation of a coupling field, allowing fluorescence spectra to probe phase-dependent coherences directly. Weak repumping fields may be employed to prevent population trapping in long-lived states and maintain steady-state conditions.

While conceptually related to fluorescence-based readout in trapped-ion systems, the present implementation operates in a thermal, Doppler-broadened regime where observables are ensemble-averaged. Fluorescence therefore acts not as a projective measurement but as a spectroscopic witness of coherence and entanglement distributed across velocity classes. In this sense, the protocol extends precision quantum-control techniques to realistic vapor systems, consistent with the theoretical framework developed above: optical pumping prepares the initial state, phase-stable multi-field driving establishes the dressed manifold, velocity selection controls Doppler effects, and fluorescence readout maps the resulting coherence and entanglement structure into experimentally accessible spectra.

\section{System and Theoretical Model}
\label{sec.3}

The atomic configuration under consideration is a four-level system in the $(V+\Theta)$ configuration, schematically illustrated in Fig.~(\ref{scheme}). It combines a V-type subsystem formed by the ground and two intermediate states with a $\Theta$-type subsystem connecting the intermediate and Rydberg states. This configuration supports both electromagnetically induced absorption (EIA) and electromagnetically induced transparency (EIT) pathways and provides a suitable platform for investigating fluorescence, entanglement, and Doppler-induced coherence effects.

The atom--field interaction Hamiltonian is written as

\begin{equation}
\label{DF1}
H=H_0+H_{int}(t),
\end{equation}

where

\begin{equation}
\label{DF2}
H_0=\sum_{i=1}^{4} \omega_i|i\rangle\langle i|,
\end{equation}

is the unperturbed Hamiltonian in the bare-state basis
\(
|1\rangle,|2\rangle,|3\rangle,|4\rangle
\).
The interaction Hamiltonian in the Schr\"odinger picture is
\small
\begin{equation}
\label{DF3}
H_{int}=\hbar\big(\Omega_P e^{i\omega_Pt}|1\rangle\langle 2|
+\Omega_C e^{i\omega_Ct}|1\rangle\langle 3|
+\Omega_D e^{i\omega_Dt}|2\rangle\langle 4|
+h.c.\big),
\end{equation}

where
\(
\Omega_P
\),
\(
\Omega_C
\),
and
\(
\Omega_D
\)
denote the probe, first-control, and second-control Rabi frequencies, respectively. They are related to the electric-field amplitudes and transition dipole moments through
\(
\Omega_j=\mu_{ij}E_j/2\hbar
\).

After transforming to the interaction picture and applying the rotating-wave approximation (RWA), the Hamiltonian becomes
\small
\begin{equation}
\label{DF4}
H= \frac{\hbar}{2}
\left[
\begin{array}{cccc}
0& \Omega_P & \Omega_C & 0\\
\Omega^{\ast}_P& A & 0 & \Omega_D\\
\Omega^{\ast}_C& 0 & B & 0\\
0& \Omega^{\ast}_D & 0 & C
\end{array}
\right],
\end{equation}

where

\begin{equation}
\label{DF5}
\begin{gathered}
A=2\Delta_P,
\\
B=2\Delta_C,
\\
C=2(\Delta_P+\Delta_D).
\end{gathered}
\end{equation}

The diagonal terms represent the effective detunings in the rotating frame. In particular, excitation of the Rydberg state
\(
|4\rangle
\)
occurs through the sequential pathway
\(
|1\rangle\rightarrow|2\rangle\rightarrow|4\rangle
\),
giving rise to the effective two-photon detuning
\(
\Delta_{2ph}=\Delta_P+\Delta_D
\).
Consequently,
\(
C=2(\Delta_P+\Delta_D)
\)
governs the resonance condition of the upper transition and plays a central role in the interference and Doppler-sensitive features discussed later.

The detunings are defined as

\begin{equation}
\label{DF6}
\Delta_{P,C,D}
=
\omega_{21,31,42}
-
\omega_{P,C,D},
\end{equation}

where
\(
\omega_{ij}
\)
are the atomic transition frequencies and
\(
\omega_P,\omega_C,\omega_D
\)
are the laser frequencies. The principal resonance conditions are therefore
\(
\Delta_P=\Delta_C=\Delta_D=0
\)
and the two-photon resonance
\(
\Delta_P+\Delta_D=0
\).
Under Doppler broadening these conditions become velocity dependent,
\(
\Delta_j(v)=\Delta_j\oslash k_jv~ (\oslash ~\text{discusses the sign of propagation geometry})
\),
leading to velocity-selective interference and modified multiphoton resonances.

Dissipation is incorporated through the Gorini--Kossakowski--Sudarshan--Lindblad master equation

\begin{equation}
\label{DF7}
\frac{d\rho}{dt}
=
-\frac{i}{\hbar}[H,\rho]
+
\sum_\mu
\left(
L_\mu \rho L_\mu^\dagger
-\frac12
\{
L_\mu^\dagger L_\mu,\rho
\}
\right),
\end{equation}

where
\(
L_\mu
\)
are collapse operators describing spontaneous emission and dephasing processes.
For the present system the dominant decay channels are
\(
L_{21}=\sqrt{\Gamma_2}|1\rangle\langle2|
\),
\(
L_{31}=\sqrt{\Gamma_3}|1\rangle\langle3|
\),
and
\(
L_{42}=\sqrt{\Gamma_4}|2\rangle\langle4|
\).
Substitution into Eq.~(\ref{DF7}) yields the relaxation matrix
%\footnotesize
\small
\begin{equation}
\label{DF8}
\mathcal L(\rho)
=
\frac12
\begin{pmatrix}
2(\Gamma_2\rho_{22}
+
\Gamma_3\rho_{33})
&
-\Gamma_2\rho_{12}
&
-\Gamma_3\rho_{13}
&
-\Gamma_4\rho_{14}
\\
-\Gamma_2\rho_{21}
&
2(\Gamma_4\rho_{44}
-
\Gamma_2\rho_{22})
&
-\gamma_{23}\rho_{23}
&
-\gamma_{24}\rho_{24}
\\
-\Gamma_3\rho_{31}
&
-\gamma_{32}\rho_{32}
&
-2\Gamma_3\rho_{33}
&
-\gamma_{34}\rho_{34}
\\
-\Gamma_4\rho_{41}
&
-\gamma_{42}\rho_{42}
&
-\gamma_{43}\rho_{43}
&
-2\Gamma_4\rho_{44}
\end{pmatrix},
\end{equation}
%\normalsize
%
where
\(
\gamma_{ij}
=
(\Gamma_i+\Gamma_j)/2
\)
and
\(
\Gamma_i
\)
denotes the population decay rate from level
\(
|i\rangle
\).

For simplicity, spontaneous decay from the Rydberg state is assumed to occur predominantly through the channel
\(
|4\rangle\rightarrow|2\rangle
\).
Additional cascade and blackbody-induced transitions are neglected, as their inclusion would only introduce extra Lindblad terms without altering the general theoretical framework.

Expanding Eq.~(\ref{DF7}) in the atomic basis yields the coupled optical Bloch equations
\allowdisplaybreaks
%
%\footnotesize
\small
\begin{align}
\label{DF9}
\dot{\rho}_{11}&=\Gamma_2 \rho_{22}+\Gamma_3 \rho_{33}+\frac{i}{2}\big(\Omega^{\ast}_C
\rho_{13}-\Omega_C \rho_{31}\big)+\frac{i}{2}\big(\Omega^{\ast}_P
\rho_{12}-\Omega_P \rho_{21}\big), \\
\label{DF10}
\dot{\rho}_{22}&=-\Gamma_2 \rho_{22}+\Gamma_4 \rho_{44}+\frac{i}{2}\big(\Omega_P
\rho_{21}-\Omega^{\ast}_P \rho_{12}\big)+\frac{i}{2}\big(\Omega^{\ast}_D
\rho_{24}-\Omega_D \rho_{42}\big),\\
\label{DF11}
\dot{\rho}_{33}&=-\Gamma_3 \rho_{33}+\frac{i}{2}\big(\Omega_C \rho_{31}-\Omega^{\ast}_C \rho_{13}\big),\\
\label{DF12}
\dot{\rho}_{44}&=-\Gamma_4 \rho_{44}+\frac{i}{2}\big(\Omega_D \rho_{42}-\Omega^{\ast}_D \rho_{24}\big),\\
\label{DF13}
\dot{\rho}_{12}&=\varLambda_{12}\rho_{12}-\frac{i}{2}\big(\Omega_C 
\rho_{32}-\Omega^{\ast}_D \rho_{14}\big)+\frac{i}{2}\Omega_{P} \big(\rho_{11}-\rho_{22}\big)= \dot{\rho}_{21}^{\ast},\\
\label{DF14}
\dot{\rho}_{13}&=\varLambda_{13} \rho_{13}+\frac{i}{2}\Omega_C \big(\rho_{11}- \rho_{33}\big)
-\frac{i}{2}\Omega_P \rho_{23}
= \dot{\rho}_{31}^{\ast},\\
\label{DF15}
\dot{\rho}_{14}&=\varLambda_{14}\rho_{14}+\frac{i}{2}\big(\Omega_D \rho_{12}-\Omega_P \rho_{24}
-\Omega_C \rho_{34}\big)=\dot{\rho}_{41}^{\ast},\\
\label{DF16}
\dot{\rho}_{23}&=\varLambda_{23}\rho_{23}+\frac{i}{2}\big(\Omega_C
\rho_{21}-\Omega^{\ast}_{P} \rho_{13}-\Omega_D \rho_{43}\big)= \dot{\rho}_{32}^{\ast},\\
\label{DF17}
\dot{\rho}_{24}&=\varLambda_{24} \rho_{24}+\frac{i}{2}\Omega_D \big(\rho_{22}- \rho_{44}\big)
-\frac{i}{2}\Omega^{\ast}_P \rho_{14}= \dot{\rho}_{42}^{\ast},\\
\label{DF18}
\dot{\rho}_{34}&=\varLambda_{34} \rho_{34}+\frac{i}{2}\big(\Omega_D \rho_{32}-\Omega^{\ast}_C \rho_{14}\big)= \dot{\rho}_{43}^{\ast},
\end{align}
whereas star on the superscript denote the complex conjugation, and the complex coefficients $\varLambda_{ij}$ encapsulate the 
decay rates and detuning frequencies as 
%\footnotesize
\small
\begin{equation}
\label{DF19}
\begin{gathered}
\varLambda_{12}=\frac{1}{2}\big(-\Gamma_2+iA\big), 
\varLambda_{13}=\frac{1}{2}\big(-\Gamma_3+iB\big), 
\varLambda_{14}=\frac{1}{2}\big(-\Gamma_4+iC\big), 
\\
\varLambda_{23}=\frac{1}{2}\big(-\gamma_{23}+i[B - A]\big),  
\varLambda_{24}= \frac{1}{2}\big(-\gamma_{24}+i[C - A]\big), 
\\
\varLambda_{34}=\frac{1}{2}\big(-\gamma_{34}+i[C - B]\big).
\end{gathered}
\end{equation}
%\normalsize

\section{Populations and coherences}
\label{Sec.3_1}

The steady-state populations and coherences relevant to the weak-probe regime are derived in Appendix~\ref{app:A}. The resulting probe coherence, which determines the linear susceptibility through
\(
\chi\propto\rho_{12}^{\mathrm{ss}},
\)
is given by Eq.~(\ref{DF28}) and forms the basis for the absorption analysis.

In a thermal vapor, atomic motion along the laser propagation direction introduces velocity-dependent detunings according to \(\Delta_{P,C,D}\rightarrow\Delta_{P,C,D}(v)=\Delta_{P,C,D}\oslash k_{P,C,D}v,\) where \(k_{P,C,D}=2\pi/\lambda_{P,C,D}\) are the corresponding wave vectors and \(\oslash\) indicates the propagation geometry. Accordingly, a frequency shift (Doppler shift) enter each detuning with a sign determined by the beam direction \cite{RRPBM23}. As a result, all density-matrix elements become velocity dependent. The experimentally observable susceptibility is therefore obtained by averaging the steady-state susceptibility $\chi$ [\ref{DF31}] over the Maxwell--Boltzmann velocity distribution  \cite{BEMA11, RRPBM23},

\begin{equation}
\label{DF33}
\chi=
\frac{2N|\mu_{12}|^2}
{\epsilon_0\hbar\Omega_P}
\rho^{\mathrm{ss}}_{12_{Dop}}
(\Delta_P,\Delta_C,\Delta_D),
\end{equation}

where  \cite{MESCHEDE07, BEMA11, NOMO11, RRPBM23, SSRSS24}

\begin{equation}
\label{DF34}
\rho^{\mathrm{ss}}_{12_{Dop}}
=
\frac{1}{\sqrt{\pi}u}
\int_{-\infty}^{\infty}
\rho_{12}^{\mathrm{ss}}(v)
e^{-v^2/u^2}
dv.
\end{equation}
Here \(u=\sqrt{2k_B T/m}\) denotes the most probable speed of the Maxwell--Boltzmann distribution \cite{JLHBRM21}, with \(T\) the vapor temperature, \(m\) the atomic mass, and \(k_B\) Boltzmann's constant. The weighting function \(\frac{1}{\sqrt{\pi}u} e^{-v^2/u^2}\) represents the normalized velocity distribution \cite{SANAHU96, JLHBRM21}. Equation~(\ref{DF34}) establishes the connection between microscopic velocity-dependent coherence and the macroscopic optical response measured experimentally.

\section{Resonance Fluorescence Spectrum}
\label{sec.4}
Following the same weak-probe methodology applied to the populations and coherences, the resonance fluorescence spectrum can be derived using the quantum regression theorem. The full derivation is provided in Appendix~\ref{app:B}, resulting in an expression, Eq.~\eqref{DF40}, that links the fluorescence spectrum $S(\omega)$ directly to the same steady-state coherences that govern absorption. Already, Eq. (\ref{DF40}) addresses the Doppler-free power spectrum. In moving atomic ensembles, however, the fluorescence becomes velocity-dependent. Straightforwardly as in susceptibility, the thermally averaged spectrum \cite{WAGA97} reads
\begin{equation}
\label{DF44}
S_{Dop}(\omega, \Delta_P, \Delta_C, \Delta_D)=\frac{1}{\sqrt{\pi} u\Omega_P} \int_{-\infty}^{\infty}S(\omega, v) ~ e^{-v^2/u^2} dv, 
\end{equation}
by which, the total fluorescence arises as a convolution of the microscopic two-time coherence spectrum with the Maxwell-Boltzmann velocity distribution~\cite{BEMA11}. The Doppler-dependent power spectrum is given by $S(\omega, v)$.

\section{Negativity as an Effective Entanglement Quantifier}
\label{sec.5}

The driven four-level V+$\Theta$ system develops nonclassical correlations through coherent laser coupling and dissipative dynamics. To quantify these correlations, we introduce an effective bipartite decomposition of the four-dimensional Hilbert space, \(\mathcal H_4\simeq\mathcal H_A\otimes\mathcal H_B\), where each subsystem is represented by an effective qubit. The bare atomic states are mapped onto the computational basis according to \(|1\rangle\rightarrow|00\rangle, \quad |2\rangle\rightarrow|01\rangle, \quad |3\rangle\rightarrow|10\rangle, \quad |4\rangle\rightarrow|11\rangle\).

Under this isomorphism, the steady-state density matrix can be treated as an effective
\(2\otimes2\)
mixed state, allowing standard entanglement measures to be applied. This mapping does not represent two physically separated particles but rather an effective partition of the internal atomic Hilbert space, a widely adopted approach in multilevel quantum systems and finite-dimensional quantum-information studies
\cite{YOUNORI05,ZHEGUO00,FZLGX03,YACHHA04,AMGUJA05,ATETO17,ATETO20,ATETO23,RANOMA16,MODEFAMO18,TAGMOJMEH22,MODETA22}.

The steady-state entanglement is quantified through the negativity

\begin{equation}
\mathcal{N}^{AB}(\rho_{\mathrm{ss}})
=
\frac{\|\rho_{\mathrm{ss}}^{T_A}\|_1-1}{2}
=
\sum_i
\frac{|\lambda_i|-\lambda_i}{2},
\label{DF45}
\end{equation}

where
\(
\rho_{\mathrm{ss}}^{T_A}
\)
is the partial transpose with respect to subsystem
\(A\),
\(
\|\cdot\|_1
\)
denotes the trace norm,
and
\(
\lambda_i
\)
are the eigenvalues of
\(
\rho_{\mathrm{ss}}^{T_A}
\). Rearranging
\(
\rho_{\mathrm{ss}}
\)
in the computational basis
\(
\{
|00\rangle,
|01\rangle,
|10\rangle,
|11\rangle
\}
\),
the partial transpose becomes
\small
\begin{equation}
\label{DF47}
\rho^{T_A}_{\mathrm{ss}} = \begin{pmatrix}
\rho_{11}^{\mathrm{ss}} & \rho_{12}^{\mathrm{ss}} & \rho_{31}^{\mathrm{ss}} & \rho_{32}^{\mathrm{ss}} \\
\rho_{21}^{\mathrm{ss}} & 0 & \rho_{41}^{\mathrm{ss}} & 0 \\
\rho_{13} & \rho_{41}^{\mathrm{ss}} & \rho_{33}^{\mathrm{ss}} & \rho_{34}^{\mathrm{ss}} \\
\rho_{23}^{\mathrm{ss}} & 0 & \rho_{43}^{\mathrm{ss}} & 0
\end{pmatrix}.
\end{equation}
The negativity is then obtained from the negative eigenvalues of
\(
\rho_{\mathrm{ss}}^{T_A}
\)

\begin{equation}
\label{DF48}
\mathcal N^{AB}(\rho_{\mathrm{ss}})
=
\sum_{\lambda_i<0}
|\lambda_i|,
\end{equation}

which is equivalent to Eq.~(\ref{DF45}).

In the presence of Doppler broadening, the velocity dependence of the detunings renders the steady-state density matrix velocity dependent,
\(
\rho_{\mathrm{ss}}
\rightarrow
\rho_{\mathrm{ss}}(v)
\),
and consequently modifies the partially transposed density matrix and its eigenvalue spectrum. The experimentally relevant entanglement measure is therefore obtained through Maxwell--Boltzmann averaging,

\begin{equation}
\label{DF49}
\mathcal N_{Dop}^{AB}
=
\frac{1}{\sqrt{\pi}u}
\int_{-\infty}^{\infty}
\mathcal N^{AB}
\!\left(
\rho_{\mathrm{ss}}(v)
\right)
e^{-v^2/u^2}
\,dv.
\end{equation}

Equation~(\ref{DF49}) quantifies the persistence of nonclassical correlations in thermal Rydberg media and enables direct comparison with Doppler-averaged absorption and fluorescence observables.

\section{Stratonovich--Weyl (SW) Wigner function}
\label{sec.5_1}

The Stratonovich--Weyl (SW) Wigner function provides a complete phase-space representation of the density matrix, treating populations and coherences on an equal footing. Its derivation for the steady state of the present four-level system, together with the corresponding Doppler-averaged form, is presented in Appendix~\ref{app:C}.

The resulting SW function
\(
W(\psi)
\),
given by Eq.~(\ref{DFE6}),
serves as a phase-space representation of the atomic state and provides a useful reference for interpreting fluorescence and entanglement signatures.

Upon averaging over the Maxwell--Boltzmann velocity distribution, the Doppler-averaged SW function becomes

\begin{equation}
\label{DFE7}
W_{Dop}
(\Delta_P,\Delta_C,\Delta_D,\psi)
=
\int_{-\infty}^{\infty}
W
(v,
\psi
)
e^{-v^2/u^2}
dv.
\end{equation}

The function
\(
W_{Dop}
\)
therefore incorporates both coherent quantum dynamics and thermal motion, providing a complementary phase-space description of the Doppler-broadened atomic response.

\section{Results and Discussion}
\label{sec.6}

Numerical simulations are performed for the EIA signal
$\mathrm{Im}[\rho_{12}^{\mathrm{ss}}\Gamma_2/\Omega_P]$,
the resonance fluorescence spectrum $S(\omega)$,
the negativity $\mathcal{N}(\rho_{\mathrm{ss}})$,
and the Stratonovich--Weyl (SW) Wigner function
$W(\rho_{\mathrm{ss}})$.
For each observable, velocity-resolved density maps are presented together with the corresponding Doppler-averaged spectra. These maps reveal how propagation geometry influences the location of dressed-state resonances within the Maxwell--Boltzmann velocity distribution, thereby determining whether the Doppler-integrated response is enhanced near the velocity center or redistributed toward the velocity wings.

The Doppler averaging is performed over the velocity interval
$-200 \le v \le 200~\mathrm{m\,s^{-1}}$
using 800 sampling points.
The calculations employ parameters corresponding to
$^{87}$Rb atoms.
The probe field is chosen sufficiently weak to satisfy the weak-probe approximation,
$\Omega_P/(2\pi)=0.001~\mathrm{MHz}$,
while the decay rate is fixed at
$\Gamma_2/(2\pi)=6~\mathrm{MHz}$.
Unless stated otherwise, both control fields are assumed resonant,
$\Delta_C=\Delta_D=0$.
The wavelength-matched configuration corresponds to
$\lambda_P=780~\mathrm{nm}$,
$\lambda_C=795~\mathrm{nm}$,
and
$\lambda_D=776~\mathrm{nm}$,
whereas the alternative cases represent wavelength-mismatched conditions.

\subsection*{Dressed-state interpretation of the Doppler response}

The numerical results are most naturally interpreted within a dressed-state framework.
The control fields $\Omega_C$ and $\Omega_D$ split the bare atomic levels into two coupled dressed manifolds associated with the transitions
$|1\rangle\leftrightarrow|3\rangle$
and
$|2\rangle\leftrightarrow|4\rangle$.
Consequently, the weak probe field interrogates transitions between these manifolds, producing four resonance branches whose locations determine the structures observed in absorption, fluorescence, entanglement, and phase-space observables.

Diagonalization of the control-field subspaces yields the generalized Rabi frequencies
\[
R_{C,D}=
\frac12
\sqrt{\Omega_{C,D}^{2}
    +
    \Delta_{C,D}^{2}},
\]
from which the probe resonances follow as
\begin{equation}
\label{DF51}
\begin{gathered}
\Delta_P=
\Delta_{CD}
\pm
R_{DC},
\qquad
R_{DC}=
R_D
\mp
R_C,
\end{gathered}
\end{equation}
where
\[
\Delta_{CD}=
\frac{\Delta_C-\Delta_D}{2}.
\]

For moving atoms, the laser detunings acquire Doppler shifts according to
$\Delta_j \rightarrow \Delta_j(v)$,
and the resonance condition becomes \cite{BBBN20, KTWZZ07}
\begin{equation}
\label{DF52}
\begin{gathered}
\Delta_P(v)=
\Delta_{CD}(v)
\pm
R_{DC}(v),
\
R_{DC}(v)=
R_D(v)
\mp
R_C(v).
\end{gathered}
\end{equation}
In the weak-probe regime
($\Omega_P\rightarrow0$),
the probe acts only as a spectroscopic monitor and does not modify the dressed-state energies.
Consequently, Eq.~(\ref{DF52}) fully determines the resonance locations observed in the numerical simulations.

A velocity-insensitive central resonance therefore depends on the interplay among the laboratory-frame detunings, control-field strengths, optical wavelengths, and propagation geometry.
For resonant driving
($\Delta_C=\Delta_D=0$),
the velocity dependence enters through
\begin{equation}
\label{DF55}
R_{C,D}(v)=
\frac12
\sqrt{
    \Omega_{C,D}^{2}
    +
    k_{C,D}^{2}v^{2}
}.
\end{equation}

For the parameters considered throughout this work, the thermally relevant velocity classes satisfy
$
|k_jv|
\gg
\Omega_C,\Omega_D
$
over most of the Maxwell--Boltzmann distribution.
The system therefore operates in the large-Doppler regime, where the resonance locations are governed primarily by the wave-vector combinations entering Eq.~(\ref{DF52}).

Expanding Eq.~(\ref{DF55}) in the limit
$
|k_jv|
\gg
\Omega_j
$
gives
\begin{equation}
\label{DF577}
R_{DC}(v)=
\frac{v}{2}
\Bigg[
k_D
\left(
1+\frac{\Omega_D^2}{2k_D^2v^2}
\right)
\mp
k_C
\left(
1+\frac{\Omega_C^2}{2k_C^2v^2}
\right)
\Bigg].
\end{equation}
Equation~(\ref{DF577}) explicitly shows that the resonance ridges are determined primarily by combinations of the optical wave vectors, while the control-field strengths provide higher-order corrections. As a result, propagation geometry and wavelength mismatch become the dominant mechanisms governing the location and shape of Doppler-resolved resonances.

The preceding analysis demonstrates that the observable response is controlled principally by the Doppler-dependent detuning combination $\Delta_{CD}(v)$ and the wave-vector combinations entering $R_{DC}(v)$. Consequently, propagation geometry, wavelength mismatch, and control-field strengths jointly determine whether dressed-state resonances overlap near the center of the velocity distribution or are displaced toward the velocity wings. The following sections examine how these mechanisms manifest themselves in absorption, fluorescence, entanglement, and SW phase-space observables.

\subsection{Effect of the control detunings in the presence of Doppler shift}

To elucidate the role of control-field detunings, we examine the influence of the Raman and upper-transition detunings, ($\Delta_C$) and ($\Delta_D$), on the Doppler-resolved and Doppler-averaged observables. Since the qualitative behavior is largely independent of propagation geometry, all fields are chosen co-propagating and directed opposite to the atomic motion, such that
\begin{equation}
\label{DF667}
\Delta_{P, C, D}^{(---) \atop (+++)}=\Delta_{P, C, D}\mp k_{P, C, D}~ v.
\end{equation}
Where the geometry sign $\oslash$ is replaced by $\oslash={(-)\atop (+)}\equiv(\mp)$. The corresponding resonance conditions become
\begin{equation}
\label{DF6667}
\begin{gathered}
\Delta_P^{(---) \atop (+++)}(v) - \Delta_C^{(---) \atop (+++)}(v) = (\mp)\Delta_C(\pm)(k_P - k_C) v, 
\\
\Delta_P^{(---) \atop (+++)}(v) - \Delta_D^{(---) \atop (+++)}(v)= (\mp)\Delta_D(\pm)(k_P + k_D) v,
\end{gathered}
\end{equation}
%\begin{equation}
%\label{DF6667}
%\Delta_P^{(---) \atop (+++)}(v) - \Delta_C^{(---) \atop (+++)}(v) = (\mp)\Delta_C(\pm)(k_P - k_C) v, \quad
%\Delta_P^{(---) \atop (+++)}(v) - \Delta_D^{(---) \atop (+++)}(v)= (\mp)\Delta_D(\pm)(k_P + k_D) v,
%\end{equation}
yielding the velocity-dependent resonances
\begin{equation}
\label{DF662}
\Delta_P^{(---) \atop (+++)} =
\begin{cases}
\frac{1}{2}(\Delta_C-\Delta_D)(\pm)k v \pm k v\Big(\frac{\Omega_D^2-\Omega_C^2}{4k^2 v^2}\Big),\\
\frac{1}{2}(\Delta_C-\Delta_D)(\pm)k v \pm k v \Big(1+\frac{\Omega_D^2+\Omega_C^2}{4k^2 v^2}\Big).
\end{cases}
\end{equation}

The architectures ${(\circ \circ \circ)\atop (\bullet \bullet \bullet)}$ are taken just for the purposes of generality. Note, each sign in the bracket maps the same sign in the corresponding geometry. However, sweeping the group of signs results only in flipping of the observable distribution across the plane parameters $(\Delta_P, v)$, but doesn't affect the Doppler averaging. The range of velocities $v$ is taken from a 1D Boltzmann distribution with a
transverse temperature $T$, that is measured from the width of the Doppler broadened spectrum \cite{HBHKFBWJ}. 

\begin{figure*}[!t]
    \centering
    \vspace{-0.2cm}
    %--- First figure ---
    \begin{minipage}{0.23\linewidth}
        \centering
        {\footnotesize (A) EIA \(\mathrm{Im}(\rho_{12}\Gamma_2/\Omega_P)\)}\\[-2pt]
        \includegraphics[width=5.cm,height=17.5cm]{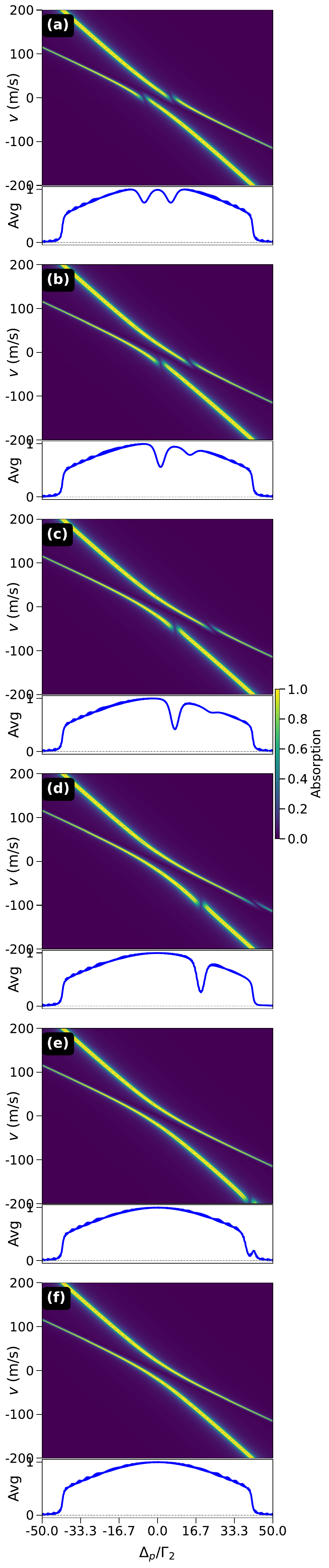}
    \end{minipage}
    \hspace{0.1cm}
    %--- Second figure ---
    \begin{minipage}{0.23\linewidth}
        \centering
        {\footnotesize (B) Fluorescence $S(\omega)$}\\[-2pt]
        \includegraphics[width=5.cm,height=17.5cm]{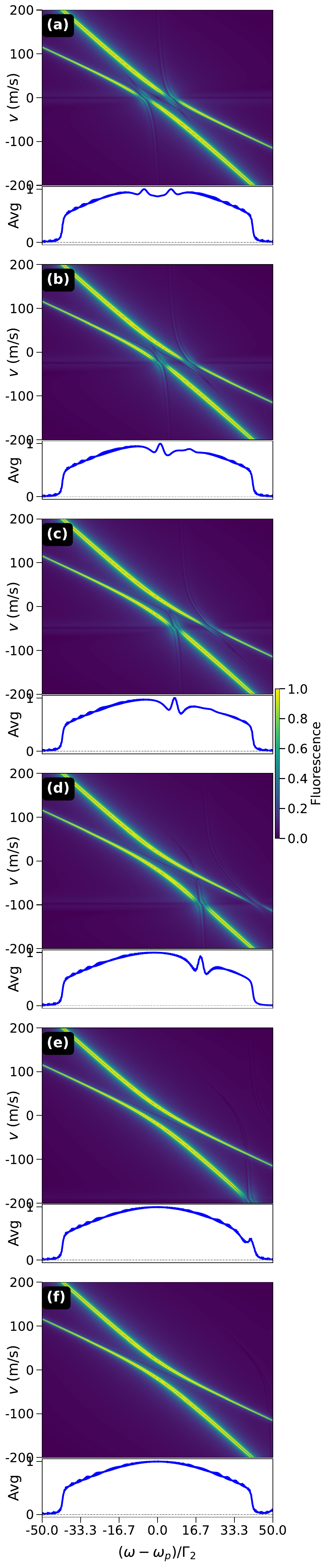}
    \end{minipage}
    \hspace{0.1cm}
    %--- Third figure ---
    \begin{minipage}{0.23\linewidth}
        \centering
        {\footnotesize (C) Negativity $\mathcal{N}(\rho)$}\\[-2pt]
        \includegraphics[width=5.cm,height=17.5cm]{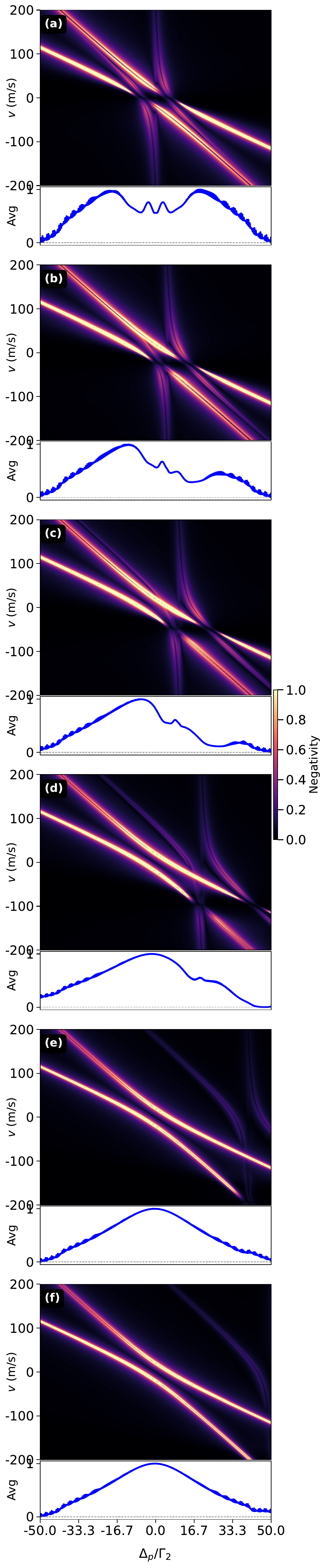}
    \end{minipage}
    \hspace{0.1cm}
    %--- Third figure ---
    \begin{minipage}{0.23\linewidth}
        \centering
        {\footnotesize (D) SW W-f $W(\rho)$}\\[-2pt]
        \includegraphics[width=5.cm,height=17.5cm]{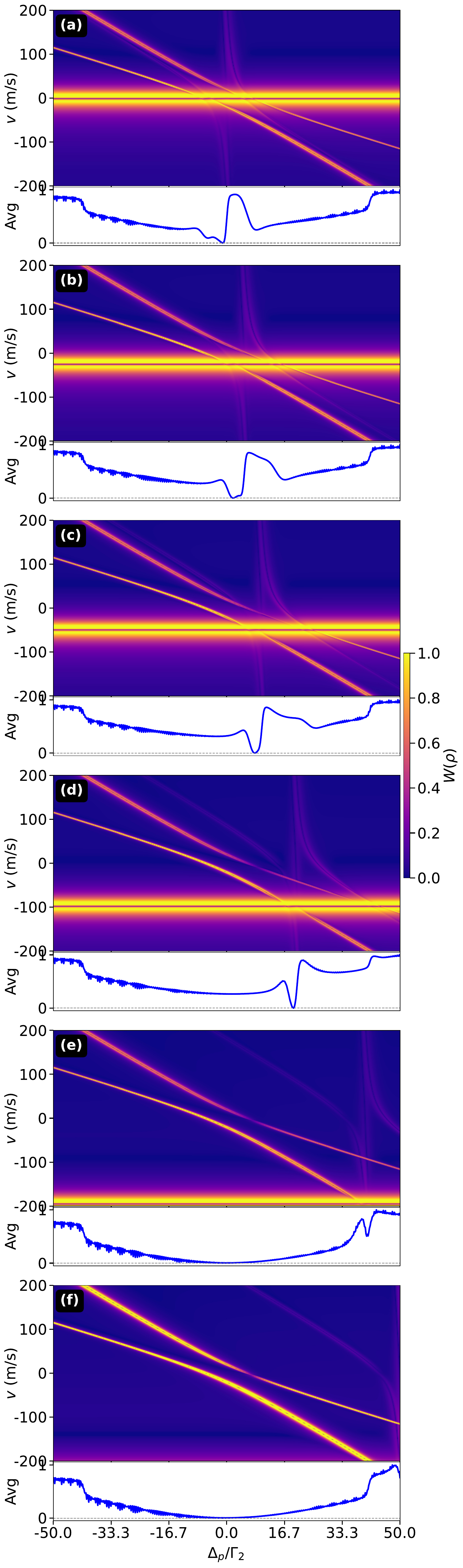}
    \end{minipage}
    \vspace{-0.1cm}
    \caption{
        Doppler-resolved maps of
        (A) normalized absorption
        $\mathrm{Im}\{\rho_{12}\Gamma_2/\Omega_P\}$,
        (B) fluorescence $S(\omega,v)$,
        (C) negativity
        $\mathcal N^{AB}_{\rm Dop}$,
        and
        (D) SW Wigner function
        $W(\rho)$,
        for $\Delta_D=0$,
        $\Omega_C=3$,
        $\Omega_D=12$,
        and
        $\Delta_C=
        0,\,
        5,\,
        10,\,
        20,\,
        40,\,
        50\,\Gamma_2$
        from (a) to (f).
        All fields co-propagate and are directed opposite to the atomic motion.
    }
    \label{fig.00}
\end{figure*}

\begin{figure*}[!t]
    \centering
    \vspace{-0.2cm}
    %--- First figure ---
    \begin{minipage}{0.23\linewidth}
        \centering
        {\footnotesize (A) EIA \(\mathrm{Im}(\rho_{12}\Gamma_2/\Omega_P)\)}\\[-2pt]
        \includegraphics[width=5.cm,height=17.5cm]{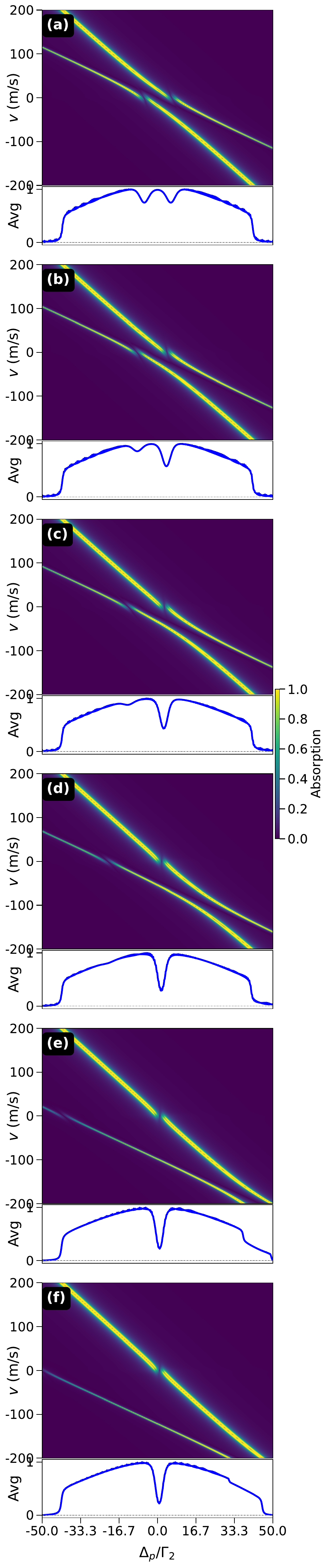}
    \end{minipage}
    \hspace{0.1cm}
    %--- Second figure ---
    \begin{minipage}{0.23\linewidth}
        \centering
        {\footnotesize (B) Fluorescence $S(\omega)$}\\[-2pt]
        \includegraphics[width=5.cm,height=17.5cm]{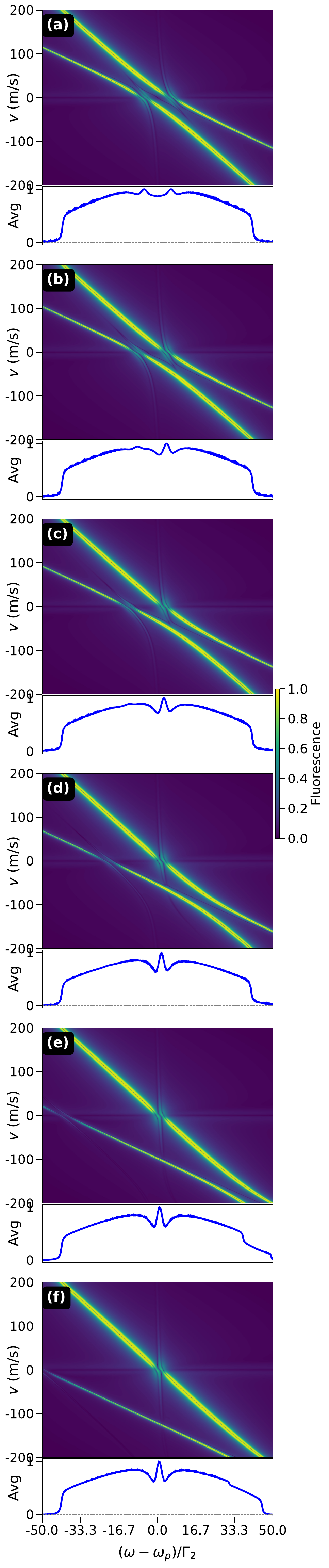}
    \end{minipage}
    \hspace{0.1cm}
    %--- Third figure ---
    \begin{minipage}{0.23\linewidth}
        \centering
        {\footnotesize (C) Negativity $\mathcal{N}(\rho)$}\\[-2pt]
        \includegraphics[width=5.cm,height=17.5cm]{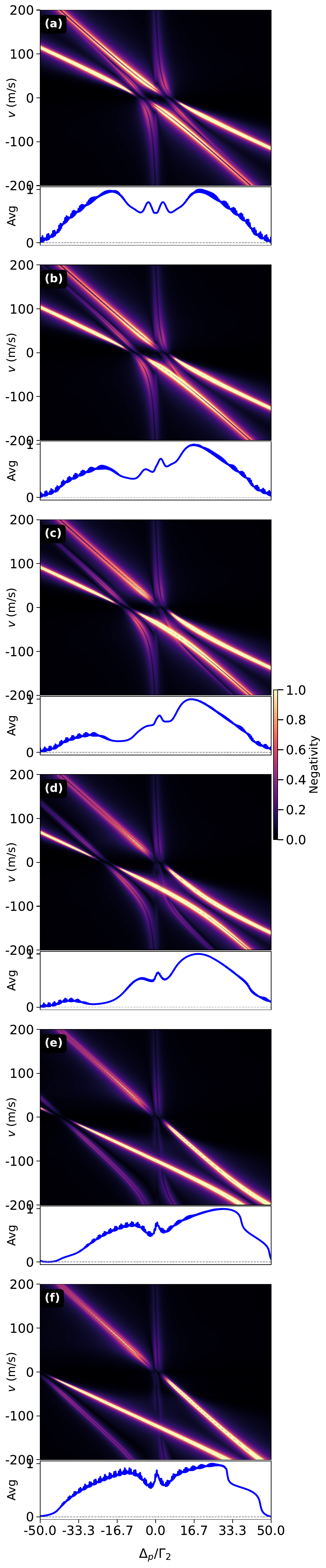}
    \end{minipage}
    \hspace{0.1cm}
    %--- Third figure ---
    \begin{minipage}{0.23\linewidth}
        \centering
        {\footnotesize (D) SW W-f $W(\rho)$}\\[-2pt]
        \includegraphics[width=5.cm,height=17.5cm]{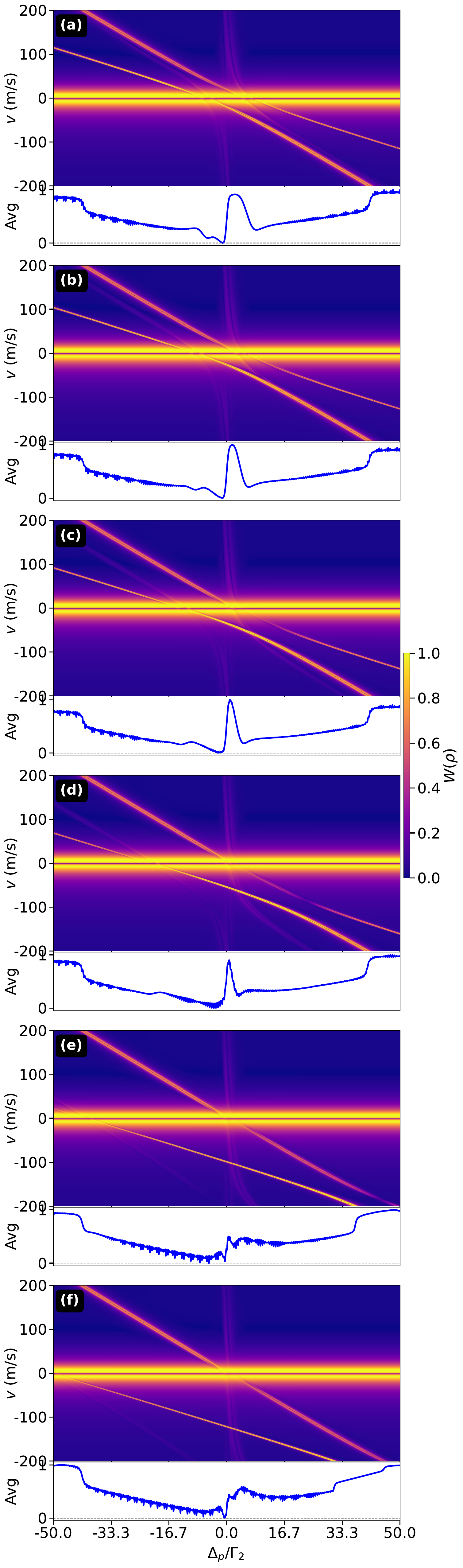}
    \end{minipage}
    \vspace{-0.1cm}
    \caption{
        Same as Fig.~\ref{fig.00}, but for
        $\Delta_C=0$
        and
        $\Delta_D=
        0,\,
        5,\,
        10,\,
        20,\,
        40,\,
        50\,\Gamma_2$
        from (a) to (f).
    }
    \label{fig.01}
\end{figure*}

Throughout this subsection, ($\Omega_C=3$) and ($\Omega_D=12$), while the control detunings are varied as shown in Figs.~(\ref{fig.00}) and (\ref{fig.01}). The velocity distribution is assumed Maxwell--Boltzmann, corresponding to a thermal vapor temperature extracted from the Doppler linewidth \cite{HBHKFBWJ}. For visualization purposes, the velocity-resolved maps of SW $W(\rho)$ distributions are plotted using gradient-based rendering with logarithmic normalization to enhance dynamic range. This procedure inherently obscures the sign of the SW function, mapping both positive and negative values onto a positive scale. At exact resonance $(\Delta_C=\Delta_D=0)$, all observables exhibit symmetric Doppler-resolved branches centered around ($\Delta_P=0$). These structures originate from balanced interference among the dressed-state pathways \cite{AGARWAL12} and lead to nearly symmetric Doppler-averaged spectra. The corresponding resonance ridges follow directly from Eq.~(\ref{DF662}), reflecting equal contributions from positive and negative velocity classes. 

The influence of $\Delta_C$ and $\Delta_D$ differs markedly. Increasing $\Delta_C$ generates velocity-class selection through the condition ($\Delta_C+kv\approx0$), yielding the resonant velocity $v_{\rm res}\approx-\Delta_C/k$ with $k=(k_C+k_D)/2$. Consequently, the resonance ridges in Fig.~(\ref{fig.00}) shift systematically across velocity space, producing increasingly asymmetric Doppler-resolved structures and corresponding peak imbalances in the Doppler-averaged spectra. Physically, the detuning modifies the competition between the dressed pathways $R_{CD}=R_D\mp R_C$ [Eq.~(\ref{DF51})], progressively favoring one dressed manifold while suppressing the other. As a result, avoided crossings migrate toward $\Delta_P\simeq\Delta_C$, and the spectral weight becomes concentrated in selected velocity groups. Here, maximal destructive interference (absorption dips) and maximal population cycling (strong emission) coincide, implying large, phase-matched $|\rho_{ij}^{\mathrm{ss}}|$ and $|\rho_{ii}^{\mathrm{ss}}|$, which drive negative eigenvalues of $\rho^{T_A}$ and enhanced entanglement. The entanglement shoulders around the probe resonance -- set by the resonance condition $\Delta_C + kv \approx 0$ -- exhibit maximal negativity and coincide with the deepest negative regions of the WS distribution. At larger detuning, this correspondence weakens, and entanglement must be inferred jointly from absorption and fluorescence [cf. Figs.~(\ref{fig.00}a, and (\ref{fig.00}b-d, columns (A, B, C)]. The entanglement strength scales with the magnitude of the negative values and follows their shift toward positive $\Delta_p$. The Doppler-averaged fluorescence -- like absorption -- varies monotonically with $\Delta_p$ and shows no clear correlation with entanglement, despite nearly identical spectral shifts across all components. 

% Large figure with [t] placement - allows text to flow around it

This peak-dip interchange reflects detuning-controlled redistribution of coherence among dressed states. The jitters at large $|\Delta_P|$ arise from phase-incoherent Doppler averaging of rapidly varying, velocity-selective dressed-state coherences, in contrast to the phase-locked, Doppler-compensated regime that produces smooth central spectral features\cite{COHEN98, FLEILUK00}.

In contrast, increasing $\Delta_D$ mainly redistributes spectral weight without introducing strong velocity selection, i.e, reduced sensitivity to Doppler-induced redistribution. Because $k_C\approx k_D$, the Doppler contribution to the associated Raman condition remains weak, and the resonance branches preserve their overall location while changing amplitude. As shown in Fig.~(\ref{fig.01}), one dressed pathway is gradually suppressed, causing the spectrum to evolve toward an effectively V-type response \cite{ATETO26} with reduced avoided-crossing behavior and broader Doppler-averaged profiles.

The SW Wigner function exhibits characteristic interference strips governed by phase-sensitive coherences $W(\rho) \sim \sum\limits_{i,j} \rho_{ij} \, \mathcal{K}_{ij}$. The strip position   follows the velocity-selection condition $v_{\rm res}\approx-\Delta_C/k$ where Doppler shift compensates the detuning effects. Such shifts vanish for finite $\Delta_D$ due to wavelength matching (Fig.~\ref{fig.01}). This mechanism underlies velocity-class selection in Rydberg electrometry and EIT sensing~\cite{SSKLP12,HGSA14}. Consequently, increasing $\Delta_C$ shifts the coherence-dominated regions toward negative velocities, whereas varying $\Delta_D$ produces comparatively weak displacement. After Doppler averaging, the oscillatory SW contributions partially cancel, leaving reduced but still identifiable coherence signatures \cite{YXWLLZWD24} which are absent in fluorescence. Their strongest features occur near velocity classes satisfying the multiphoton resonance conditions, where coherent superpositions are enhanced and interference among dressed pathways becomes maximal \cite{FLIM05,MOJA07,PMGWJA10,ADAMS19,HICOSCWI84}.

SW negativity is linked to non-Gaussian correlations and entanglement in atomic systems~\cite{MCZHJISE15,WACLPATR17,XLGGTHW22}. Maximal negativity occurs when dressed-pathway superpositions are strongest, i.e., when quasi-probability oscillations are most pronounced. Accordingly, the displacement of the SW interference strip with $\Delta_C$ reflects redistribution of entanglement across velocity classes. In the large-detuning limit, Doppler compensation [Eq.~(\ref{DF667})] becomes optimal, leading to near-perfect coincidence among observables: extended SW negativity, long-range entanglement, and direct correspondence between absorption and fluorescence.

Overall, fluorescence probes population redistribution, whereas SW and negativity probe coherence -- encoding its phase structure and non-separability, respectively. The observed peak-dip interchange and spectral asymmetry thus directly reflect detuning-controlled redistribution of coherence and entanglement among Doppler-selected velocity classes, linking microscopic dynamics to experimentally accessible observables.

\subsection{Counter-propagating second control}

Here, to establish a baseline for comparison, we focus on the standard configuration where the probe and the first control fields are co-propagating, while both counter-propagate with respect to the second control field \cite{VIAJNA16}. The corresponding Doppler-shifted interaction geometry is given by
\begin{equation}
\label{DF32}
\begin{gathered}
\Delta_{P, C}^{(++-)\atop (--+)}(v)=\Delta_{P, C} (\pm) k_{P, C} v,
\\
\Delta_D^{(-++)\atop (+--)}(v)=\Delta_D (\mp) k_D v,
\end{gathered}
\end{equation}

Throughout this study, all numerical simulations are performed employing the top architectural configuration. The positive (negative) signs correspond to fields addressing upshifted (downshifted) frequencies experienced simultaneously by an atom moving toward the probe beam with a Doppler shift of $2\pi v/\lambda_i$ \cite{RRPBM23, PBRBSA23}. On full resonance, and in accordance with the geometry defined in Eq.~(\ref{DF32}), the prefactor $\Delta_{CD}$ is expressed as:
\begin{equation}
\label{DF58}
\Delta_{CD}^{(++-)\atop (--+)}(v) = K^{(++-) \atop (--+)} v
\end{equation}
where
\begin{equation}
\label{DF59}
K^{(++-) \atop (--+)} = (\mp) k_P (\pm) \frac{k_C + k_D}{2}, 
\end{equation}

This framework dictates the shifts of the Electro-Magnetically Induced Absorption (EIA) positions corresponding to two-photon resonances within the V and $\Theta$ subsystems across different velocity groups $v$ as \cite{AMKAPAKA18}
\begin{equation}
\label{DF60}
\begin{gathered}
\Delta_P^{(++-)\atop (--+)}(v) - \Delta_C^{(++-)\atop (--+)}(v) = (\mp)(k_P - k_C) v,
\\
\Delta_P^{(++-)\atop (--+)}(v) + \Delta_D^{(++-)\atop (--+)}(v) = (\mp)(k_P - k_D) v,
\end{gathered}
\end{equation}
%
%\begin{equation}
%\label{DF60}
%\Delta_P^{(++-)\atop (--+)}(v) - \Delta_C^{(++-)\atop (--+)}(v) = (\mp)(k_P - k_C) v, \quad\text{and}\quad   \Delta_P^{(++-)\atop (--+)}(v) + \Delta_D^{(++-)\atop (--+)}(v) = (\mp)(k_P - k_D) v,
%\end{equation}
These equations define straight lines in the $(\Delta_P , v)$ plane, whose slopes and orientations depend exclusively on the Doppler phase-matching conditions. It is apparent that Doppler-free vertical ridges emerge under perfect wavelength matching \cite{FLIM05, JKLS02}. Nevertheless, eliminating the velocity prefactor is not entirely sufficient to guarantee velocity-independent amplitudes according to the combinations in Eqs.~(\ref{DF52}-\ref{DF577}), even under Rabi frequency matching. To establish the intended baseline references, the system behavior under weak control fields is examined first.

\subsubsection{\textbf{Weak controls: Effect of control field intensities and wavelengths' mismatch}}

\begin{figure*}[!t]
    \centering
    \vspace{-0.2cm}
    %--- First figure ---
    \begin{minipage}{0.23\linewidth}
        \centering
        {\footnotesize (A) EIA \(\mathrm{Im}(\rho_{12}\Gamma_2/\Omega_P)\)}\\[-2pt]
        \includegraphics[width=5.cm,height=13.5cm]{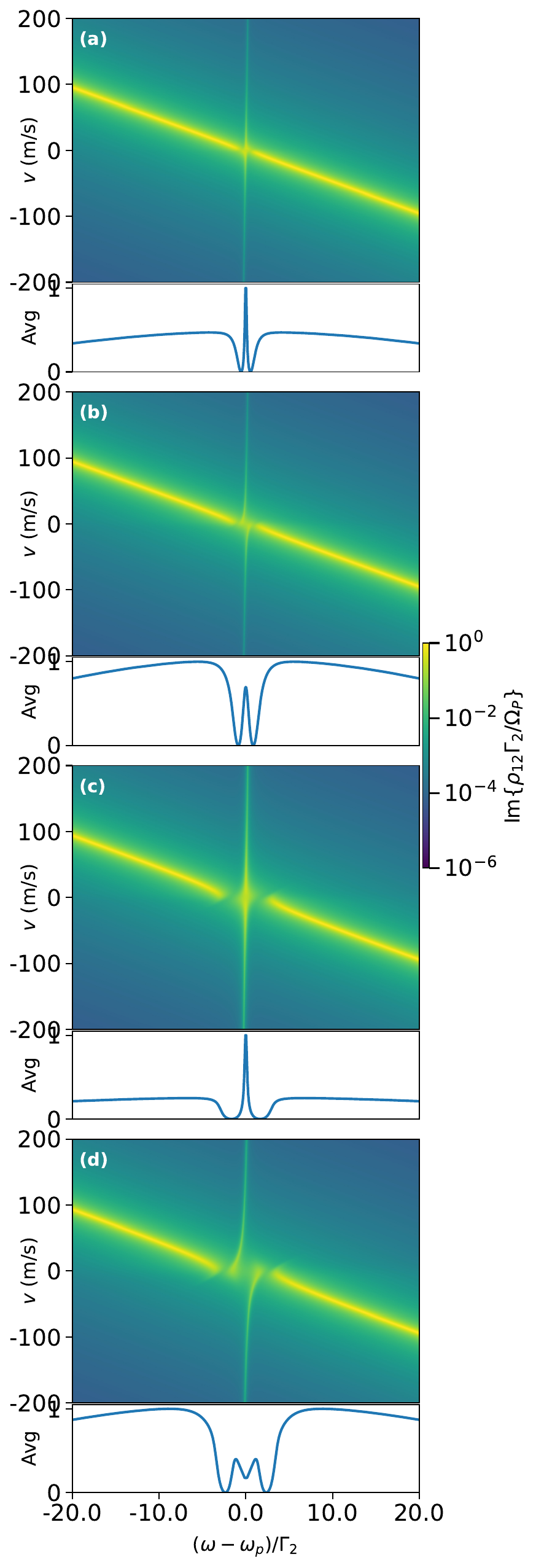}
    \end{minipage}
    \hspace{0.1cm}
    %--- Second figure ---
    \begin{minipage}{0.23\linewidth}
        \centering
        {\footnotesize (B) Fluorescence $S(\omega)$}\\[-2pt]
        \includegraphics[width=5.cm,height=13.5cm]{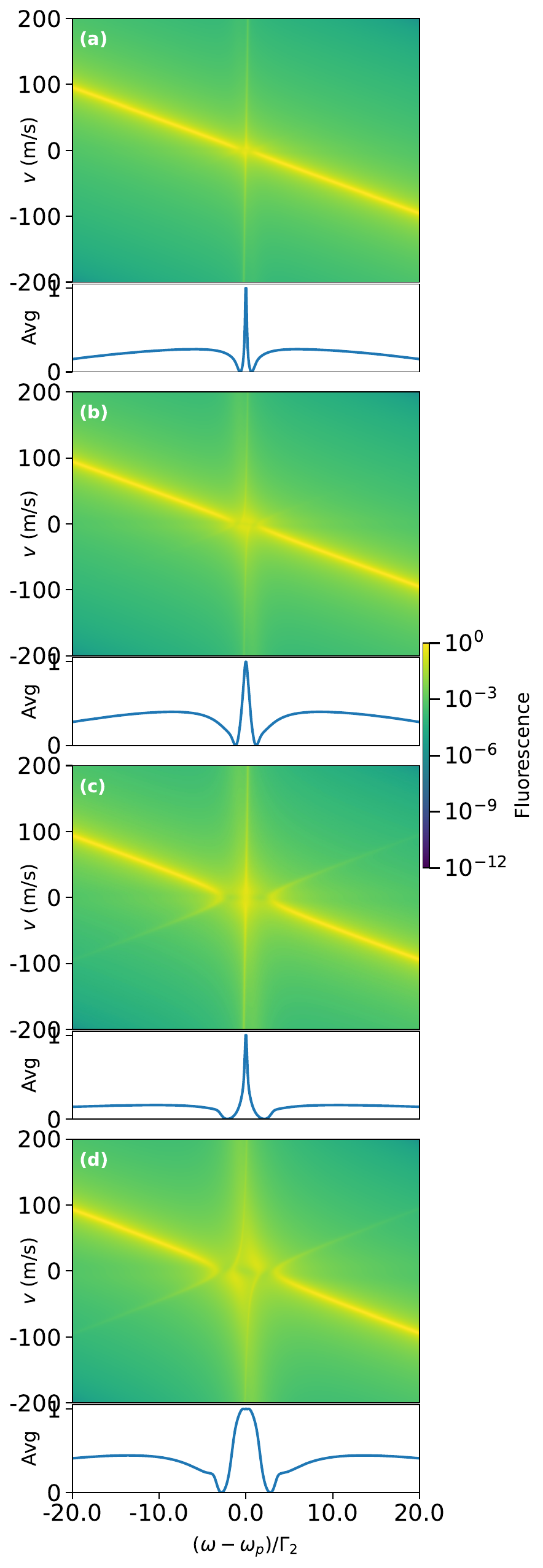}
    \end{minipage}
    \hspace{0.1cm}
    %--- Third figure ---
    \begin{minipage}{0.23\linewidth}
        \centering
        {\footnotesize (C) Negativity $\mathcal{N}(\rho)$}\\[-2pt]
        \includegraphics[width=5.cm,height=13.5cm]{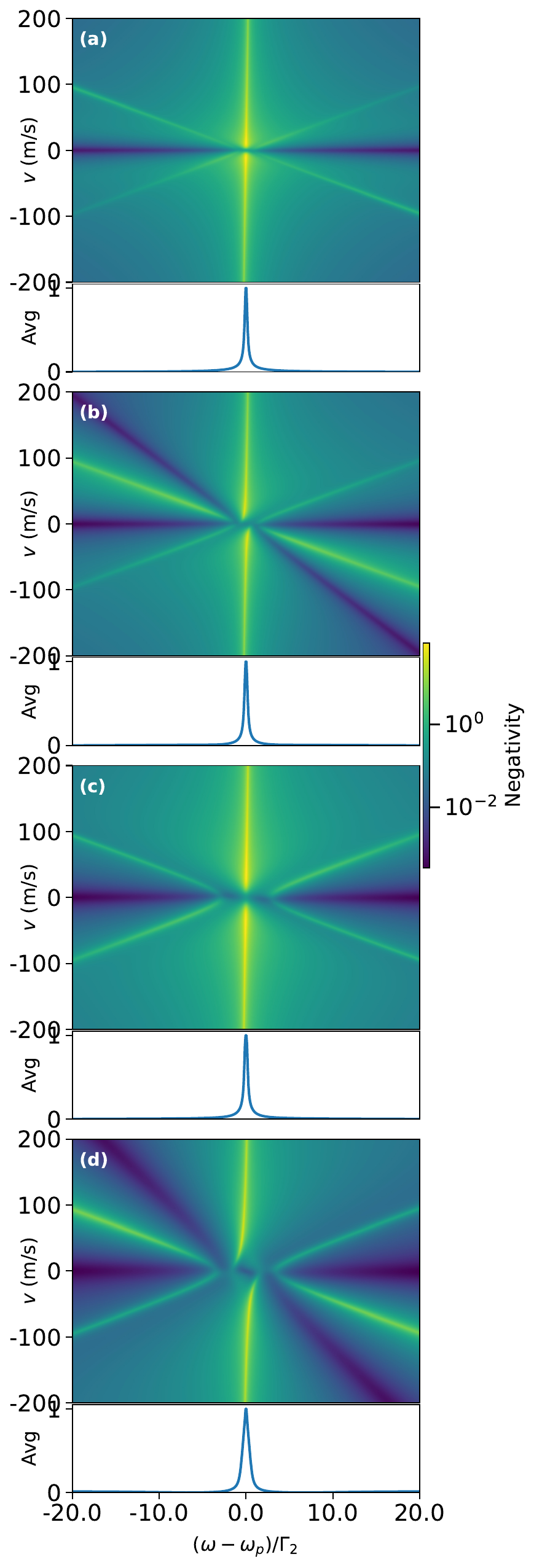}
    \end{minipage}
    \hspace{0.1cm}
    %--- Third figure ---
    \begin{minipage}{0.23\linewidth}
        \centering
        {\footnotesize (D) SW W-f $W(\rho)$}\\[-2pt]
        \includegraphics[width=5.cm,height=13.5cm]{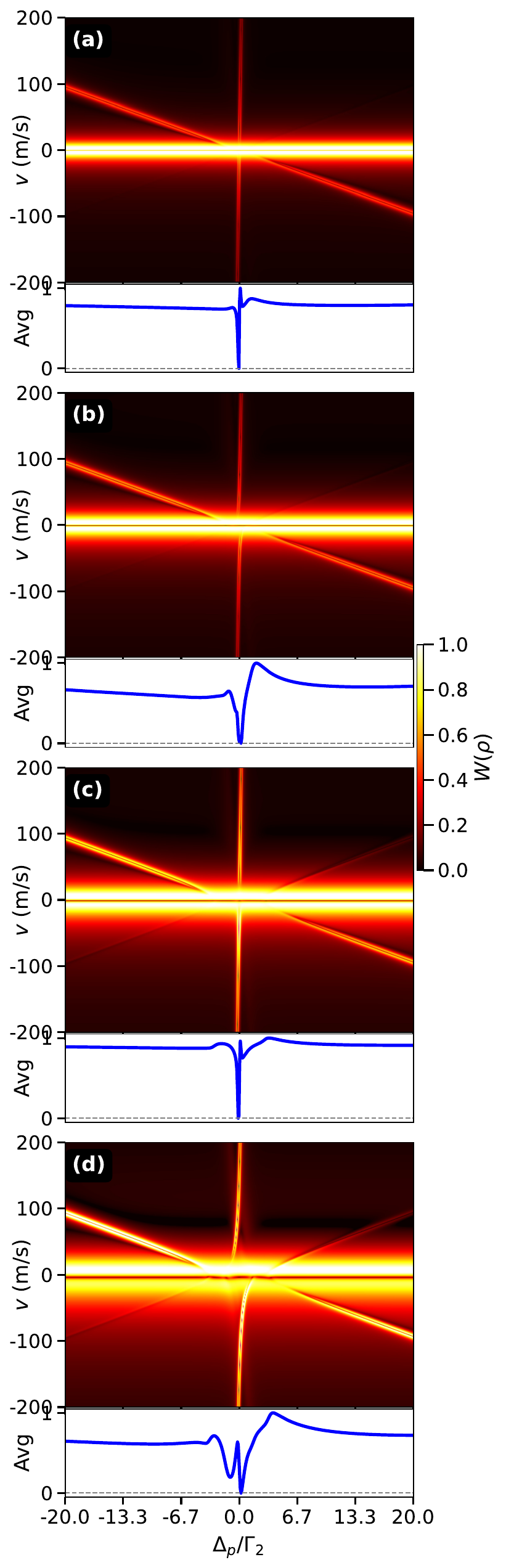}
    \end{minipage}
    \vspace{-0.1cm}
    \caption{Velocity-resolved maps (integrands) and Doppler-averaged profiles for absorption (A), fluorescence (B), negativity (C), and Wigner function (D) under varying intensities $(\Omega_C, \Omega_D)$: (a) $(\Gamma_2, \Gamma_2)$, (b) $(2\Gamma_2, \Gamma_2)$, (c) $(3\Gamma_2, 3\Gamma_2)$, and (d) $(5\Gamma_2, 2\Gamma_2)$. Balanced fields isolate a velocity-insensitive coherence ridge at $\Delta_P=0$, maximizing EIA contrast and phase-space correlations, whereas intensity asymmetry splits the resonance channels and induces severe dephasing.}
    \label{fig.135}
\end{figure*}

\begin{figure*}[!t]
    \centering
    \vspace{-0.2cm}
    %--- First figure ---
    \begin{minipage}{0.23\linewidth}
        \centering
        {\footnotesize (A) EIA \(\mathrm{Im}(\rho_{12}\Gamma_2/\Omega_P)\)}\\[-2pt]
        \includegraphics[width=5.cm,height=13.5cm]{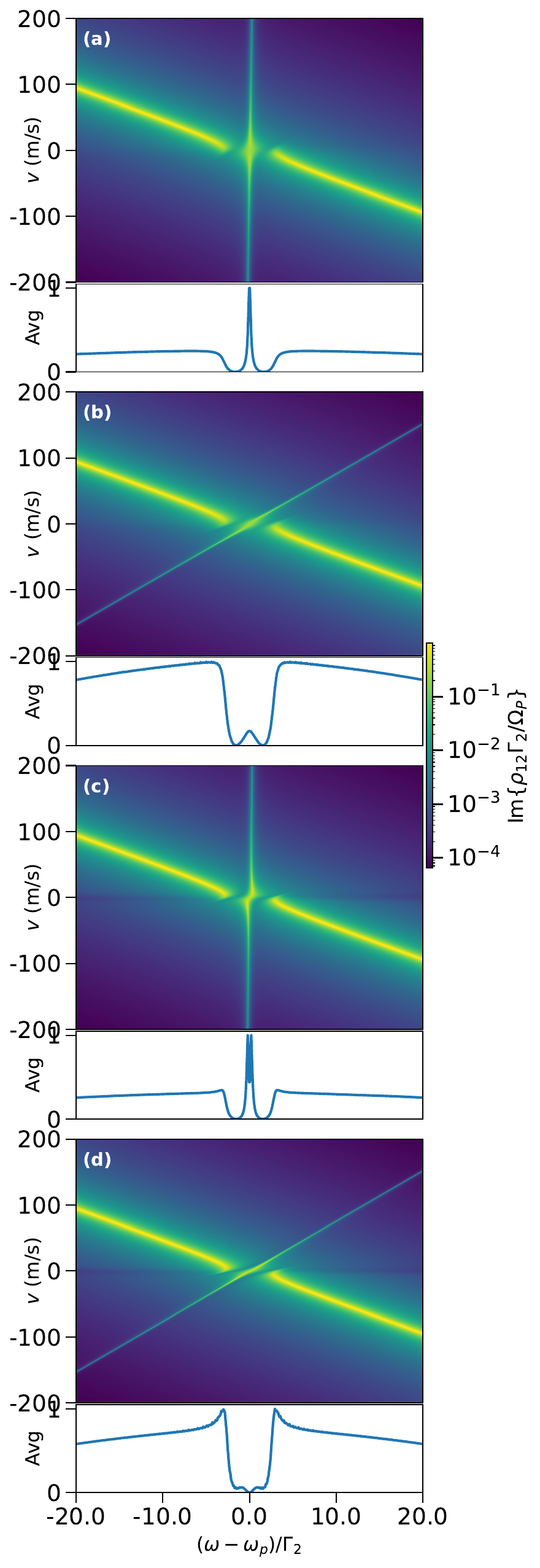}
    \end{minipage}
    \hspace{0.1cm}
    %--- Second figure ---
    \begin{minipage}{0.23\linewidth}
        \centering
        {\footnotesize (B) Fluorescence $S(\omega)$}\\[-2pt]
        \includegraphics[width=5.cm,height=13.5cm]{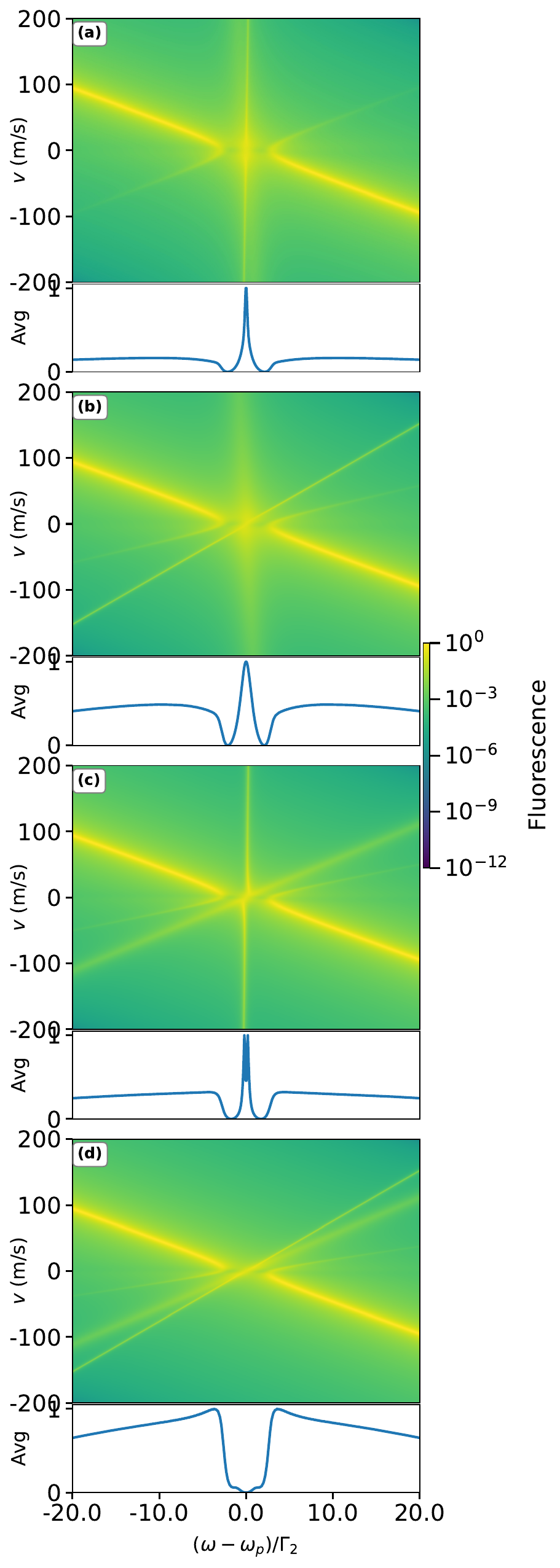}
    \end{minipage}
    \hspace{0.1cm}
    %--- Third figure ---
    \begin{minipage}{0.23\linewidth}
        \centering
        {\footnotesize (C) Negativity $\mathcal{N}(\rho)$}\\[-2pt]
        \includegraphics[width=5.cm,height=13.5cm]{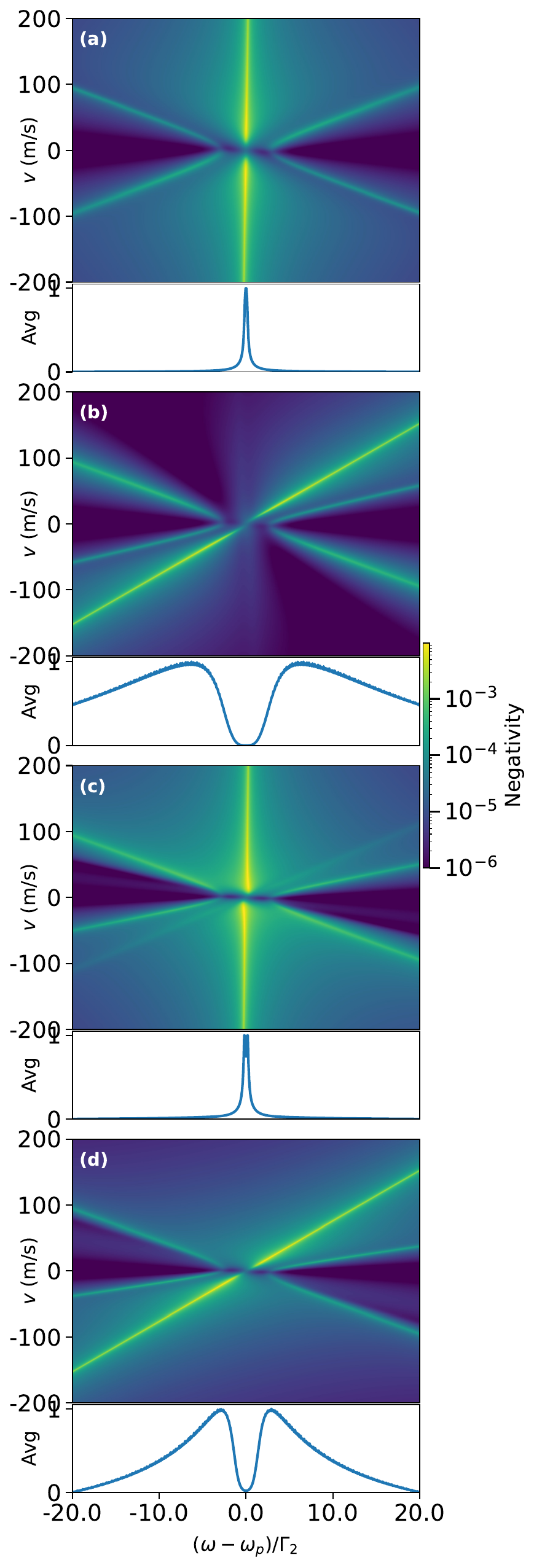}
    \end{minipage}
    \hspace{0.1cm}
    %--- Third figure ---
    \begin{minipage}{0.23\linewidth}
        \centering
        {\footnotesize (D) SW W-f $W(\rho)$}\\[-2pt]
        \includegraphics[width=5.cm,height=13.5cm]{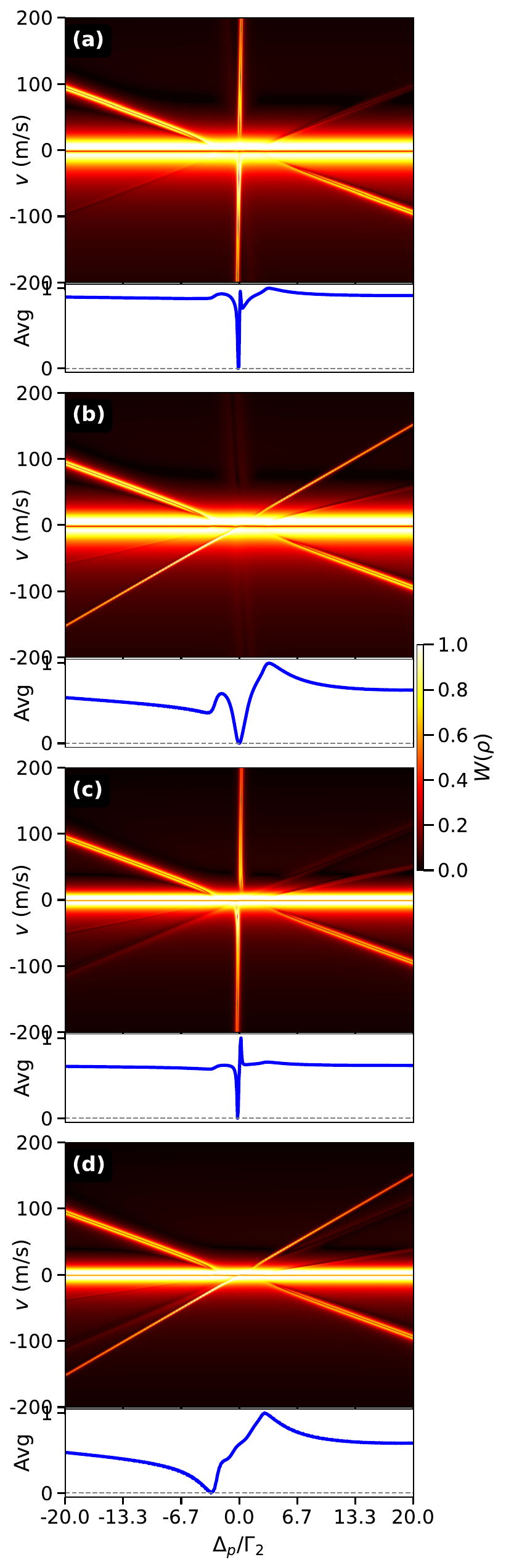}
    \end{minipage}    
    \vspace{-0.1cm}
    \caption{Influence of wavelength configuration on the four observables for fixed intensities $(\Omega_C, \Omega_D)=(3\Gamma_2, 12\Gamma_2)$: (a) fully matched, (b) $\lambda_D$ mismatch, (c) $\lambda_C$ mismatch, and (d) complete mismatch. A $\lambda_C$ mismatch suppresses the Raman pole generating an AT doublet, while a $\lambda_D$ mismatch activates a linear velocity term that rotates the resonance trajectories, smearing the ensemble quantum correlations.}
    \label{fig.246}
\end{figure*}

Figures~\ref{fig.135} and \ref{fig.246} summarize the velocity-resolved and Doppler-averaged responses of four complementary observables in the weak-probe regime, $\Omega_P\ll\Omega_C,\Omega_D$: the probe absorption $\mathrm{Im}(\rho_{12}\Gamma_2/\Omega_P)$ [column (A)], resonance fluorescence $S(\omega)$ [column (B)], quantum entanglement quantified through the negativity $\mathcal{N}(\rho)$ [column (C)], and the SW Wigner function $W(\rho)$ [column (D)], all plotted as functions of the normalized probe detuning $\Delta_P/\Gamma_2$. The velocity-resolved panels correspond to the integrands entering the Doppler averages defined in Eqs.~(\ref{DF31}), (\ref{DF40}), (\ref{DF45}), and (\ref{DFE6}). Examining these observables simultaneously provides a direct connection between the microscopic steady-state coherences and populations and their optical, quantum-correlation, and phase-space manifestations.

\paragraph{\textbf{Balanced configurations and wavelength matching:}}

Under perfect wavelength matching ($k_P \approx k_C \approx k_D \equiv k$), the linear Doppler contribution $K^{(++-) \atop (--+)}$ in Eq.~(\ref{DF59}) vanishes, leaving only the amplitude-dependent combinations $R_{CD}$ in Eq.~(\ref{DF577}). Consequently, the four resonance branches of Eq.~(\ref{DF52}) reduce to

\begin{equation}
\label{DF61}
\Delta_P^{(++-) \atop (--+)} =
\begin{cases}
\displaystyle \pm kv\Big(\frac{\Omega_D^2-\Omega_C^2}{4k^2v^2}\Big), \\[2mm]
\displaystyle \pm kv\Big(1+\frac{\Omega_D^2+\Omega_C^2}{4k^2v^2}\Big).
\end{cases}
\end{equation}

For balanced control fields ($\Omega_C=\Omega_D=\Omega$), the first branch collapses to $\Delta_P\simeq0$, whereas the second branch remains explicitly velocity dependent. The resulting resonance trajectories become effectively reduced to three branches, one of which remains nearly independent of atomic velocity around line center. Physically, the Doppler shifts entering the effective two-photon detunings cancel to a large extent, allowing a broad range of velocity classes to satisfy the resonance condition simultaneously \cite{TMAPPR19}. Such collective Doppler compensation enables coherent accumulation throughout the thermal ensemble and constitutes the fundamental origin of the robust line-center structures observed throughout Figs.~\ref{fig.135} and \ref{fig.246}.

When the resonance branches interfere, the second branch dominates and generates the dressed-state trajectories \(\Delta_P \approx \pm kv\Big(1+\frac{\Omega^2}{2k^2v^2}\Big)\), 
which are present for all velocity classes. In the rotated $(\Delta_P,v)$ plane, these trajectories follow rectangular-hyperbola-like curves satisfying $\Delta_P \mp kv \approx \alpha\mathcal{O}(kv)$. Their asymptotic behavior approaches $\Delta_P \approx (-k\pm\sqrt{\alpha}k)v$ and, for sufficiently large Doppler shifts, reduces approximately to $\Delta_P\approx kv$. These asymptotes generate the tilted bright-state ridges visible throughout the velocity-resolved maps. Although similar dressed-state structures occur in both single- and dual-control configurations, the single-control case approaches the two-photon resonance more rapidly because of the larger effective denominator in Eq.~(\ref{DF61}). Consequently, the dressed-state ridges remain more strongly localized around the vertical resonance, producing narrower Doppler-averaged transparency windows than those obtained with two controls; compare Figs.~\ref{fig.135}a and \ref{fig.135}c [column (A)].

The same Doppler-compensated coherence manifests differently in the four observables. In the absorption maps [column (A)], it appears as a pronounced vertical ridge centered at $\Delta_P=0$, visible in Figs.~\ref{fig.135}a,c and \ref{fig.246}a. Because a wide range of velocity classes contribute constructively to the same resonance condition, the corresponding Doppler-averaged spectra exhibit high-contrast EIA resonances \cite{VIAJNA16,MIKAHA23}. 
Similarly, because quantum entanglement relies nonlinearly on both populations and coherences ($\rho_{ii}^{\mathrm{ss}}\rho_{jj}^{\mathrm{ss}} \gtrsim \left|\rho_{ij}^{\mathrm{ss}}\right|^2$), the $\mathcal{N}(\rho_{\mathrm{ss}})$ maps (Column C) mirror this zero-velocity robustness, yielding sharp, Doppler-free peaks at line center where phase coherence is optimally preserved during successive absorption-emission cycles \cite{WLZJQM13,PBRAS24,NSKVR04}.

The SW Wigner function [column (D)] reveals this behavior even more directly and clearly. Interestingly the appearance of a broad horizontal ridge centered around $v\approx0$ throughout the entire detuning range in both SW and $\mathcal{N}(\rho_{\mathrm{ss}})$  [column (C)] maps. By contrast, this central coherent structure does not appear explicitly in
absorption [column (A)] and fluorescence [column (B)]. The physical interpretation and output correlate to the specific  observable. The SW $W(\rho)$ preserves the complete phase-space structure of the steady state, revealing that the horizontal ridge therefore represents coherent accumulation of both populations and coherences under nearly simultaneous probe and control resonances. Negativity demonstrates that quantum correlations remain robust whenever Doppler-induced dephasing is sufficiently weak due to the nonlinear dependence of the negativity on both populations and coherences. In this case, entanglement survives only when coherence preservation and population redistribution remain simultaneously favorable. Accordingly, the central entanglement region may be regarded as the quantum-correlation counterpart of the Doppler-compensated coherence ridge observed in the SW representations.  Consequently, the underlying velocity-insensitive coherent accumulation remains only indirectly visible in the optical spectra, while it is retained explicitly in $W(\rho)$ and partially preserved in $\mathcal{N}(\rho)$. 

Absorption and fluorescence probe only specific projections of the density matrix. Absorption isolates the antisymmetric component of $\rho_{12}^{\mathrm{ss}}$, whereas fluorescence predominantly probes the symmetric radiative contribution. Both observables therefore emphasize the Doppler-shifted resonance conditions \(\Delta_P \pm kv \approx 0\), which appear as tilted velocity-selective ridges in the $(\Delta_P,v)$ plane. Consequently, the underlying velocity-insensitive coherent accumulation remains only indirectly visible in the optical spectra, while it is retained explicitly in $W(\rho)$ and partially preserved in $\mathcal{N}(\rho)$. 

Already, in the weak-probe regime, $\Omega_P\ll\Omega_C,\Omega_D$, the resonance fluorescence  [column (B)] originates primarily from bright-state transitions \cite{FLIM05,ARUN07,VINA15}. The fluorescence maps therefore appear more diffuse than the corresponding absorption maps  [column (A)] and retain residual dressed-state signatures that are absent in $\mathrm{Im}(\rho_{12}\Gamma_2/\Omega_P)$; see Figs.~\ref{fig.135} and \ref{fig.246} [column (B)]. Nevertheless, under wavelength matching, the bright-state resonance remains aligned across the thermal velocity distribution, preserving phase coherence during repeated absorption-emission cycles and producing a pronounced central fluorescence peak \cite{WLZJQM13,PBRAS24}. This fluorescence maximum closely follows the EIA resonance observed in the absorption spectra \cite{VINA15,NSKVR04}. The persistence or degradation of the central fluorescence ridge therefore reflects the competition between bright-state coherence and Doppler-induced  dephasing. After Doppler averaging, however, residual phase dispersion broadens and weakens the fluorescence peak, as evident in subpanels~\ref{fig.135}b,d and \ref{fig.246}b,d [column (B)]~\cite{NSKVR04,WLZJQM13}, while the AT splitting remains observable, particularly in Fig.~\ref{fig.246}c [column (B)]. 

\paragraph{\textbf{Intensity asymmetry ($\Omega_C\neq\Omega_D$):}}

When the control strengths become unequal, the degeneracy of Eq.~(\ref{DF61}) is lifted and all four resonance branches remain distinct. The quadratic velocity dependence then becomes increasingly important for slow atoms ($v\approx0$), causing mutual repulsion and fragmentation of the central resonance structure and hyperbola-like curves, as observed in Figs.~\ref{fig.135}b,c [columns (A-D)]. The separation between the split branches grows with both the ratio $\Omega_C/\Omega_D$ and the factor $k^2v^2$, resulting in broader Doppler-averaged linewidths and reduced absorption contrast, as shown in Fig.~\ref{fig.135}d [columns (A-D)].

Physically, these results indicate that $\Omega_D$ predominantly mediates coherence transfer, whereas $\Omega_C$ controls the population redistribution responsible for resonance quality and contrast \cite{PBRBSA23}. The same imbalance disrupts the smooth bright-state structures in the fluorescence response and strongly suppresses nonlocal quantum correlations. Since Doppler averaging acts as an efficient low-pass filter for entanglement, the fragmented velocity-resolved contributions overlap only weakly after thermal integration. Consequently, finite negativity survives only within localized regions where transient coherence enhancement overcomes the strong decay-induced mixing. The phase-space distribution exhibits a similar fragmentation, confirming that intensity imbalance perturbs the coherent overlap among different velocity classes rather than completely destroying the resonance condition.

\paragraph{\textbf{Wavelength mismatch regimes:}}

The spectra obtained under different wavelength combinations reveal an even richer behavior arising from the interplay between Doppler-broadened two-photon resonances and the velocity-dependent dressed states described by Eq.~(\ref{DF577}). The wavelength mismatch activates additional Doppler-sensitive pathways and profoundly modifies the interference landscape \cite{MISI18}. Noteworthy is that, the fully matched configuration, $\lambda_P\approx\lambda_C\approx\lambda_D\equiv\lambda$, the response follows directly the behavior discussed above for balanced control strengths, yielding similar velocity-resolved and Doppler-averaged structures; compare Figs.~\ref{fig.135}a,c and \ref{fig.246}a [column (A)].

When the mismatch occurs in $\lambda_C$, the Raman resonance condition $\Lambda_{32}\approx0$ is displaced along the trajectory \(\Delta_P \approx (\mp)(k_P-k_C)v\). Along this trajectory, the probe denominator becomes \(\Lambda_{12}\approx-\frac{\Gamma_2}{2}+i(k_Cv)\), driving the probe transition away from resonance. Since the Raman resonance remains embedded within the probe denominator [Eq.~(\ref{DF30})], efficient overlap between the Raman pole and the probe absorption pole is lost. Consequently, the central resonance channel cannot form efficiently, leading to strong suppression of the coherent response, as observed in Figs.~\ref{fig.246}b,c [column (A)]. After Doppler averaging, the off-resonant velocity classes dominate over the exact resonance region, producing level splitting and the characteristic AT doublet structure. Physically, the Doppler compensation mechanism remains partially preserved for many velocity groups, but the Raman and probe resonances no longer coincide sufficiently to support coherent accumulation at line center.

A qualitatively different behavior emerges when the mismatch occurs in $\lambda_D$, while $\lambda_P\approx\lambda_C$. In this case, the Raman resonance remains approximately centered around $\Delta_P\approx0$, maintaining strong overlap with the probe resonance \(
\Lambda_{12}\approx -\frac{\Gamma_2}{2}+i(k_Pv)\). Consequently, the inclined resonance structures remain visible in Figs.~\ref{fig.246}b,d [column (A)]. However, the linear Doppler contribution $\Delta_{CD}^{(++-) \atop (--+)}(v)$ becomes active through \(K^{(++-) \atop (--+)}=\frac{(\mp)k_P(\pm)k_D}{2}\), which rotates the two-photon resonance trajectories around $\Delta_P=0$. The magnitude of this rotation increases with the degree of wavelength mismatch. As a result, velocity-dependent phase accumulates nonuniformly throughout the ensemble, violating the resonance symmetry associated with Eq.~(\ref{DF61}) and progressively smearing the central resonance structure.  After Doppler averaging, this behavior manifests as spectral broadening accompanied by reduced resonance amplitudes in the averaged spectra [Fig.~\ref{fig.135}b,d, column (A)].

In this regime, interference between the resonance branches \(\Delta_P(v)=\pm\frac{k_D-k_P}{2}v,\qquad\Delta_P(v)=\pm\frac{k_C-k_P}{2}v\), renders the two-photon resonance highly velocity selective \cite{RKLP18,BZBZBS16}. The resulting dephasing amplifies velocity dispersion, destroys Doppler compensation, and confines the entangled regions to narrow intervals of probe detuning \cite{UCRAW13}. The velocity-resolved negativity maps clearly show that entanglement survives only where competing resonance branches overlap efficiently; see Fig.~\ref{fig.246}b,d [column (C)]. Near $v\approx0$, however, the two-photon resonance and dressed-state branches no longer overlap strongly, preserving a transparency window around line center and suppressing entanglement in that region \cite{MISI18}. A narrow residual AT splitting nevertheless remains visible because of the persistent dressed-state level splitting.

This behavior differs fundamentally from the intensity-asymmetry case. Unequal control strengths perturb the resonance topology while still allowing a subset of velocity classes to satisfy the two-photon resonance condition. By contrast, wavelength mismatch activates uncompensated velocity-dependent phase accumulation that directly destroys the Doppler-cancellation mechanism. Consequently, resonance structures remain partially localized when $\Omega_C\neq\Omega_D$, whereas wavelength mismatch produces much stronger spectral smearing and degradation across all observables; compare Figs.~\ref{fig.135}b,d and \ref{fig.246}b,d.

Away from $v\approx0$, the dressed-state manifolds \(\Delta_P\approx\pm kv\Big(1+\frac{\Omega^2}{2k^2v^2}\Big)\) appear in all observables and correspond to velocity-dependent interference between AT dressed states. Because these off-axis branches accumulate phase nonuniformly across different velocity classes, their contributions largely cancel after Doppler averaging. Consequently, the Doppler-robust central structures dominate the averaged response. Since all optical coherences contain Doppler-dependent factors of the form \(\Lambda_{mn}=-\frac{\Gamma}{2}\pm i(\Delta+kv)\), coherence suppression occurs primarily near Doppler-shifted multiphoton resonance conditions such as \(\Delta_P(\pm)k_Pv\approx0,\qquad\Delta_P(\pm)(k_P+k_C-k_D)v\approx0\), which define the tilted resonance ridges in the $(\Delta_P,v)$ plane.

Entanglement survives only along restricted segments of these resonance manifolds where coherence preservation and population redistribution remain simultaneously optimized. After Doppler averaging, this produces pronounced central signatures in the observable spectra: a negative dip in $W(\rho)$ and a sharp peak in $\mathcal{N}(\rho)$ near $\Delta_P\approx0$, coinciding with the EIA maximum in the absorption spectrum and the central fluorescence peak. Although the Doppler-averaged negativity differs qualitatively from the absorption response, it closely follows the fluorescence behavior; compare Fig.~\ref{fig.135} [columns (A), (B), and (C)]. Physically, this occurs because negativity remains finite only within narrow velocity intervals, causing Doppler averaging to operate as a stringent low-pass filter for entanglement. The collective results therefore demonstrate that maintaining wavelength matching and balanced control strengths is essential for preserving Doppler-compensated coherence, maximizing optical contrast, and sustaining quantum correlations throughout the thermal ensemble.

\subsubsection{\textbf{Strong control field intensities}}
\begin{figure*}[!t]
    \centering
    \vspace{-0.1cm}
    %--- First figure ---
    \begin{minipage}{0.23\linewidth}
        \centering
        {\footnotesize (A) EIA \(\mathrm{Im}(\rho_{12}\Gamma_2/\Omega_P)\)}\\[-2pt]
        \includegraphics[width=5.5cm,height=20.cm]{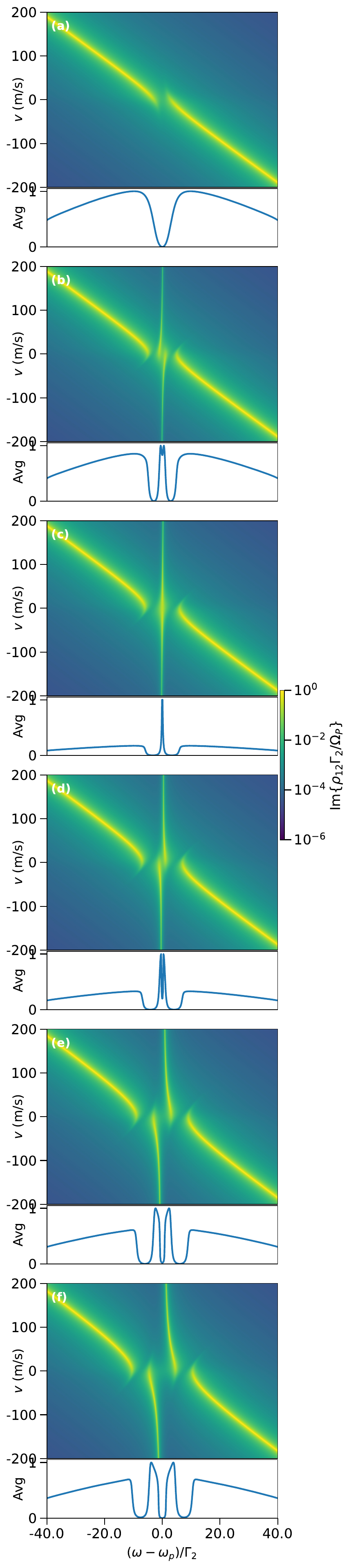}
    \end{minipage}
    \hspace{0.1cm}
    %--- Second figure ---
    \begin{minipage}{0.23\linewidth}
        \centering
        {\footnotesize (B) Fluorescence $S(\omega)$}\\[-2pt]
        \includegraphics[width=5.5cm,height=20.cm]{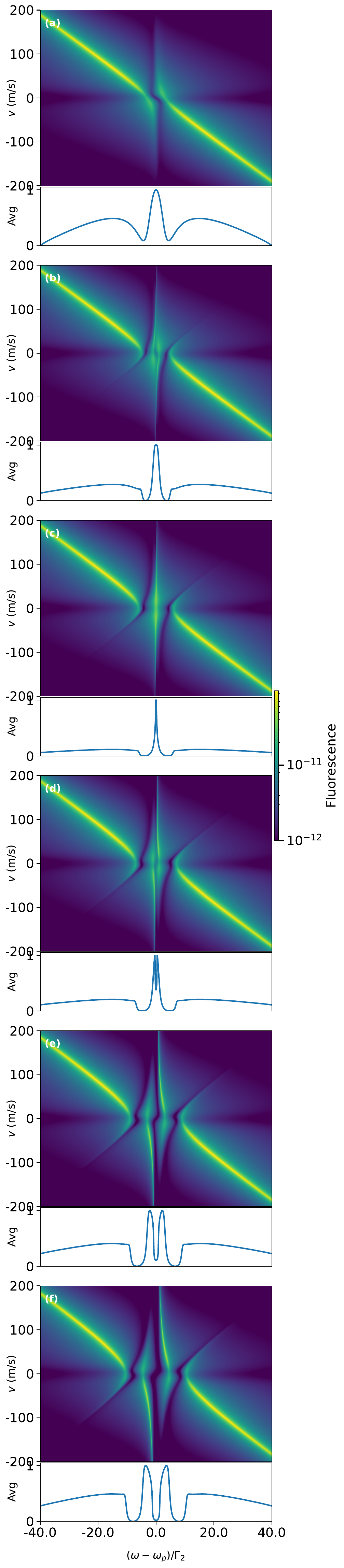}
    \end{minipage}
    \hspace{0.1cm}
    %--- Third figure ---
    \begin{minipage}{0.23\linewidth}
        \centering
        {\footnotesize (C) Negativity $\mathcal{N}(\rho)$}\\[-2pt]
        \includegraphics[width=5.5cm,height=20.cm]{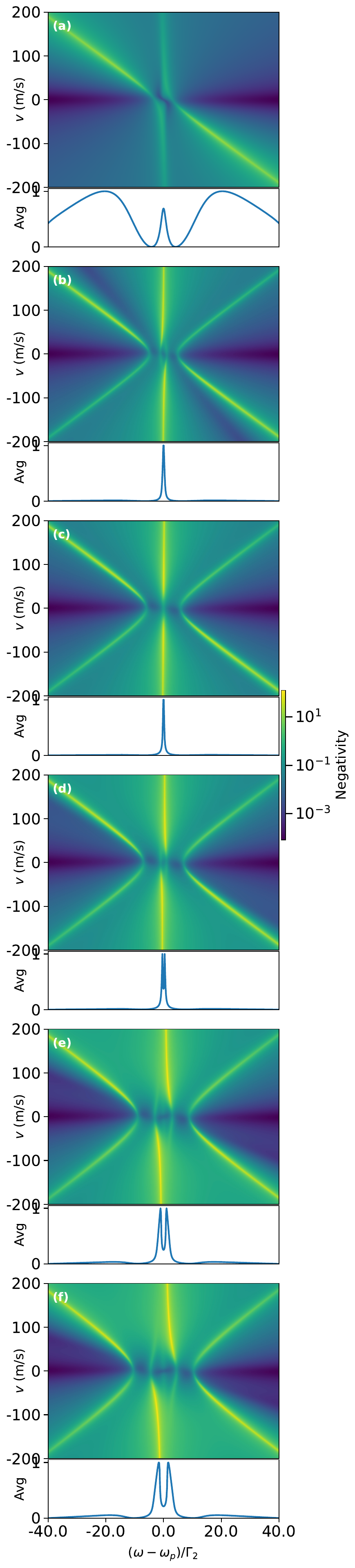}
    \end{minipage}
    \hspace{0.1cm}
    %--- Third figure ---
    \begin{minipage}{0.23\linewidth}
        \centering
        {\footnotesize (D) SW W-f $W(\rho)$}\\[-2pt]
        \includegraphics[width=5.5cm,height=20.cm]{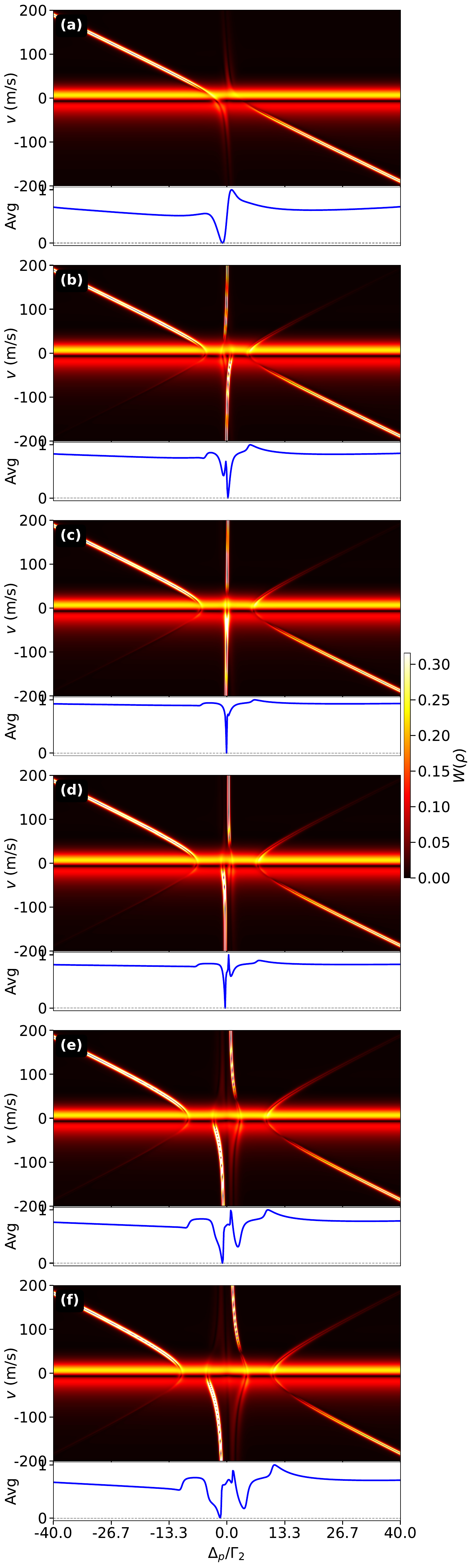}
    \end{minipage}            
    \vspace{-0.1cm}
    \caption{(A): Doppler-resolved normalized absorption $\mathrm{Im}\{\rho_{12}\Gamma_2/\Omega_P\}$, against the probe detuning $\Delta_p$ (B): Doppler-resolved fluorescence \(S(\omega, v)\) maps against the normalized detuning $(\omega-\omega_p)/\gamma_2$. (C): Doppler-resolved negativity maps \(\mathcal{N}_{\mathrm{Dop}}^{AB}(\Delta_P, v)\) against the probe detuning $\Delta_p$. (D) Doppler-resolved SW Wigner function $W(\rho)$ against the probe detuning $\Delta_p$. All plots are pictured for varying ($\Omega_C, \Omega_D$) as: (a) ($6\Gamma_2, 0$), (b) ($6\Gamma_2, 4\Gamma_2$), (c) ($6\Gamma_2, 6\Gamma_2$), (d) ($6\Gamma_2, 8\Gamma_2$), (e) ($6\Gamma_2, 12\Gamma_2$), (f) ($6\Gamma_2, 15\Gamma_2$).}
    \label{fig.789}
\end{figure*}

As the control field intensities surpass the atomic decay rates, the system enters the strong AT regime. Here, the dynamics are governed by a dressed-state framework rather than quantum interference, where strong driving field interactions optimize coherent superpositions between the ground and excited state manifolds \cite{BWVARI22}. This transition yields clear spectroscopic AT splitting across all observables, creating a unified mapping of the system states. Figure~\ref{fig.789} outlines the joint behaviors under these heavily strongly coupled conditions with the wavelength matching maintained.

The velocity-resolved manifolds in Fig.~\ref{fig.789} (Column A) show that scaling up the Rabi coupling radically relocates the velocity bands satisfying the resonance criteria. In the single-control limit ($\Omega_D \to 0$), the dressed states align closely with the standard $\Lambda + \Theta$ profiles \cite{AMKAPAKA18}. However, under full dual-control operation, Eq.~(\ref{DF61}) reveals a complete modification: rather than a broad, uniform Maxwellian distribution contributing indiscriminately, the system isolates narrow, highly selective velocity bands that carry the non-classical signatures. 

When the constructive conditions of Eq.~(\ref{DF61}) are simultaneously satisfied, they map straight, highly illuminated slanted trajectories across the $(\Delta_P, v)$ plane. This structural change generates two macroscopic consequences: it yields broad side-wings fed by the wide Maxwellian background, while establishing clear transparency windows rooted in the averaged residuals of the off-resonant velocity groups \cite{FLIM05}. These residual features define fundamental performance caps for atomic vapor quantum memories, while the steep positive dispersion slope remaining at line center offers excellent potential for slow-light propagation \cite{AMKAPAKA18}.

Activating $\Omega_D$ drives an intense cross-channel interference that forces the resonance trajectories to track a rotated rectangular hyperbola described by $\Delta_P^{(++-) \atop (--+)} =\pm x \big(1+\frac{a}{x}\big)$ (Figs.~\ref{fig.789}b-f, Column A). The ideal operating symmetric configuration is realized when the ratio $\Omega_C/\Omega_D = 1$. Here, the absorption peak resulting from the cooperative interference of the V and $\Theta$ channels is locked precisely at line center (Fig.~\ref{fig.789}c, Column A) via perfect spatial overlap across all velocity groups. 

This tracking locks the maximum local amplitude inside the high-density core of the Maxwell velocity distribution. As the probe detuning shifts outward, incoherent paths dominate, creating symmetric absorption minima. When the ratio $\Omega_C/\Omega_D$ is decreased, this central peak collapses into a growing transparency minimum, reflecting a constructive quantum interference engineered by the secondary optical pathway ($\ket{2}\to\ket{4}$) that switches the medium between AT splitting, EIA, or EIT behavior depending on field configuration and propagation geometries \cite{AMKAPAKA18}.

The corresponding fluorescence signals (Column B) follow a tightly correlated physical track but exhibit sharper velocity localization. Because fluorescence is fundamentally driven by population factors, it exhibits a higher robustness against pure phase dephasing compared to the fragile coherences measured by absorption. Strong Rabi fields effectively overcome the finite Raman linewidth mask, reducing background fluctuations and narrowing the active bright-state emission ridges.

In general, this detailed comparative analysis between the absorption and fluorescence demonstrates how Doppler averaging converts microscopic velocity-group coherence into macroscopic spectral broadening~\cite{ARUN07}, establishing the fluorescence-based electrometry as a coherence-sensitive yet thermally robust alternative to EIT-based sensing. That is, maintaining wavelength and strength matching minimizes Doppler dephasing and maximizes the bright-state population, thereby enhancing fluorescence-based detection sensitivity~\cite{HSGDA17,GSPH22,PBRAS24}, hence improving the performance of photon sources \cite{LLKM16, KLBTT18}. 

For entanglement, opposite to the weak controls case, the rotated hyperbola profile of the velocity-resolve maps of negativity \( \mathcal{N}_{\mathrm{Dop}}^{AB}(\rho^{\mathrm{ss}})\) is highly enhanced under the strong dressing, compare subpanels (b, c, d, Figs. \ref{fig.135} and \ref{fig.789}, column (C)). Off-diagonal changes qualitatively alter the dressed structure, modifying the eigenvalues abruptly. When $\Omega_D = 0$, the rotated hyperbola $\Delta_P^{(++-) \atop (--+)} =\pm x \big(1+\frac{a}{x}\big)$ effectively interpret the velocity-resolved maps; for $\Omega_D \neq 0$, the hyperbolic bending of entanglement ridges in the $(\Delta_P,v)$ plane emerges distinctly, which can be expressed schematically as
\begin{equation}
\label{DF62}
(\Delta_P + k_P v)^2 \simeq \Omega_C^2 + \Omega_D^2 + \mathcal{O}((k v)^2). 
\end{equation}

Then, the entanglement ridges deviate from straight Doppler-shifted lines and bend outward symmetrically with increasing $|v|$. Generally, in the presence of either one or two controls, the finding show that EIA induces correlations between the two subsystems, while EIT suppresses these correlations effectively. However, due to the  delocalization, the Doppler-averaged profile appears significant only for narrow ranges of \(v\) around $\Delta_P \approx 0$ in similar to the weak controls case, where multiple steady coherences reach sufficient amplitude simultaneously. Already, the residual velocity-dependent slices is responsible for weakening the side shoulders when Doppler-averaging is performed, a noticeable signature of the noneffective contributions of fast atoms due to the out-of-phase correlations. As indicated before, because negativity depends on a competition between coherence-induced terms and population-dominated terms, different dressed branches cross the entanglement threshold at different detunings and velocities. This naturally produces the fragmented, multi-ridge structure observed in panels (\ref{fig.789}b-f, column (C)). For this reason, the hyperbolic features become particularly pronounced in the entanglement maps more than in the absorption and fluorescence spectra because negativity acts as a sharp threshold detector for dressed-state crossings.

The SW function \(W(\rho)\) in Fig.~\ref{fig.789} [column (D)] reveals the phase-space origin of the strongly dressed dynamics. Unlike the weak-field regime's broad horizontal ridge near $v\approx0$ indicating global Doppler-compensated coherence, the strong-field regime breaks this into curved high-density ridges governed by Eq.~(\ref{DF62}).  These manifolds define velocity-selective phase-matching channels in which the generalized detuning \(\Delta_P+k_Pv\) is compensated by the large control-field Rabi frequencies, producing localized coherence accumulation. This restructuring directly reflects in the entanglement dynamics [column (C)], where the negativity \(\mathcal{N}_{\mathrm{Dop}}^{AB}\) becomes finite only along these curved resonance manifolds, while separability dominates elsewhere. Thus, similarly to the weak-field case but now in a strongly dressed configuration, entanglement remains confined to regions where \(W(\rho)\) exhibits strong localization and anisotropic phase-space structure.

The avoided crossings observed in all optical observables appear in \(W(\rho)\) as ridge deformation and splitting, originating from rapid variations in the steady-state coherences \(\rho_{ij}^{\mathrm{ss}}\). These structures are resolved most clearly in the entanglement maps, where they generate abrupt transitions between vanishing and finite negativity. This indicates that entanglement is formed predominantly near the boundaries separating competing dressed-state pathways. Consequently, the geometry of \(W(\rho)\) directly encodes the phase-space loci of entanglement generation.

As the control strengths increase further, the slanted resonance ridges identified in the weak-field regime evolve into hyperbolically curved manifolds satisfying the strong-field balance between Doppler shifts and the dressed-state splitting \(\sqrt{\Omega_C^2+\Omega_D^2}\) [Eq.~(\ref{DF62})]. Coherence redistributes among multiple velocity-selective channels carrying generally out-of-phase contributions. Doppler averaging then produces partial cancellation rather than coherent accumulation: the averaged SW function develops weaker and redistributed negative regions, whereas the Doppler-averaged negativity remains enhanced because of the strongly localized entanglement. This explains the broader, flatter, multi-peaked 2D spectra relative to the weak regime.

The numerical results reveal that increasing control strengths does not merely modify spectral widths, but drives a qualitative transition in the underlying coherence topology: Doppler-compensated linear manifolds evolve into velocity-selective dressed-state channels. Since absorption, fluorescence, negativity and SW phase-space distributions probe different projections of the same steady-state density matrix, comparing their responses across weak and strong control regimes provides insight into which observables remain experimentally robust and which retain direct sensitivity to entanglement generation. In phase space and correlation coordinates, the strong-field SW mappings (Column D) mirror these hyperbolic structures explicitly, manifesting deep negative dips at line center that accurately coincide with the localized peaks of the negativity maps (Column C). Thus, under strong driving, the phase space non-classicality is highly robust, concentrating quantum correlations into specific velocity channels that directly match the macroscopically observed optical profiles.

\subsection{Counter-propagating First control}

In contrast to the previous configuration, we now consider an interaction geometry where the first control field ($k_C$) is chosen to be counter-propagating relative to the probe field. The corresponding Doppler-shifted detunings are given by:
\begin{equation}
\label{DF64}
\begin{gathered}
\Delta_{P, D}^{(+-+) \atop (-+-)}=\Delta_{P, D}(\pm) k_{P, D} v,
\\
\Delta_C^{(+-+) \atop (-+-)}=\Delta_C(\mp) k_C v,
\end{gathered}
\end{equation}
which dictates the following multi-photon resonance conditions
\begin{equation}
\label{DF644}
\begin{gathered}
\Delta_P^{(+-+) \atop (-+-)}(v) + \Delta_C^{(+-+) \atop (-+-)}(v) = (\mp)(k_P + k_C) v,
\\
\Delta_P^{(+-+) \atop (-+-)}(v) + \Delta_D^{(+-+) \atop (-+-)}(v)= (\mp)(k_P + k_D) v.
\end{gathered}
\end{equation}
%
%\begin{equation}
%\label{DF644}
%\Delta_P^{(+-+) \atop (-+-)}(v) + \Delta_C^{(+-+) \atop (-+-)}(v) = (\mp)(k_P + k_C) v, \quad\text{and}\quad  \Delta_P^{(+-+) \atop (-+-)}(v) + \Delta_D^{(+-+) \atop (-+-)}(v)= (\mp)(k_P + k_D) v,
%\end{equation}
This framework fundamentally contradicts the phase-matching condition defined in Eq.~(\ref{DF60}), leading to velocity-dependent resonance pathways that deviate significantly from Eq.~(\ref{DF61}):
\begin{equation}
\label{DF641} 
\Delta_P^{(+-+) \atop (-+-)} =
\begin{cases}
(\mp)2k v \pm k v\Big(\frac{\Omega_D^2-\Omega_C^2}{4k^2 v^2}\Big),\\[2mm]
(\mp)2k v \pm k v \Big(1+\frac{\Omega_D^2+\Omega_C^2}{4k^2 v^2}\Big),
\end{cases}
\end{equation}

\begin{figure*}[!t]
    \centering
    \vspace{-0.2cm}
    \begin{minipage}{0.23\linewidth}
        \centering
        {\footnotesize (A) EIA \(\mathrm{Im}(\rho_{12}\Gamma_2/\Omega_P)\)}\\[-2pt]
        \includegraphics[width=5.5cm,height=20.cm]{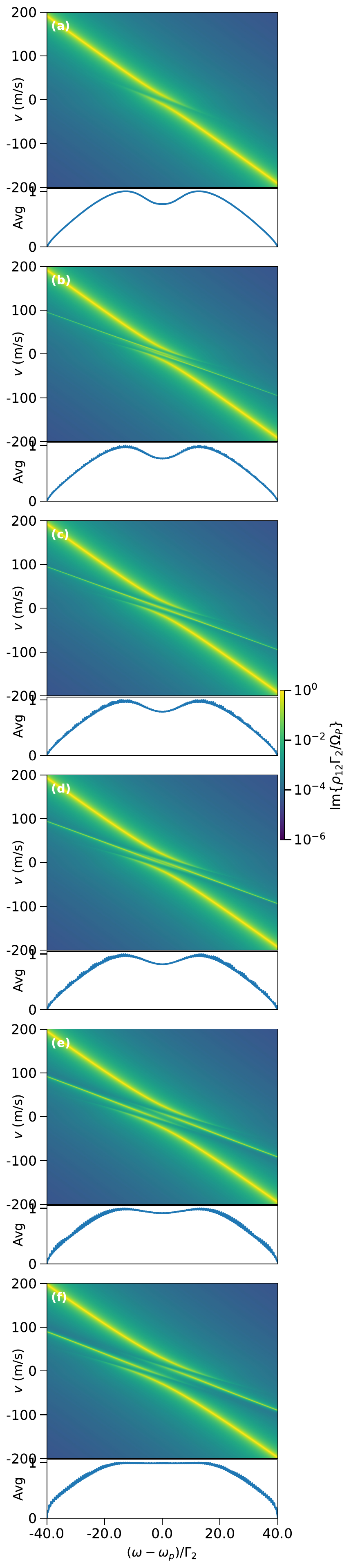}
    \end{minipage}
    \hspace{0.1cm}
    \begin{minipage}{0.23\linewidth}
        \centering
        {\footnotesize (B) Fluorescence $S(\omega)$}\\[-2pt]
        \includegraphics[width=5.5cm,height=20.cm]{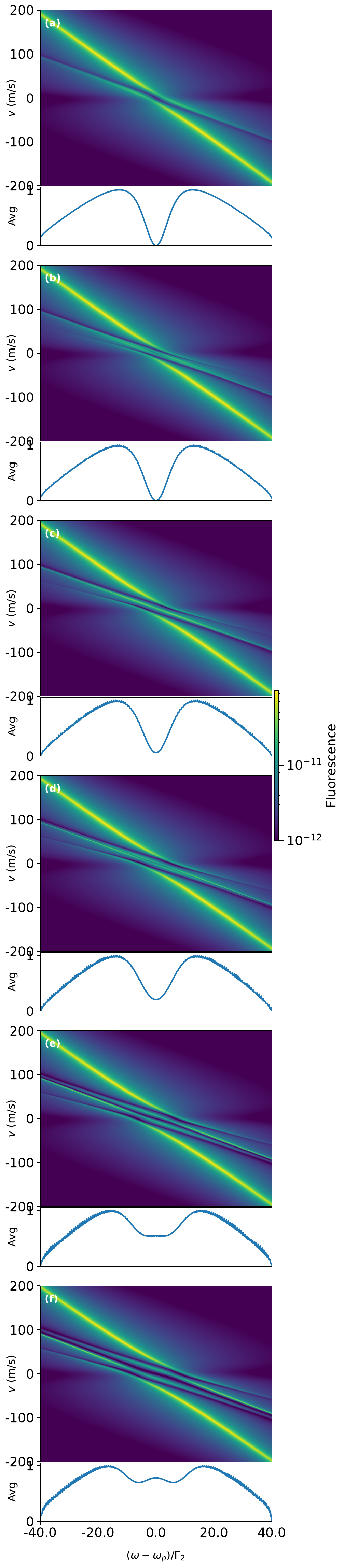}
    \end{minipage}
    \hspace{0.1cm}
    \begin{minipage}{0.23\linewidth}
        \centering
        {\footnotesize (C) Negativity $\mathcal{N}(\rho)$}\\[-2pt]
        \includegraphics[width=5.5cm,height=20.cm]{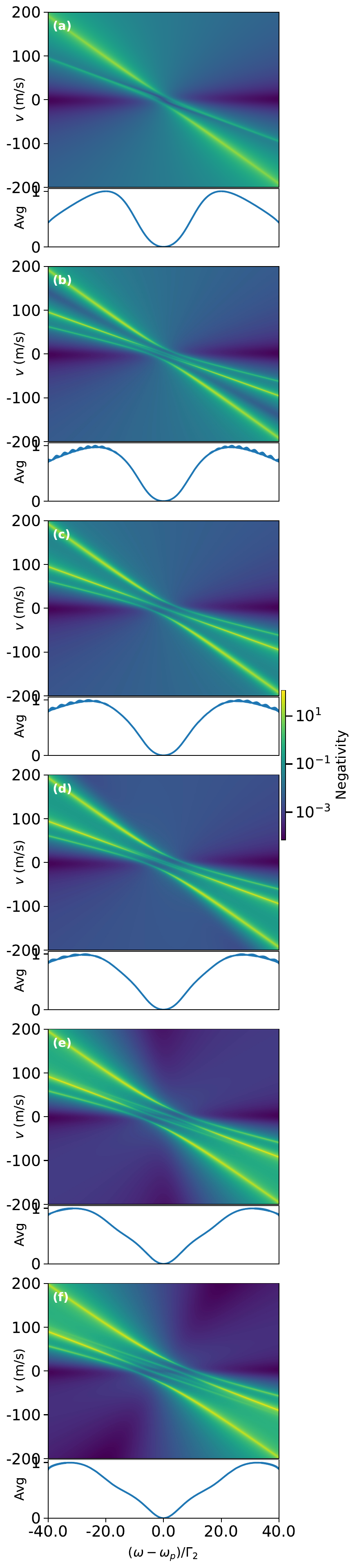}
    \end{minipage}
    \hspace{0.1cm}
    \begin{minipage}{0.23\linewidth}
        \centering
        {\footnotesize (D) SW W-f $W(\rho)$}\\[-2pt]
        \includegraphics[width=5.5cm,height=20.cm]{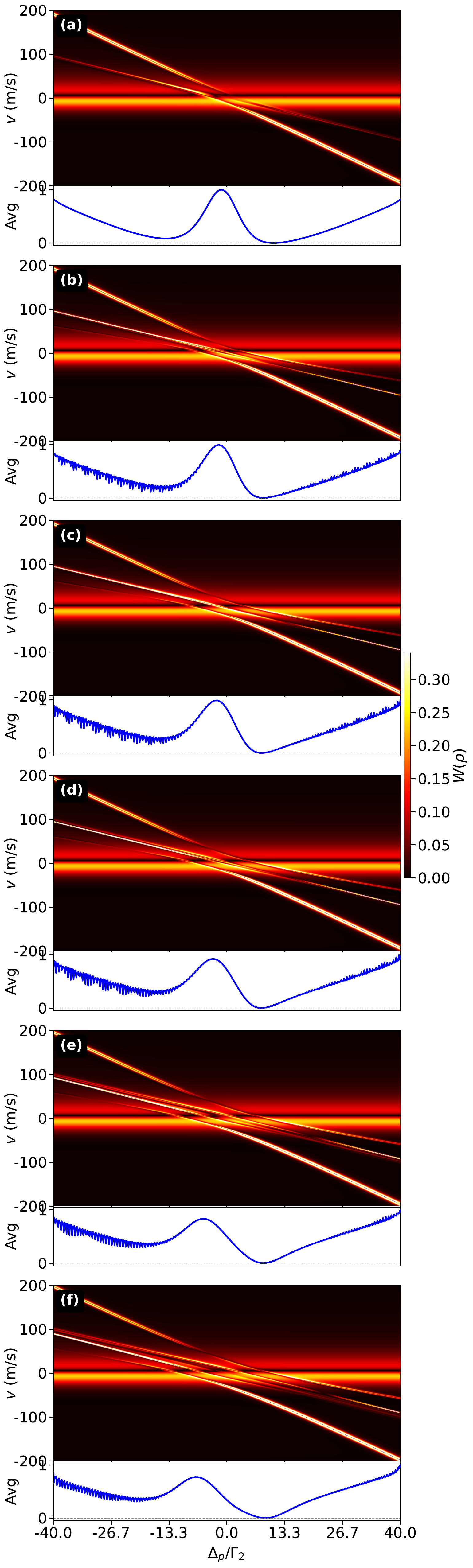}
    \end{minipage}    
    \vspace{-0.1cm}
    \caption{\label{fig.101112} Multi-observable mapping under a counter-propagating first control field geometry ($k_C$) for varying intensities $(\Omega_C, \Omega_D)$ from (a) to (f). The counter-propagating geometry introduces an unavoidable linear Doppler term $(\mp)2kv$, forcing the state configurations into highly tilted asymptotic branches. This geometric fragmentation isolates fluorescence into sharp emission lines but subjects the phase-space Wigner function and negativity to severe thermal phase cancelation during Doppler averaging.}
\end{figure*}

Equation~(\ref{DF641}) highlights the absolute dominance of linear Doppler terms over the dressed-state configurations compared to the co-propagating control setup. Even under exact Rabi frequency matching ($\Omega_C=\Omega_D$), the central resonance ridge cannot maintain verticality; instead, it becomes heavily tilted with a slope of $2k$ in the $(\Delta_P, v)$ plane, driven by the linear prefactor $(\mp)2k v$. This induces a strong spatial coincidence between the multi-photon resonances and the dressed-state trajectories, thoroughly altering the evolution of all four observables under thermal averaging.

\paragraph{\textbf{Asymptotic Geometric Regimes and Linearity Dominance:}}

For a single control field ($\Omega_D=0$), the central resonance ridge persists as a background structure, causing the Doppler-averaged profile to exhibit a shallow, reduced transparency dip [Fig.~\ref{fig.101112}a, Columns (A-D)]. When the second control field $\Omega_D$ is activated, the central ridge remains heavily slanted due to the velocity-dependent dressing of the nested $\mathrm{V}+\Theta$ channels \cite{BBBN20,VINA15,HSKGC18} [Fig.~\ref{fig.101112}b-f, Columns (A-D)]. 

Diagonalization of the exact Hamiltonian yields the same quadratic constraint as the previous configuration. However, within the experimentally relevant large-Doppler regime ($|k v| \gg \Omega_C,\Omega_D$), the constant dressing term $\Omega_C^2+\Omega_D^2$ becomes entirely subdominant in shaping the ridge orientations. Consequently, the exact hyperbola is tightly approximated by its asymptotic arms:
\begin{equation}
\label{DF642}
\Delta_P \simeq \left(-k_P \pm \sqrt{\alpha}\,k\right)v.
\end{equation}
These asymptotic branches map out straight, strongly tilted tracks in the $(\Delta_P,v)$ plane, whose slopes are governed strictly by the Doppler geometry and the dressed-state mixing coefficient $\alpha$ (Fig.~\ref{fig.101112}, Column A). Because the condition $|k v|\gg \sqrt{\Omega_C^2+\Omega_D^2}$ is satisfied across nearly the entire thermal velocity profile, the hyperbola is visually indistinguishable from straight tilted lines, while its localized curvature near $v=0$ is completely suppressed by thermal effects [Fig.~\ref{fig.101112}b-e, Columns (A, B)]. Therefore, the prominent tilted ridges observed in the velocity-resolved maps do not signal a weak-Doppler interaction but are explicit manifestations of the asymptotic limits of the exact hyperbolic dispersion relation.

\paragraph{\textbf{Coherence Fragmentation and Observable Splitting:}}

The introduction of the second control field $\Omega_D$ shifts these asymptotic arms outward in the $(\Delta_P,v)$ plane via the parametric threshold $|k v|\gg \sqrt{\Omega_C^2+\Omega_D^2}$ [Fig.~\ref{fig.101112}d-f, Columns (A-D). This driving produces two immediate consequences: the central resonance splits into multiple distinct branches corresponding to uncoupled bright dressed states, and the spectral separation between these branches expands parametrically as:
\begin{equation}
\Delta\lambda \sim 2\sqrt{\Omega_C^2+\Omega_D^2},
\end{equation}
This scaling explains the systematic broadening of the ridge gaps observed across all velocity-resolved scans with increasing values of $\Omega_D$  [Fig.~\ref{fig.101112}e, f, Columns (A-D). 

Because absorption, fluorescence, entanglement, and phase-space density are all governed by the identical underlying dressed-state Hamiltonian, this hyperbolic dispersion profile provides a unified explanation for the tilted and split ridge networks appearing across all four domains. However, the precise manifestation remains highly observable-dependent. 

For weak to moderate driving fields [Figs.~\ref{fig.101112}a-c, Column B], the fluorescence response mirrors the absorption profiles, tracking the dominant tilted branches. Yet, fluorescence exhibits an enhanced spatial localization because spontaneous radiative decay channels selectively amplify specific dressed-state transitions. In other words, the fluorescence resolves these branches into sparse, high-contrast emission ridges because it interrogates individual dressed-state decay channels rather than the mean probe coherence. This distinction, well known in resonance-fluorescence theory (see, e.g.,  \cite{BREMAL11}), explains why Fig. (\ref{fig.101112}, column (B)) displays sharper, more fragmented structures than Fig. (\ref{fig.101112}, column (A)) despite sharing the same underlying Doppler-shifted dressed spectrum. As $\Omega_D$ scales up [Figs.~\ref{fig.101112}d-f, Column B], the AT splitting embedded within $\mathrm{det}(\mathbf{Q})$ -- and weighted by the population sum $Q_{11}^{-1}R_1+Q_{13}^{-1}R_3$ -- resolves into multiple, well-isolated poles whose velocity projections no longer overlap. 

In stark contrast, the state variables in phase space undergo complete geometric reorganization. The phase-space SW distribution $W(\rho)$ [Column D] reveals a complete breakdown of Doppler compensation. While the previous geometry was dominated by a phase-synchronized vertical manifold near $v \approx 0$ that survived averaging, the current counter-propagating configuration introduces the linear term $(\mp)(k_P+k_C)v$, eliminating vertical localization entirely. The SW maps reorganize into strongly tilted, disjointed linear tracks following the asymptotic arms of Eq.~(\ref{DF642}). Coherence is thus fragmented along multiple independent Doppler-shifted pathways rather than concentrated within a collective, velocity-robust manifold.

This fragmentation is closely tracked by the entanglement maps [Column C]. For weak to moderate control strengths ($\Omega_D$) [Fig.~\ref{fig.101112}(a-d), column (C)], each velocity-resolved negativity ridge coincides precisely with its counterparts in absorption (Fig.~\ref{fig.101112}, column (A)) and fluorescence (Fig.~\ref{fig.101112}, column (B)), because a single bright dressed state dominates the system dynamics and preserves a shared velocity selectivity. However, as $\Omega_D$ scales upward, this structural harmony degrades. The intense control field stimulates the highly nonlinear coherence conditions required for non-local states ($\rho_{ii}^{\mathrm{ss}}\rho_{jj}^{\mathrm{ss}} \gtrsim \left|\rho_{ij}^{\mathrm{ss}}\right|^2$) but drives them away from resonance. Due to the linear Doppler dominance near line center ($\Delta_P \approx 0$), slow atoms cannot maintain sufficient coherence cross-correlations between the atomic subsystems, causing the entanglement ridges to terminate abruptly and demonstrating that the correlation-spectroscopy correspondence is partial rather than universal.

\paragraph{\textbf{Thermal Averaging and Phase Cancellation:}}

The physical impact of this linear Doppler dominance becomes most evident after performing the Maxwell-Boltzmann thermal convolution. In the absorption spectra, the widening of the central transparency window is a direct consequence of the outward ridge splitting, while its depth is determined by the overlap region between these branches [Figs.~\ref{fig.101112}a-e, Column A]. When the intensity ratio $\Omega_C/\Omega_D$ exceeds approximately one-half, the separation between the dressed branches matches or exceeds the Doppler width, allowing multiple velocity classes to satisfy near-transparency conditions simultaneously over a wide detuning sweep. The Doppler-averaged absorption $\rho_{12_{Dop}}^{\mathrm{ss}}$  Eq.(\ref{DF34}) then accumulates overlapping transparency contributions from several split branches, yielding an effectively broadened window. Thus, the suppressed absorption signal near line center arises from ladder-induced destructive interference in the probe coherence rather than from isolated dressed eigenvalues.

Conversely, the Doppler-averaged fluorescence remains sharply structured and highly resistant to smearing [Figs.~\ref{fig.101112}a-e, Column B]. In the absorption response, overlapping ridges cancel out due to the subtractive weight difference $Q_{11}^{-1}R_1-Q_{13}^{-1}R_3$, blurring into a wide, shallow feature [Fig.~\ref{fig.101112}f, Column A]. In fluorescence, however, the thermal averaging integrates localized spectral poles rather than broad susceptibilities, meaning individual dressed-state decay channels are successfully resolved [Fig.~\ref{fig.101112}f, Column B]. 

Finally, the thermal averaging acts as a destructive phase filter for the quantum state variables. Because the fragmented, tilted ridges of $W(\rho)$ [Fig.~\ref{fig.101112}, Column C] and $\mathcal{N}(\rho)$ [Fig.~\ref{fig.101112}, Column D] intersect the Maxwell velocity distribution over highly restricted intervals and with rapidly varying phases, they experience severe out-of-phase cancellation during integration. This produces broadly smeared, low-contrast averaged SW profiles and heavily suppresses the central integrated negativity $\mathcal{N}(\rho)$. Consequently, while local quantum entanglement remains exceptionally robust along individual velocity-selected channels, its macroscopically integrated contribution is severely penalized due to the geometric elimination of velocity-insensitive phase alignment.

\subsection{Mixed-propagating fields in the direction or reverse to moving atoms}

We now turn our attention to the configuration defined by Eqs.~(\ref{DF667}-\ref{DF662}), where all three optical fields are assumed to be co-propagating with respect to one another, but propagate in a direction counter to the atomic velocity vector. Under complete multi-photon resonance ($\Delta_C=\Delta_D=0$), this specific propagation geometry establishes the following resonance criteria
\begin{equation}
\label{DF666}
\begin{gathered}
\Delta_P^{(---) \atop (+++)}(v) - \Delta_C^{(---) \atop (+++)}(v) = (\pm)(k_P - k_C) v,
\\
\Delta_P^{(---) \atop (+++)}(v) - \Delta_D^{(---) \atop (+++)}(v)= (\pm)(k_P + k_D) v.
\end{gathered}
\end{equation}
%
%\begin{equation}
%\label{DF666}
%\Delta_P^{(---) \atop (+++)}(v) - \Delta_C^{(---) \atop (+++)}(v) = (\pm)(k_P - k_C) v, \quad\text{and}\quad  \Delta_P^{(---) \atop (+++)}(v) - \Delta_D^{(---) \atop (+++)}(v)= (\pm)(k_P + k_D) v,
%\end{equation}
This framework is quantitatively distinct from the conditions derived in Eqs.~(\ref{DF60}) and (\ref{DF644}), yielding dressed-state resonance trajectories of the form:
\begin{equation}
\label{DF661} 
\Delta_P^{(---) \atop (+++)} =
\begin{cases}
(\pm)k v \pm k v\Big(\frac{\Omega_D^2-\Omega_C^2}{4k^2 v^2}\Big),\\[2mm]
(\pm)k v \pm k v \Big(1+\frac{\Omega_D^2+\Omega_C^2}{4k^2 v^2}\Big), 
\end{cases}
\end{equation}

Comparing Eq.~(\ref{DF661}) to the previously analyzed counter-directed setup in Eq.~(\ref{DF641}) reveals a significantly weaker linear Doppler prefactor ($(\pm)kv$ versus $(\mp)2kv$). This reduction in linear Doppler vulnerability profoundly changes how the atomic medium maps microscopic velocity profiles onto macroscopic optical signals and quantum states.

\paragraph{\textbf{Hyperbolic Bending and Suppression Thresholds:}}

\begin{figure*}[!t]
    \centering
    \vspace{-0.2cm}
    \begin{minipage}{0.23\linewidth}
        \centering
        {\footnotesize (A) EIA \(\mathrm{Im}(\rho_{12}\Gamma_2/\Omega_P)\)}\\[-2pt]
        \includegraphics[width=5.5cm,height=20.cm]{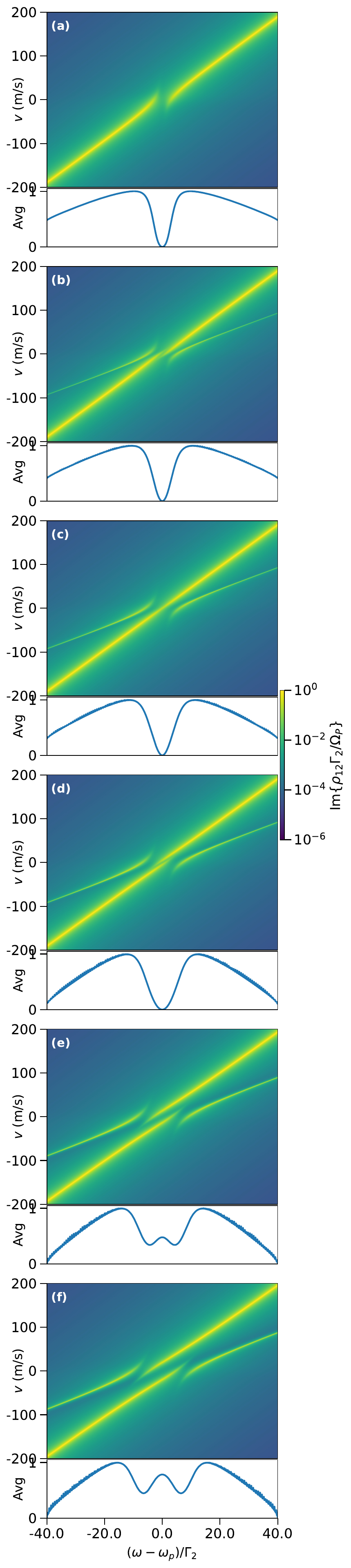}
    \end{minipage}
    \hspace{0.1cm}
    \begin{minipage}{0.23\linewidth}
        \centering
        {\footnotesize (B) Fluorescence $S(\omega)$}\\[-2pt]
        \includegraphics[width=5.5cm,height=20.cm]{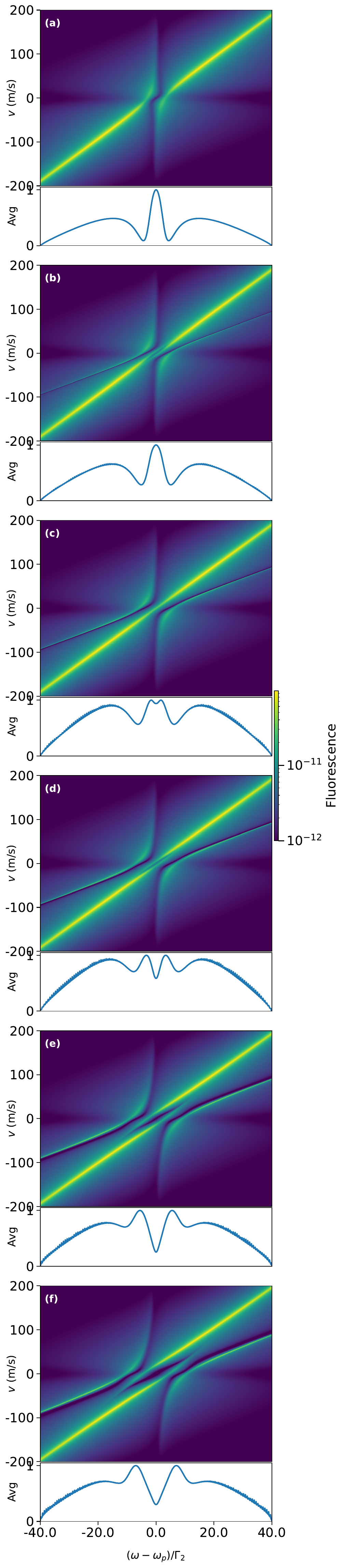}
    \end{minipage}
    \hspace{0.1cm}
    \begin{minipage}{0.23\linewidth}
        \centering
        {\footnotesize (C) Negativity $\mathcal{N}(\rho)$}\\[-2pt]
        \includegraphics[width=5.5cm,height=20.cm]{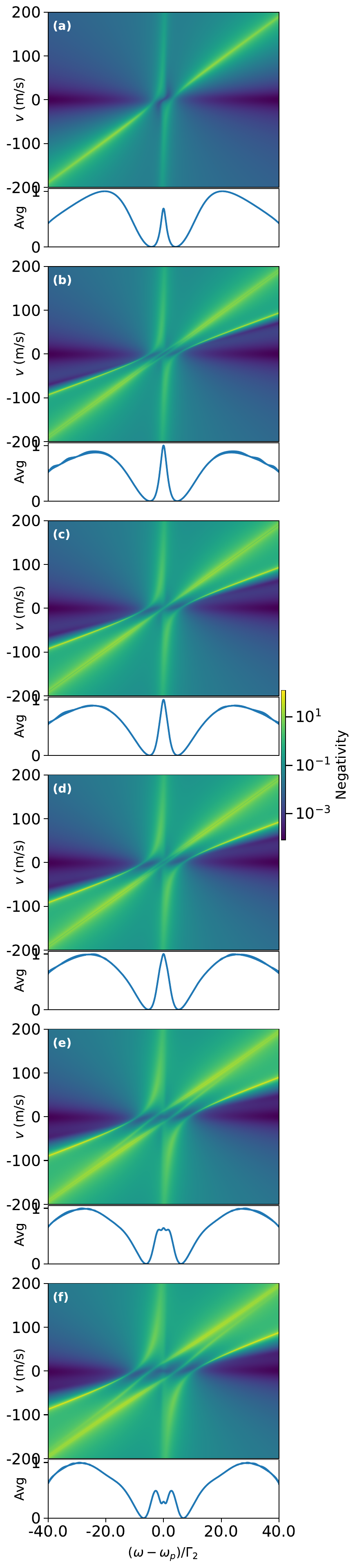}
    \end{minipage}
    \hspace{0.1cm}
    \begin{minipage}{0.23\linewidth}
        \centering
        {\footnotesize (D) SW W-f $W(\rho)$}\\[-2pt]
        \includegraphics[width=5.5cm,height=20.cm]{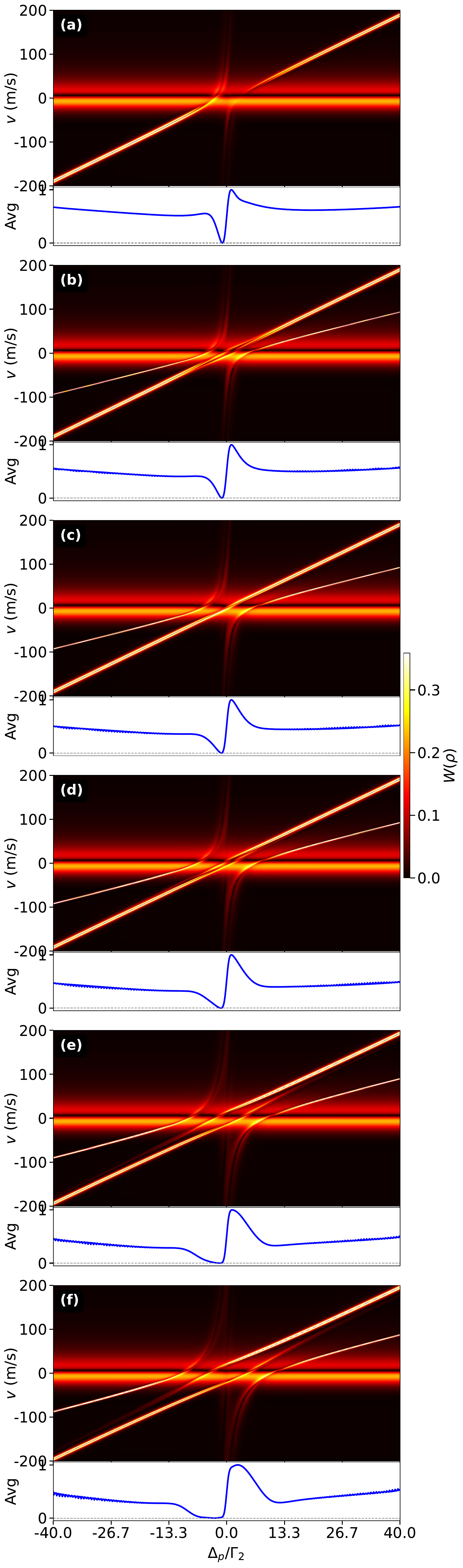}
    \end{minipage}        
    \vspace{-0.1cm}
    \caption{\label{fig.131415} Multi-observable mapping under a mixed-propagating geometry for varying intensities $(\Omega_C, \Omega_D)$ from (a) to (f). The weaker linear Doppler prefactor allows the resonance tracks to preserve their hyperbolic curvature around line center. This promotes simultaneous population and coherence accumulation at $\Delta_P \approx 0$, driving visible EIA-EIT oscillations and sustaining broad, on-resonance quantum entanglement under thermal averaging.}
\end{figure*}

Owing to the weaker linear Doppler prefactor, the velocity-resolved profiles in Fig.~\ref{fig.131415} diverge markedly from those in Fig.~\ref{fig.101112}. In the single-control limit ($\Omega_D=0$), this geometry shows a unified physical behavior with the counter-propagating second control setup [compare Fig.~\ref{fig.789}a, Column A and Fig.~\ref{fig.131415}a, Column A]. Specifically, the suppressed impact of the linear term preserves a steep hyperbolic bending of the resonance branches around line center, contrasting sharply with the rigidly straight lines of Eq.~(\ref{DF641}).

This persistent curvature forces an immediate, localized spatial overlap between the multi-photon resonance trajectories and the dressed-state channels. Because the trajectories remain curved, standard destructive interference channels are no longer broad enough to thoroughly suppress the macroscopic susceptibility. 

For small to moderate secondary control strengths ($\Omega_D$), the tilted central ridge carries insufficient optical brightness. As a result, the Doppler-averaged absorption profile successfully preserves a highly localized transparency window directly at line center ($\Delta_P=0$), fundamentally distinguishing this geometry from the counter-propagating first control case  Eq. (\ref{DF64}). Furthermore, as $\Omega_D$ scales up, an early spatial splitting of the slanted central ridges occurs, accompanied by a rapid parameter sweeping of the dressed-state positions across the velocity distribution.

\paragraph{\textbf{Coherence Overlap and Spectral Oscillations:}}

As the system enters intense dual-control regimes, the wide structural separation engineered between the two-photon resonance tracks and the dressed-state boundaries induces high-frequency macroscopic oscillations alternating between EIA and EIT behaviors [Figs.~\ref{fig.131415}e,f, Column A]. 

This state-mixing manifests in the fluorescence response through distinct physical conditions. In the susceptibility response (absorption), poles are weighted by the subtractive difference $Q_{11}^{-1}R_1-Q_{13}^{-1}R_3$, which favors signal smoothing. In contrast, resonance fluorescence amplitudes are governed by the population sum $Q_{11}^{-1}R_1+Q_{13}^{-1}R_3$, which acts to isolate individual radiative pathways. In the single-control limit ($\Omega_D=0$), this additive weighting creates a perfect overlap between the bright dressed states and the multi-photon resonance tracks. Under these conditions, the multi-photon pathways completely dominate the emission dynamics, maximizing the Doppler-averaged fluorescence. This creates two narrow transparency dips symmetrically sandwiched between a sharp central emission peak and two side shoulders [Fig.~\ref{fig.131415}a, Column B], mirroring the balanced configuration of Fig.~\ref{fig.789}a.

When $\Omega_D$ is turned on, the two-photon resonance and dressed-state trajectories begin to uncouple spatially, altering the thermal averaging profile [Fig.~\ref{fig.131415}b, Column B]. Continuously increasing $\Omega_D$ expands this uncoupling gap and fragments the central ridge. After Doppler averaging, this fragmentation splits the central emission peak into a doublet, whose depth and structural width are regulated by the intensity ratio $\Omega_C/\Omega_D$. Minimizing the $\Omega_C/\Omega_D$ ratio drives deep transparency dips into the emission spectrum [Figs.~\ref{fig.131415}d-f, Column B], showing how control fields can open narrow, high-contrast transmission channels within a fluorescence profile.

\paragraph{\textbf{Phase Synchronization and On-Resonance Correlation Dynamics:}}

The reduced linear Doppler prefactor also alters the quantum correlation landscape. At first glance, the mixed-propagating configuration introduces a relative phase dispersion among the atomic coherences. This phase variance broadens the velocity-resolved entanglement ridges [Column C], as individual density-matrix eigenvalues find independent velocity intervals where multiple relative phase combinations of $\rho_{ij}^{\mathrm{ss}}$ satisfy the negativity condition. 

Crucially, despite the extra coherence pathways created by $\Omega_D$, the slanted multi-photon resonance tracks function as the primary engine for non-local correlations. This interaction renders the entanglement far less velocity-selective than in previous geometries. Multiple broad and narrow velocity groups contribute cooperatively, allowing different coherence phases to align constructively and satisfy the partial-transposition negativity criteria ($\rho_{ii}^{\mathrm{ss}}\rho_{jj}^{\mathrm{ss}} \gtrsim \left|\rho_{ij}^{\mathrm{ss}}\right|^2$). Consequently, the non-zero negativity spans wider velocity segments, stabilizing a broad, robust macroscopic entanglement window. Once the slanted two-photon ridge splits under intense driving, the central negativity peak undergoes amplitude reduction and rapid oscillations [Figs.~\ref{fig.131415}e,f, Column C], matching the trends observed in both absorption and fluorescence, see Figs. (\ref{fig.131415}, columns (A, B)).

This cooperative phase alignment reshapes the phase-space topography. The SW function $W(\rho)$ [Column D] displays a phase-space organization that differs fundamentally from all prior geometries, indicating a partial, highly favorable breakdown of Doppler compensation. In the fully compensated geometry (Fig.~\ref{fig.789}), the avoided crossing is driven by the two-photon resonance strip, which delays coherence accumulation; this suppresses entanglement at $\Delta_P \approx 0$ and forces it to emerge only on the off-resonant wings where population and coherence conditions balance. 

In the present mixed-propagating configuration, the avoided crossing is instead governed by velocity-dependent dressed-state pathways. The resonance structure becomes inherently Doppler-sensitive, triggering immediate coherence redistribution at exact probe resonance. Consequently, atomic populations and optical coherences accumulate simultaneously at line center ($\Delta_P\approx 0$), enabling strong quantum entanglement to build concurrently at resonance and across the side wings. This shifts the Doppler-averaged Wigner distribution from a flat profile into a highly curved, EIT-like bending profile centered around resonance.

\section{Doppler-odd projection as an entanglement proxy}
\label{sec.7_2}

The SW quasiprobability distribution provides a complete phase-space representation of finite-dimensional quantum states, encoding populations, coherences, and interference structures within a unified framework~\cite{TILMA16,RUNDLE17}. Because both the SW distribution and negativity \(\mathcal{N}^{AB}\) originate from the same steady-state density operator,
\(
\rho_{\mathrm{ss}},
\)
their Doppler-resolved structures may exhibit correlated behavior.

\begin{figure*}[!t]
    \centering
    \vspace{-0.2cm}
    %--- First figure ---
    \begin{minipage}{0.23\linewidth}
        \centering
        {\footnotesize (A) SW(\(\Delta_C\)), \(\Delta_{P, C, D}^{(---) \atop (+++)}\)}\\[-2pt]
        \includegraphics[width=3.7cm,height=15.6cm]{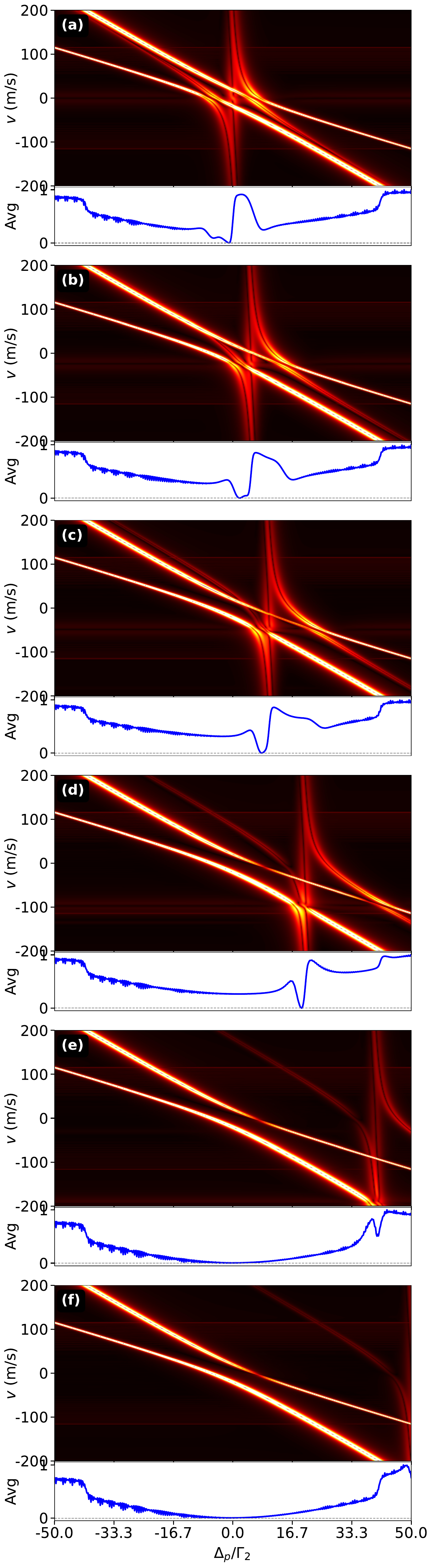}
    \end{minipage}
    \hspace{-1.1cm}
    %--- Second figure ---
    \begin{minipage}{0.23\linewidth}
        \centering
        {\footnotesize (B) SW(\(\Delta_D\)), \(\Delta_{P, C, D}^{(---) \atop (+++)}\)}\\[-2pt]
        \includegraphics[width=3.5cm,height=15.5cm]{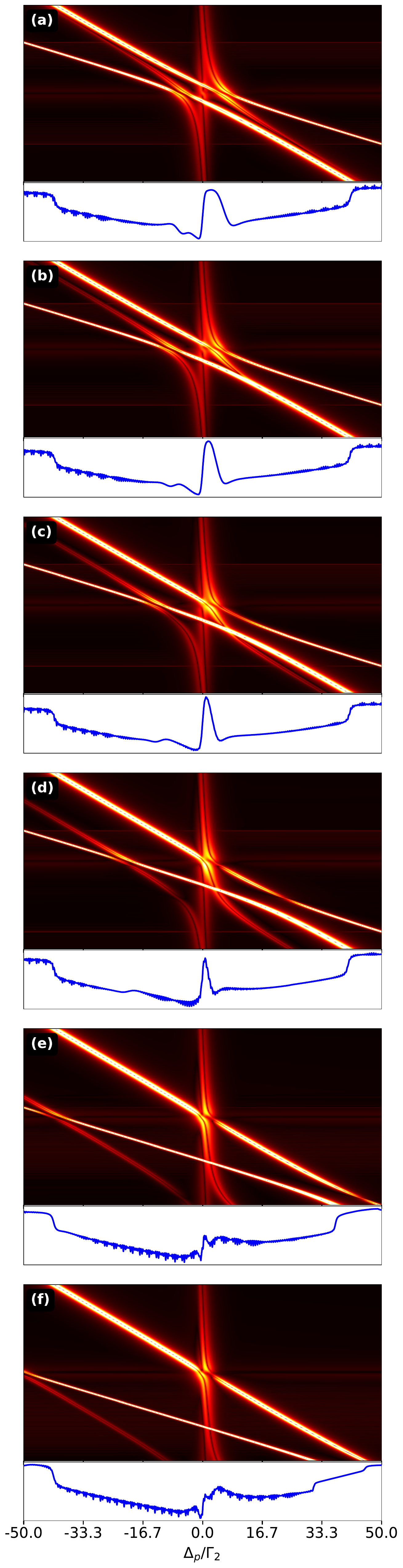}
    \end{minipage}
    \hspace{-1.2cm}
    %--- Third figure ---
    \begin{minipage}{0.23\linewidth}
        \centering
        {\footnotesize (C) SW(\(\Omega\)), \(\Delta_{P, C}^{(++-) \atop (--+)}\)}\\[-2pt]
        \includegraphics[width=3.5cm,height=15.5cm]{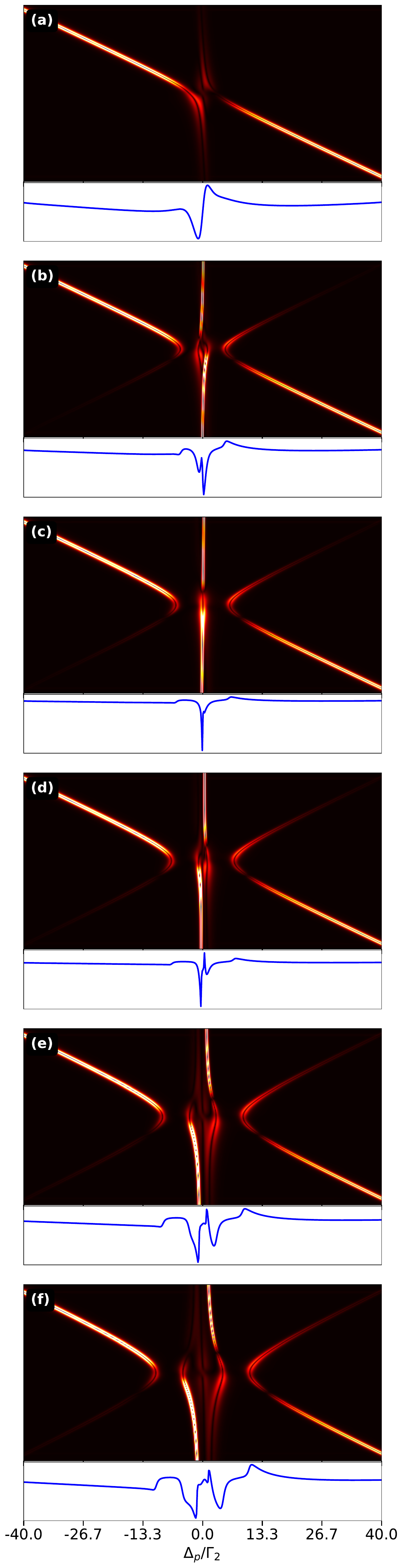}
    \end{minipage}
    \hspace{-1.25cm}
    %--- Third figure ---
    \begin{minipage}{0.23\linewidth}
        \centering
        {\footnotesize (D) SW(\(\Omega\)), \(\Delta_{P, D}^{(+-+) \atop (-+-)}\)}\\[-2pt]
        \includegraphics[width=3.5cm,height=15.5cm]{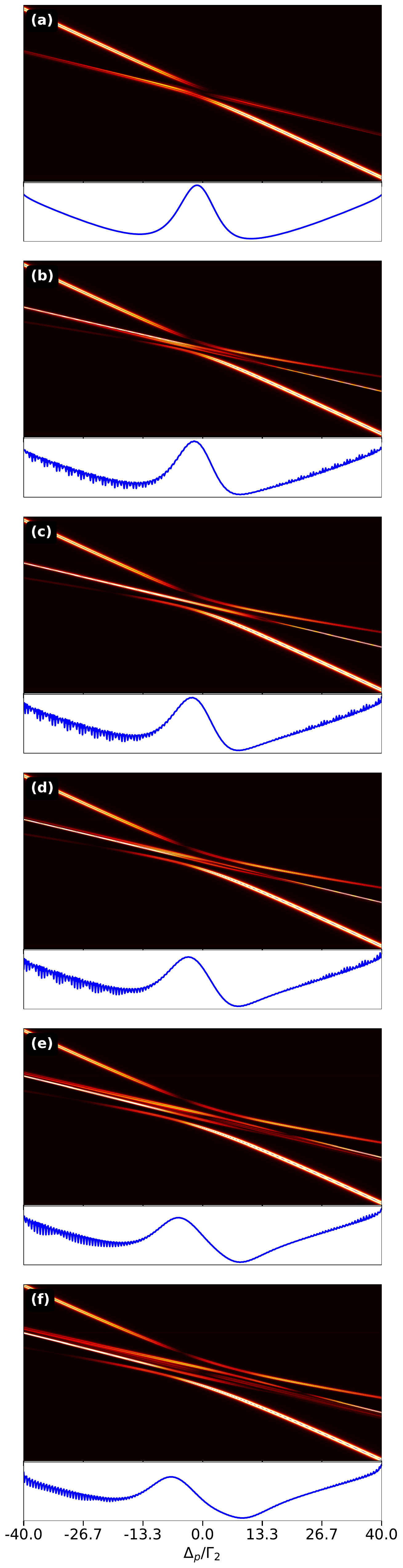}
    \end{minipage}        
    \hspace{-1.2cm}
    %--- Third figure ---
    \begin{minipage}{0.23\linewidth}
        \centering
        {\footnotesize (E) SW(\(\Omega\)), \(\Delta_{P, C, D}^{(---) \atop (+++)}\)}\\[-2pt]
        \includegraphics[width=3.6cm,height=15.5cm]{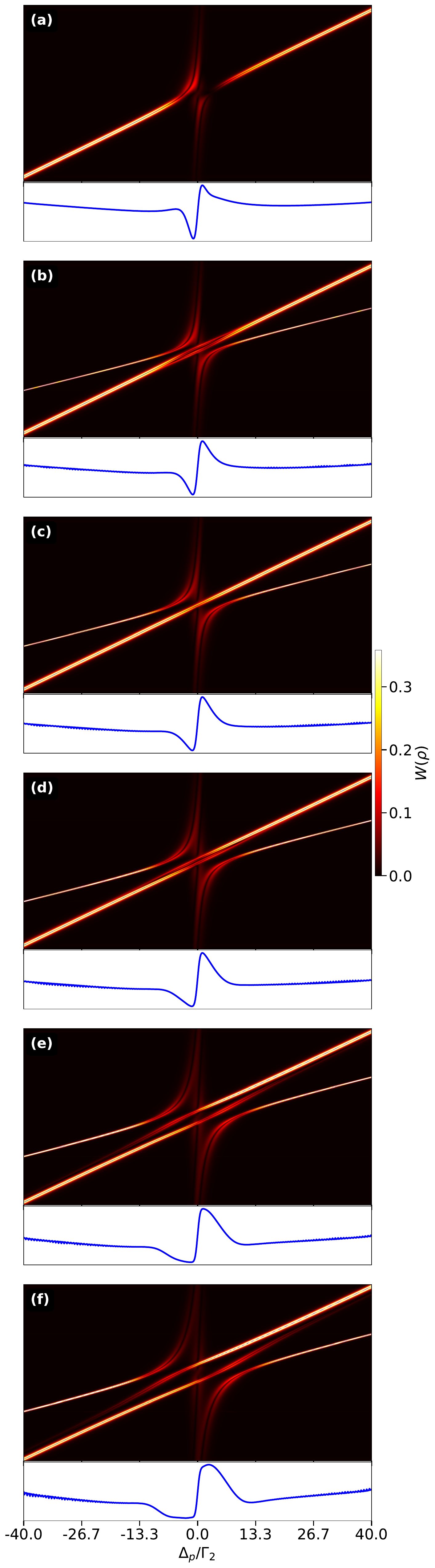}
    \end{minipage}
    \vspace{-0.3cm}
    \caption{Doppler-resolved SW Wigner function $W(\rho)$ against the probe detuning $\Delta_p$. The columns (A)-(E) reflect the horizontal-strip-filtered landscapes in Figs. (\ref{fig.00}, \ref{fig.01}), \ref{fig.789}-\ref{fig.131415} column (D), respectively.}
    \label{fig.161718}
\end{figure*}

A systematic comparison between \(\mathcal{N}^{AB}\) and SW distributions reveals pronounced similarities across coherence-dominated regions, including avoided crossings, slanted resonant branches, and detuning-dependent interference features. Nevertheless, these correspondences are not exact: residual discrepancies emerge in velocity-symmetric regions near
\(
v\approx0,
\)
where Doppler-even population dynamics contribute significantly while entanglement remains comparatively insensitive.

This distinction suggests that SW quasiprobabilities contain both entanglement-relevant and entanglement-irrelevant information. Isolating the former motivates decomposition into Doppler-even and Doppler-odd sectors.

Assuming local smoothness of the SW distribution with respect to atomic velocity,
the Doppler dependence may be expanded near \(v=0\) as \(SW(v,\Delta_P)=SW(0,\Delta_P)+v\left.
\frac{\partial SW}{\partial v}\right|_{v=0}+\mathcal O(v^2)\). The corresponding even and odd Doppler sectors become \(SW_{\rm even}=\big(SW(v)+SW(-v)\big)/2, \quad SW_{\rm odd}=\big(SW(v)-SW(-v)\big)/2\). The even contribution is dominated by velocity-symmetric populations and broad background structures, whereas the odd component preferentially emphasizes coherence-driven asymmetries induced by multiphoton interference.

To suppress this contribution, we subtract the $\Delta_P$-average at each velocity,
\[
SW(v,\Delta_P) \;\rightarrow\; SW(v,\Delta_P) - \langle SW(v,\Delta_P) \rangle_{\Delta_P},
\]
thereby removing $\Delta_P$-uniform backgrounds. Operationally, this eliminates the low-frequency sector of the Doppler-even component \(SW_{\mathrm{even}}(v)=\langle SW(v,\Delta_P)\rangle_{\Delta_P}\), while preserving velocity-asymmetric, coherence-dominated features. Motivated by the observed correspondence between coherence-dominated SW structures and negativity, we introduce the Doppler-odd SW component as a phenomenological estimator,

\begin{equation}
\label{Estimator}
E_{SW}
\equiv
SW_{\rm odd}(v,\Delta_P),
\end{equation}
which serves as an experimentally accessible indicator of entanglement-sensitive coherence rather than a strict entanglement monotone. We emphasize that this constitutes an empirical mapping rather than a strict equivalence: regions of strong SW oscillations consistently coincide with maxima of entanglement.

Since SW quasiprobabilities can be experimentally reconstructed via phase-sensitive techniques such as homodyne tomography and parity-based measurements~\cite{LVOVSKY09,LEIBFRIED96}, this correspondence provides a practical route for entanglement estimation in complex systems where full state reconstruction or eigenvalue computation is prohibitive.

\subsection{Validation of the SW-based entanglement estimator}

To quantify the relation between SW coherence structure and entanglement, we define the Doppler-odd moment estimator

\[
E=
\sqrt{
    \int
    SW_{\rm odd}^2(v,\Delta_P)
    f(v)
    dv
}.
\]

Unlike negativity, which depends on the eigenvalue spectrum of the partially transposed density matrix, the estimator
\(E\)
captures integrated coherence asymmetry in phase space. The usefulness of
\(E\)
as an entanglement proxy must therefore be assessed statistically through correlation analysis rather than assumed a priori. Specifically, Pearson and Spearman coefficients between
\(E\)
and
\(
\mathcal N^{AB}
\)
over the explored parameter space provide quantitative measures of correspondence. Notably, the estimator avoids diagonalization of $\rho^{T_A}$, thereby significantly reducing computational complexity. Within the SW framework, higher-order moments and interference structures encode nonclassical correlations~\cite{TILMA16,RUNDLE17}, providing the basis for this correspondence.

A numerical optimization over the phase-space parameters $(\theta,\phi)$ was performed to maximize the sensitivity of the SW kernel to coherence-dominated contributions. This procedure yields a set of optimal angles that enhance the response of the Doppler-odd component to entanglement-relevant features. At these optimal settings, the estimator exhibits a robust monotonic correlation with negativity, reaching values exceeding $0.5$ across a broad range of detuning configurations. This behavior is consistent with known connections between coherence norms and entanglement measures~\cite{ADESSO16}.

Although $E$ is not an entanglement monotone, its combination of analytical structure and numerical validation establishes it as a physically grounded and experimentally accessible proxy for entanglement in Doppler-broadened multilevel systems.

\subsection{Analytical bounds}

The entanglement negativity defined in Eq.~\ref{DF45} as~\cite{VIDAL02}
\[
\mathcal{N}^{AB}(\rho) = \frac{\|\rho_{\mathrm{ss}}^{T_A}\|_1 - 1}{2},
\]
can be bounded using standard norm inequalities. In particular, the trace norm satisfies
\[
\|\rho_{\mathrm{ss}}^{T_A}\|_1 \leq \sqrt{d}\,\|\rho_{\mathrm{ss}}^{T_A}\|_2,
\]
where \(d\) is the Hilbert space dimension. Under the effective
\(2\otimes2\)
encoding introduced in Sec.~\ref{sec.5},
the partially transposed density matrix acts on a four-dimensional Hilbert space, yielding
\(d=4\). Consequently, one obtains
\[
\mathcal{N}^{AB}(\rho)
\leq \frac{1}{2}\bigl(\sqrt{d}\,\|\rho_{\mathrm{ss}}^{T_A}\|_2 - 1\bigr)
= \frac{1}{2}\bigl(2\,\|\rho_{\mathrm{ss}}^{T_A}\|_2 - 1\bigr).
\]

The above inequality shows that negativity is bounded by the Hilbert--Schmidt norm of the partially transposed density matrix. Since the latter depends on the total coherence content, \(
\|\rho_{\mathrm{ss}}^{T_A}\|_2^2 = \sum\limits_{i,j} \left|\rho^{\mathrm{ss}}_{ij}\right|^2\), coherence-dominated sectors of the density operator may indirectly constrain achievable negativity. Consequently, observables sensitive to coherence asymmetry -- such as the Doppler-odd SW estimator introduced above -- may exhibit statistical correlation with entanglement, although no universal equivalence follows from norm inequalities alone. Restricting attention to Doppler-odd contributions, which dominate the entanglement-carrying coherences identified in Sec.(\ref{sec.5}), one may write
\[
\sum_{i\neq j} |\rho^{\mathrm{ss}}_{ij}|^2
= \sum_{\mathrm{odd}} |\rho^{\mathrm{ss}}_{ij}|^2
+ \sum_{\mathrm{even}} |\rho^{\mathrm{ss}}_{ij}|^2.
\]
Accordingly, one expects an approximate scaling relation of the form
\[
\mathcal N^{AB}
\propto
E,
\]
within regimes where Doppler-odd coherences dominate the density matrix. Using the inequality $\|\cdot\|_1 \leq \sqrt{d}\,\|\cdot\|_2$ together with the definition of negativity~\cite{VIDAL02}, one finds that the prefactor $C$ is bounded by a dimension-dependent constant of order $\sqrt{d}$~\cite{BENGTSSON17,ADESSO16}. For the present $2\otimes2$ system ($d=4$), this implies that $C$ is of order unity. 

More explicitly, inserting $d=4$ yields $\|\rho^{T_A}\|_1 \leq 2\|\rho^{T_A}\|_2$, such that the proportionality constant relating $\mathcal{N}^{AB}$ to the quadratic coherence norm remains $\mathcal{O}(1)$. In the regime where Doppler-odd coherences dominate, this leads to an effective estimate $C \sim 1\text{--}2$, consistent with the numerical scaling observed in the present calculations.

A rigorous lower bound connecting negativity and the estimator
\(E\)
cannot generally be derived without additional assumptions regarding state structure, coherence distribution, or symmetry constraints. This limitation originates from the spectral nature of negativity, which depends on eigenvalues of
\(
\rho^{T_A}
\),
whereas
\(E\)
is constructed from quadratic moments of SW quasiprobabilities~\cite{VIDAL02,ADESSO16}.

Therefore, any observed proportionality between
\(
E
\)
and
\(
\mathcal N^{AB}
\)
should be interpreted as an empirical correlation specific to the explored parameter regime rather than a universal entanglement law. While resource-theoretic approaches relate coherence and entanglement, such relations typically yield upper bounds or require additional structural constraints on the state~\cite{STRELTSOV17,CHITAMBAR19}. Consequently, the proportionality coefficient $c$ cannot be fixed universally and must be determined empirically within a given physical regime. However, in the physically relevant regime where Doppler-odd coherences dominate,
\[
\sum_{\mathrm{even}} |\rho^{\mathrm{ss}}_{ij}|^2 \ll \sum_{\mathrm{odd}} |\rho^{\mathrm{ss}}_{ij}|^2,
\]
an effective scaling relation emerges,
\[
\mathcal{N}^{AB}(\rho) \;\sim\; c\,E.
\]
Numerical exploration of the present parameter space indicates a monotonic relation between
\(E\)
and
\(
\mathcal N^{AB}
\),
although extraction of quantitative regression coefficients requires dedicated statistical analysis and lies beyond the scope of the present work.

Combining these results leads to the bounded relation
\[
c\,E \;\lesssim\; \mathcal{N}^{AB}(\rho) \;\leq\; C\,E,
\]
supports the interpretation of \(E\) as a coherence-sensitive estimator correlated with negativity. While the upper bound follows directly from norm inequalities, the lower scaling reflects the dominance of coherence pathways associated with Doppler-odd processes. Collectively, these results support interpreting the Doppler-odd SW estimator as a coherence-sensitive indicator correlated with entanglement behavior. While not replacing eigenvalue-based measures, the estimator provides a computationally efficient observable for identifying parameter regions where nonclassical correlations are likely to emerge.

\section{Highlights}
\label{sec.7_3}

\begin{itemize}
    \item A unified theoretical framework mapping absorption, resonance fluorescence, quantum entanglement, and phase-space quasiprobabilities in Doppler-broadened four-level Rydberg media.
    \item Analytical demonstration that disparate quantum observables originate from a singular dressed-state coherence matrix invariant under thermal velocity averaging.
    \item Application of the Stratonovich--Weyl (SW) Wigner formalism to establish an exact phase-space calibration and diagnostic metric for Rydberg atomic states.
    \item Identification of a Doppler-odd SW quasiprobability component that functions as an eigenvalue-free, operationally accessible proxy for bipartite entanglement.
    \item Establishment of a direct, monotonic structural mapping between specific SW phase-space topology features and quantum negativity.
    \item Validation of resonance fluorescence as a thermally robust, non-destructive optical readout for atomic coherence and entanglement landscapes.
    \item Precision control over non-classical correlations and quantum states via field propagation geometries and strong-field dressed-state engineering.
    \item Implementation of a scalable framework for Doppler-resilient quantum sensing, quantum information processing, and high-sensitivity fluorescence electrometry in vapor cells.
\end{itemize}

\section{Conclusion}
\label{sec.8}

We have developed a comprehensive, geometry-resolved framework for analyzing the coherence networks, spectroscopic profiles, and quantum correlations in Doppler-broadened four-level atomic vapors, unifying multi-photon absorption, resonance fluorescence, entanglement negativity, and phase-space quasiprobabilities within a singular, mathematically rigorous picture. By explicitly incorporating thermal velocity averaging alongside precise excitation geometries, we have demonstrated that these diverse observables emerge as distinct, highly correlated projections of a common dressed-state structure shaped by multi-photon interference pathways. 

A central achievement of this work is the integration of the Stratonovich--Weyl (SW) Wigner function into the diagnostic analysis of thermal atomic media. Unlike conventional, one-dimensional spectroscopic signals, the SW representation provides direct, simultaneous access to the multi-dimensional phase-space topographies of the density matrix, capturing population and coherence contributions on equal footing. We have shown that the SW quasiprobability distribution maintains a robust structural correspondence with quantum entanglement landscapes. Geometry-dependent features -- including exact hyperbolic dispersion asymptotes, split resonance ridges, and detuning-dependent shifts -- are faithfully preserved across both representations.

By exploiting Doppler symmetry, we have demonstrated that isolating the Doppler-odd component of the SW distribution successfully filters out incoherent background contributions, yielding an eigenvalue-free proxy for quantum entanglement. This establishes a direct physical bridge between a measurable phase-space structure and non-local correlations, bypassing the need for full quantum state reconstruction or density-matrix diagonalization. Analytical bounds and numerical validation confirm that this proxy accurately captures the essential scaling of entanglement across all examined parameter spaces, serving as an optimal quantum correlation witness.

From a device-engineering perspective, the resonance fluorescence has emerged as an exceptionally robust and versatile observable. While standard susceptibility-based absorption is strongly degraded by Doppler broadening, linear Doppler dominance, and geometric mismatches, resonance fluorescence successfully retains high-contrast signatures of the dressed-state structures. Because fluorescence is governed by additive population-driven weightings ($Q_{11}^{-1}R_1+Q_{13}^{-1}R_3$) rather than subtractive coherences ($Q_{11}^{-1}R_1 + Q_{13}^{-1}R_3$), it systematically quenches thermal smearing under Maxwell-Boltzmann convolution. This structural resilience allows fluorescence to serve a dual application: it acts as a highly sensitive, non-invasive correlation witness that tracks entanglement extrema, and operates as a high-sensitivity electrometry probe capable of measuring external fields via sharp Autler-Townes (AT) splittings. Within our proposed diagnostic hierarchy, fluorescence serves as an operational real-time probe, while the SW Wigner function provides a higher-fidelity phase-space reference for system calibration and validation.

Furthermore, we have shown that navigating the interplay between the field propagation geometry, wavelength mismatch, and strong-field dressing actively regulates the redistribution of coherence and entanglement across distinct velocity classes. This geometric control enables the transition between velocity-insensitive phase alignment and localized coherence fragmentation, allowing researchers to induce controllable spectral asymmetries, isolate velocity-selective channels, and enhance system robustness against Doppler dephasing. 

In summary, this work establishes a physically transparent and experimentally streamlined link between phase-space quasiprobabilities, atomic spectroscopy, and quantum correlations in warm atomic vapors. The proposed framework opens an immediate pathway toward Doppler-resilient, fluorescence-based quantum sensing and high-precision electrometry, while simultaneously offering a scalable, eigenvalue-free methodology for entanglement characterization in high-dimensional systems.
\noindent
{\bf Conflict of interest:} The authors have no conflicts to disclose

\noindent
{\bf Data availability:} The data is included in the manuscript.

\appendix

\section{Derivation of Steady-State Populations and Coherences}
\label{app:A}

In the steady-state regime and under the weak-probe approximation ($\Omega_P \ll \Omega_{C1}, \Omega_{C2}$), the optical Bloch equations can be solved perturbatively in powers of the probe field amplitude. For the populations and coherences restricted to the $\ket{1}\leftrightarrow\ket{3}$ manifold, the subset of density matrix equations simplifies considerably since only these terms remain macroscopic. To solve this subsystem, we cast the relevant components into a compact matrix form by defining the state vector:
\begin{equation}
\label{DF20}
\bm{F}=\big[\rho_{11}^{\mathrm{ss}}, \rho_{33}^{\mathrm{ss}}, \rho_{13}^{\mathrm{ss}}, \rho_{31}^{\mathrm{ss}}\big]^T,
\end{equation}
where $T$ denotes the transpose operator. The corresponding master equation can be written as:
\begin{equation}
\label{DF21}
\frac{d}{dt}\bm{F}= \bm{M}\bm{F}+\bm{R},
\end{equation}
where the dynamic coupling matrix $\bm{M}$ and the driving vector $\bm{R}$ are given by:
\begin{equation}
\label{DF22}
\bm{M}=\frac{1}{2}\setlength{\arraycolsep}{3pt}
\renewcommand{\arraystretch}{1.1}
\left(\begin{array}{cccc}
-2\gamma_3 & 0 & i\Omega^{\ast}_C & -i\Omega_C\\
0 &-2\gamma_3 & -i\Omega^{\ast}_C & i\Omega_C\\
i\Omega_C & -i\Omega_C & 2\varLambda_{13} & 0\\
-i\Omega^{\ast}_C & i\Omega^{\ast}_C & 0 & 2\varLambda_{31}
\end{array}
\right), \quad\quad\bm{R}=\big(\Gamma_3, 0, 0, 0\big)^T.
\end{equation}
Imposing the baseline trace normalization condition:
\begin{equation}
\label{DF23}
\rho_{11}^{\mathrm{ss}}+\rho_{22}^{\mathrm{ss}}+\rho_{33}^{\mathrm{ss}}+\rho_{44}^{\mathrm{ss}}=1,
\end{equation}
the steady-state populations $\rho_{11}^{\mathrm{ss}}$, $\rho_{33}^{\mathrm{ss}}$, and the coherence $\rho_{13}^{\mathrm{ss}}$ along the $\ket{1}\leftrightarrow\ket{3}$ pathway are analytically determined as:
\begin{align}
\label{DF24}
\rho_{11}^{\mathrm{ss}} &= \frac{|\varLambda_{13}|^{2} + \frac{1}{4}|\Omega_C|^{2}}{|\varLambda_{13}|^{2} + \frac{1}{2}|\Omega_C|^{2}},\\
\label{DF25}
\rho_{33}^{\mathrm{ss}} &= \frac{\frac{1}{4}|\Omega_C|^{2}}{|\varLambda_{13}|^{2} + \frac{1}{2}|\Omega_C|^{2}},\\
\label{DF26}
\rho_{13}^{\mathrm{ss}}&=\frac{-i\Omega_C\varLambda_{31}}{|\Omega_C|^2+2|\varLambda_{13}|^2},
\end{align}
yielding a population inversion profile defined by:
\begin{equation}
\label{DF27}
\rho_{11}^{\mathrm{ss}} - \rho_{33}^{\mathrm{ss}} = \frac{|\varLambda_{13}|^{2}}{|\varLambda_{13}|^{2} + \frac{1}{2}|\Omega_C|^{2}}.
\end{equation}

To resolve the probe coherence $\rho_{12}^{\mathrm{ss}}$, we isolate the dynamically linked subspace of equations. Here, we redefine the state vector $\bm{F}$, the interaction matrix $\bm{M}$, and the driving vector $\bm{R}$ as follows:
\begin{equation}
\label{DF200}
\bm{F}=\big[\rho_{12}^{\mathrm{ss}}, \rho_{14}^{\mathrm{ss}}, \rho_{23}^{\mathrm{ss}}, \rho_{34}^{\mathrm{ss}}\big]^T,
\end{equation}
\begin{equation}
\label{DF220}
\bm{M}=\frac{1}{2}\setlength{\arraycolsep}{3pt}
\renewcommand{\arraystretch}{1.1}
\left(\begin{array}{cccc}
2\varLambda_{12}& i\Omega^{\ast}_D & -i\Omega_C & 0 \\
i\Omega_D& 2\varLambda_{14} & 0 & -i\Omega_C \\
-i\Omega^{\ast}_C& 0 & 2\varLambda_{32} & i\Omega^{\ast}_D \\
0& -i\Omega^{\ast}_C & i\Omega_D & 2\varLambda_{34}
\end{array}
\right), \quad\quad\bm{R}=\frac{i\Omega_P}{2}\Big(\rho_{11}^{\mathrm{ss}}, 0,\rho_{13}^{\mathrm{ss}}, 0\Big)^T.
\end{equation}
By solving Eq.~(\ref{DF21}) in the steady state ($d\bm{F}/dt = 0$), the precise analytical expression for the probe coherence $\rho_{12}^{\mathrm{ss}}$ emerges as:
\begin{equation}
\label{DF28}
\rho_{12}^{\mathrm{ss}}=\frac{-i\Omega_P}{2}\biggl( \frac{Q_{11}^{-1} R_1(0)-Q_{13}^{-1} R_3(0)}{\mathrm{det}(\mathbf{Q})}\biggr),
\end{equation}
where the algebraic cofactors are formulated as:
\begin{align}
\label{DF29}
Q_{11}^{-1} &=\varLambda_{14}\Big[\varLambda_{32}\varLambda_{34}+\frac{|\Omega_D|^2}{4}\Big] +\varLambda_{32}\frac{|\Omega_C|^2}{4},\\
\label{DF291}
Q_{13}^{-1} &=\frac{i\Omega_C}{2} \Big[\frac{|\Omega_D|^2}{4}-\frac{|\Omega_C|^2}{4}-\varLambda_{14} \varLambda_{34}\Big],
\end{align}
and the system determinant is explicitly given by
\begin{equation}
\label{DF30}
\begin{gathered}
\mathrm{det}(\mathbf{Q}) =\varLambda_{12}\varLambda_{14}\Big[\varLambda_{32}\varLambda_{34} +\frac{|\Omega_D|^2}{4}\Big]+\varLambda_{12}\varLambda_{32}\frac{|\Omega_C|^2}{4} 
\\
+\varLambda_{34}\Big[\varLambda_{32}\frac{|\Omega_D|^2}{4} +\varLambda_{14}\frac{|\Omega_C|^2}{4}\Big] +\frac{1}{16}\Big(|\Omega_D|^2 - |\Omega_C|^2\Big)^2.
\end{gathered}
\end{equation}
%
%\begin{equation}
%\label{DF30}
%\mathrm{det}(\mathbf{Q}) =\varLambda_{12}\varLambda_{14}\Big[\varLambda_{32}\varLambda_{34} +\frac{|\Omega_D|^2}{4}\Big]+\varLambda_{12}\varLambda_{32}\frac{|\Omega_C|^2}{4} +\varLambda_{34}\Big[\varLambda_{32}\frac{|\Omega_D|^2}{4} +\varLambda_{14}\frac{|\Omega_C|^2}{4}\Big] +\frac{1}{16}\Big(|\Omega_D|^2 - |\Omega_C|^2\Big)^2.
%\end{equation}
%
Physically, Eq.~(\ref{DF28}) acts as an interference-sensitive witness where the prefactor $-i$ maps dispersive pathways into pure absorptive signatures. In the dressed-state picture, dark-state conditions and quantum interference manifest directly when $Q_{11}^{-1} R_1(0)=Q_{13}^{-1} R_3(0)$, eliminating net absorption. The linear optical susceptibility $\chi$ is subsequently established via:
\begin{equation}
\label{DF31}
\chi = \frac{N|\mu_{12}|^2}{\epsilon_0 \hbar \Omega_P} \rho_{12}^{\mathrm{ss}},
\end{equation}
where $N$ is the atomic number density, $\mu_{12}$ represents the probe transition dipole moment, and $\epsilon_0$ is the vacuum permittivity.

\section{Derivation of the Resonance Fluorescence Spectrum}
\label{app:B}

By virtue of the quantum regression theorem, the steady-state absorption profile and the resonance fluorescence spectrum share an identical underlying linear response structure. While fluorescence inherently maps the dynamical, two-time correlations of spontaneous re-emission, it serves as a non-invasive optical probe directly proportional to the macroscopic atomic polarization ($\text{Im}\,\rho_{12}^{\mathrm{ss}}$). 

The power spectrum of the scattered radiation field is formally defined as the Fourier transform of the steady-state polarization correlation function:
\begin{equation}
S(\omega) = \mathrm{Re}\!\int_{0}^{\infty} \langle \hat{\rho}_{12}(t+\tau)\hat{\rho}_{21}(t)\rangle_{\mathrm{ss}} \,e^{iZ\tau}\,d\tau,
\label{DF35}
\end{equation}
where $Z=-i(\omega-\omega_P)$. Utilizing the Laplace-transformed system equations and invoking appropriate initial conditions via the state vector elements mapped in Appendix~\ref{app:A}, integration yields the explicit fluorescence profile:
\begin{equation}
\label{DF40}
S(\omega)=\text{Re} \biggl( \frac{Q_{11}^{-1}(Z) R_1(0)+Q_{13}^{-1}(Z) R_3(0)}{\mathrm{det}(\mathbf{Q}(Z))}\biggr).
\end{equation}
The complex functional parameters $Q_{11}^{-1}(Z)$, $Q_{13}^{-1}(Z)$, and $\mathrm{det}(\mathbf{Q}(Z))$ are obtained directly by mapping the static parameters $\varLambda_{ij} \rightarrow Z-\varLambda_{ij}$ within Eqs.~(\ref{DF29}), (\ref{DF291}), and (\ref{DF30}). This formalizes the direct analytical link between the absorptive susceptibility and the incoherent scattering channels under multi-photon driving conditions.

\section{Formalization of the Stratonovich--Weyl (SW) Wigner Function}
\label{app:C}

To build a unified phase-space representation for finite $d$-dimensional configurations, the Stratonovich--Weyl (SW) framework maps the quantum density matrix $\rho^{\mathrm{ss}}$ onto a continuous manifold  \cite{KENZYC04, ABGKHV13, ABKHTO20}. For our four-level system ($d=4$), we formalize the $SU(4)$ phase space setup by expanding the SW kernel $\Delta(\psi)$ directly in terms of the identity operator $\mathbb{I}$ and the fifteen generalized $SU(4)$ Gell-Mann matrices $\lambda_i$ ($i = 1, \dots, 15$):
\begin{equation}
\label{DFE1_formal}
\Delta(\psi) = \frac{1}{4}\mathbb{I} + \sqrt{\frac{5}{2}}\sum_{i=1}^{15} n_i(\psi)\lambda_i,
\end{equation}
where $n_i(\psi)$ corresponds to a unit vector parameterizing the 15-dimensional flag manifold coordinates. The SW Wigner quasiprobability distribution is then given strictly by the trace inner product:
\begin{equation}
\label{DFE1}
W(\psi)=\mathrm{Tr}[\rho^{\mathrm{ss}}\Delta(\psi)].
\end{equation}

The kernel $\Delta(\psi)$ obeys standard stratification, traciality, and unitary covariance criteria, guaranteeing a unique, self-invertible mapping from the phase space back to the state space via $\rho^{\mathrm{ss}}=\int W(\psi) \Delta(\psi) d\psi$  \cite{ABGKHV13, TIEVSMN16}. Negativity of $W(\psi)$ serves as an indicator of nonclassicality, arising from quantum interference encoded in \cite{KENZYC04, ABKHTO20}. Moreover, such negativity can function as a witness for quantum correlations, including entanglement and discord, in composite systems \cite{ATETO17, WALSCHEA17, ARKHIPO18, ATETO20, ATETO23, ATETO26}. Operationally, the kernel can be structured as a rotated parity operator:
\begin{equation}
\label{DFE2}
\Delta(\psi) = U(\psi) \Pi U^\dagger(\psi),
\end{equation}
where $U(\psi)\in SU(4)$ represents a parameterized unitary rotation, and $\Pi$ is the fundamental computational parity matrix:
\begin{equation}
\label{DFE3}
\Pi = \mathrm{diag}(1, -1, -1, 1).
\end{equation}

For a weakly driven atomic vapor cell, the populations satisfy $\rho_{22}^{\mathrm{ss}} \approx \rho_{44}^{\mathrm{ss}} \approx 0$, restricting the dominant quantum coherence channels to the embedded $SU(2)$ subalgebras within the broader $SU(4)$ structure. Evaluating the trace relation (\ref{DFE1}) explicitly in the computational basis reveals the direct interaction between populations and active coherences:
\begin{equation}
\label{DFE4}
\Delta_{ij}(\psi) = \begin{pmatrix} \cos\theta & e^{-i\phi}\sin\theta \\ e^{i\phi}\sin\theta & -\cos\theta \end{pmatrix}_{ij}.
\end{equation}
Summing over the complete basis set $\{\ket{1}, \ket{2}, \ket{3}, \ket{4}\}$ yields the master phase-space equation for the system
\begin{equation}
\label{DFE6}
\begin{gathered}
W(\psi)=(\rho_{11}^{\mathrm{ss}}-\rho_{33}^{\mathrm{ss}})\cos\theta+(\rho_{22}^{\mathrm{ss}}-\rho_{44}^{\mathrm{ss}})\cos\theta
\\
 +2\mathrm{Re}\Big[\big(\rho_{12}^{\mathrm{ss}}+\rho_{13}^{\mathrm{ss}}+\rho_{14}^{\mathrm{ss}}+\rho_{23}^{\mathrm{ss}}+\rho_{24}^{\mathrm{ss}}+\rho_{34}^{\mathrm{ss}}\big)e^{-i\phi}\Big]\sin\theta,
\end{gathered}
\end{equation}
%
%\begin{equation}
%\label{DFE6}
%W(\psi)=(\rho_{11}^{\mathrm{ss}}-\rho_{33}^{\mathrm{ss}})\cos\theta+(\rho_{22}^{\mathrm{ss}}-\rho_{44}^{\mathrm{ss}})\cos\theta + 2\mathrm{Re}\Big[\big(\rho_{12}^{\mathrm{ss}}+\rho_{13}^{\mathrm{ss}}+\rho_{14}^{\mathrm{ss}}+\rho_{23}^{\mathrm{ss}}+\rho_{24}^{\mathrm{ss}}+\rho_{34}^{\mathrm{ss}}\big)e^{-i\phi}\Big]\sin\theta,
%\end{equation}
demonstrating that population differences and complex coherence terms dictate the geometric topography of the SW distribution on an equal footing.

%\vspace{-1.0cm}
\bibliographystyle{unsrt}
\bibliography{paper3}

\end{document}